\documentclass[11pt]{article}
 \usepackage[left=2.5cm, right=2.5cm,top=2.5cm, bottom=2.5cm]{geometry}
\usepackage{bm, amsmath, bbold, amssymb, array, lipsum, latexsym}
\usepackage{mathtools}
\usepackage{accents}
\usepackage{setspace}
\usepackage{placeins}
\usepackage{multirow} 
\usepackage{soul}
\usepackage{natbib}
\usepackage{graphicx}
\usepackage{enumitem}
\graphicspath{{pics/}}
\usepackage{appendix}
\usepackage{algorithm2e}
\usepackage[skins]{tcolorbox}
\RestyleAlgo{ruled}
\usepackage{pdflscape}
\usepackage{float}
\usepackage{xcolor}
\usepackage{authblk}
\usepackage{colortbl}
\usepackage{rotating}

\usepackage{tikz}
\usetikzlibrary{shapes.geometric,arrows.meta}
\usetikzlibrary{positioning}
\usetikzlibrary{calc}

\tikzstyle{diam} = [diamond, aspect=2, draw, text width=6em,text centered, 
]
\tikzstyle{output} = [rectangle, draw, double, double distance=1mm,
text width=2cm, text centered, minimum height=2em, 
]
\tikzstyle{input_narrow} = [trapezium, trapezium left angle=70, trapezium right angle=110, minimum height=2em, text width=1.7em, text centered, draw
]
\tikzstyle{input_wide} = [trapezium, trapezium left angle=70, trapezium right angle=110, minimum height=2em, text width=4em, text centered, draw
]
\tikzstyle{proc} = [rectangle, rounded corners, minimum width=2cm, minimum height=1cm, text width=2cm, text centered, draw
]
\tikzstyle{param} = [circle, minimum width=2cm, minimum height=2cm, text width=1.5cm, text centered, draw, dashed
]
\tikzstyle{line} = [draw, -latex]
\tikzstyle{arrow} = [thick,->,>=stealth]

\definecolor{lightgray}{gray}{0.9}
\definecolor{lightgray1}{gray}{0.8}

\def\spacingset#1{\renewcommand{\baselinestretch}%
	{#1}\small\normalsize} \spacingset{1}

\usepackage[T1]{fontenc}
\usepackage[utf8]{inputenc}

\usepackage{amsmath}

\usepackage{hyperref}

\newcommand{\bzero}{ { \mathbf{0} }}
\newcommand{\bA}{ { \bf A }}
\newcommand{\bB}{ { \bf B }}

\newcommand{\bI}{ { \bf I }}

\newcommand{\bR}{ { \bf R }}
\newcommand{\bs}{ { \bf s }}
\newcommand{\bss}{ { \boldsymbol{s} }}

\newcommand{\bS}{ { \bf S }}

\newcommand{\bW}{ { \bf W }}

\newcommand{\bV}{ { \bf V }}

\newcommand{\bx}{ { \bf x }}
\newcommand{\bY}{ { \bf Y }}

\newcommand{\bst}{ { \boldsymbol{t} }}

\newcommand{\bsv}{ { \boldsymbol{v} }}

\newcommand{\bsV}{ { \boldsymbol{V} }}

\newcommand{\bsx}{ { \boldsymbol{x} }}
\newcommand{\bsY}{ { \boldsymbol{Y} }}
\newcommand{\bsy}{ { \boldsymbol{y} }}

\newcommand{\balpha}{ { \boldsymbol{\alpha} }}
\newcommand{\beps}{ { \boldsymbol{\epsilon} }}
\newcommand{\btheta}{ { \boldsymbol{\theta} }}
\newcommand{\bvtheta}{ { \boldsymbol{\vartheta} } }
\newcommand{\bTheta}{ { \boldsymbol{\Theta} }}
\newcommand{\bdelta}{ { \boldsymbol{\delta} }}

\newcommand{\bmu}{ { \boldsymbol{\mu} }}
\newcommand{\bphi}{ { \boldsymbol{\phi} }}

\newcommand{\bbeta}{ { \boldsymbol{\beta} }}
\newcommand{\bsigma}{ { \boldsymbol{\sigma} }}
\newcommand{\bSigma}{ { \boldsymbol{\Sigma} }}
\newcommand{\bGamma}{ { \boldsymbol{\Gamma} }}

\newcommand{\bomega}{ { \boldsymbol{\omega} }}

\newcommand{\mfr}{ \mathfrak{r} }
\newcommand{\mfR}{ \mathfrak{R} }

\newcommand{\ut}[1]{\underaccent{\tilde}{#1}}
\renewcommand{\vec}[1]{\ut{#1}}
\newcommand{\dbtilde}[1]{\widetilde{\raisebox{0pt}[0.85\height]{$\widetilde{#1}$}}}

\newcommand{\MVN}{ \text{MVN} }

\newcommand{\pacc}{ \bar{p}^*_{\text{acc}} }

\newcommand{\mcJ}{\mathcal{J}}
\newcommand{\obs}{\mathrm{obs}}
\newcommand{\ABC}{\mathrm{ABC}}

\newcommand{\E}{\mathrm{E}}
\newcommand{\model}[1]{ { \mathcal{M}_{#1} } }
\newcommand{\ECR}{ {\mathrm{ECR}} }
\newcommand{\OECS}{ {\mathrm{OECS}} }

\bibpunct{(}{)}{;}{a}{,}{,}

\providecommand{\keywords}[1]
{
  \textbf{\textit{Keywords---}} #1
}

\definecolor{darkgreen}{HTML}{009F00}

\definecolor{teal}{RGB}{0,128,128}

\title{Modeling Zero-Inflated Correlated Dental Data \\ through Gaussian Copulas and \\ Approximate Bayesian Computation}
\author[1]{Anish Mukherjee}
\author[1]{Jeremy T.\ Gaskins}
\author[2]{Shoumi Sarkar}
\author[3]{Steven Levy}
\author[2]{Somnath Datta}
\affil[1]{Department of Bioinformatics and Biostatistics, University of Louisville}
\affil[2]{Department of Biostatistics, University of Florida}
\affil[3]{Department of Preventive and Community Dentistry and Department of Epidemiology, University of Iowa}
\date{}

\begin{document}
% \spacingset{1.9} % DON'T change the spacing!
\def\spacingset#1{\renewcommand{\baselinestretch}%
{#1}\small\normalsize} \spacingset{1.5}

\maketitle

% \org{
% TO DO:
% \begin{enumerate}
%     \item Edit ABC-MCMC section to represent the algorithm we are using
%     \item Add IFS data analysis.  Finish introducing the data structure including fixed effects used and the choice of association relationship.  Some comments on model fit/mixing and a discussion of the interpretations of parameters.
%     \item Possibly include simulated dataset results, but this is less important than the others.
%     \item Brief conclusion.  Can include comments on difficulty of sampling and some of the strategies we've tried.  Also include some discussion of remaining aspects to the project.
%     \item All of the tables and other things at the end that aren't currently a part of the manuscript have to be incorporated or removed.  
% \end{enumerate}
% }

\begin{abstract}
    We develop a new longitudinal count data regression model that accounts for zero-inflation and spatio-temporal correlation across responses.  
    This project is motivated by an analysis of Iowa Fluoride Study (IFS) data, a longitudinal cohort study with data on caries (cavity) experience scores measured for each tooth across five time points. 
    To that end, we use a hurdle model for zero-inflation with two parts: the presence model indicating whether a count is non-zero through logistic regression and the severity model that considers the non-zero counts through a shifted Negative Binomial distribution  allowing overdispersion. 
    To incorporate dependence across measurement occasion and teeth, these marginal models are embedded within a Gaussian copula that introduces  spatio-temporal correlations.  
    A distinct advantage of this formulation is that it allows us to determine covariate effects with population-level (marginal) interpretations in contrast to mixed model choices. 
    Standard Bayesian sampling from such a model is infeasible, so we use approximate Bayesian computing  for inference. 
    % Our ABC strategy involves an initial MCMC-based algorithm under an independence assumption to obtain a proposal distribution for the ABC-MCMC algorithm.  
    This approach is applied to the IFS data to gain insight into the risk factors for dental caries and the correlation structure across teeth and time.
\end{abstract}


% Things to check for consistency across full manuscripts:
% \begin{itemize}
%     \item NB capitalization.  Either always use Negative Binomial or always use Negative Binomial.
%     \item non-zero or non-zero, ABC-MCMC, In the case of
%     \item $z$-transformation or $z$-transformation
% \end{itemize}

\section{Introduction}

It is  common in medical studies to collect longitudinal or clustered data with measurements belonging to an individual or group correlated over time. 
Additionally, measurements can be spatially dependent when recorded at multiple locations at the same time point. 
For instance, in long-term  dental studies, 
% we typically measure some indicator of dental health, and the measurements are often believed to have complex spatio-temporal dependence structure.
% In particular, 
scores describing  tooth health are expected to be relatively similar at adjacent time points, as are the scores of horizontally- and vertically-adjacent teeth at the same time point,
leading to a complex spatio-temporal dependence structure. 
% Teeth, which are vertically adjacent in closed mouth position, also often maintain similar dental health.
It is, therefore, important  to formulate such a dependence structure in terms of clinically-relevant adjacency relations to  assess their significance.
% would be integral part of the inference.
% Furthermore the mean structure, in a typical modeling approach, is expressed as a linear function of relevant predictors. 
% Estimating their effects at population-level would provide useful interpretation about their overall nature of association.

When the data distribution is non-normal, such as with discrete counts, there are two main modeling approaches  for incorporating dependence:
% used that have been used  in the literature
generalized linear mixed effects models (GLMM) and generalized estimating equations (GEE).
In a hierarchical GLMM framework, the non-normal data are modeled via different link functions, and the linear predictor is specified in terms of fixed and random effects. 
Modeling zero-inflated data can be accomodated in GLMM frameworks by opting for a two-component mixture model like a zero-inflated model or hurdle model. 
% A Poisson, Nengative Binomial (NB), or Conway-Maxwell-Poisson (CMP) distribution \citep{Conway1962} may be used to model the count data depending on the dispersion level.
\cite{Choo-Wosoba2016, Choo-Wosoba2018} explored both zero-inflated and hurdle models using the Conway-Maxwell-Poisson (CMP) distribution for modeling the Iowa Fluoride Study (IFS) data. 
A longitudinal CMP model with excess zeros has also been proposed by \cite{Kang2021} in a Bayesian setting, where the correlations among the caries scores for different teeth were introduced via random effects.
%\org{[describe its main feature]}

% Due to the hierarchical setup GLMMs can be easily formulated in a Bayesian setting and can be  fit using  established Markov chain Monte Carlo (MCMC) algorithms. 
% But, as discussed by \cite{Neuhaus1991} and  others, the fixed effects in this model are determined conditionally on the random effects, and in the case of a non-identity link function, their coefficient estimates do not lend themselves to population-level interpretation. 
While a GLMM (with non-identity link function) can be easily formulated in a Bayesian setting due to its hierarchical specification, the fixed effects in this model do not lend themselves to population-level interpretation  \citep{Neuhaus1991}.
On the other hand, the GEE approach models the mean response of these repeated measurements directly in terms of the marginal effects, and a working correlation matrix is used to define the dependence structure. 
This modeling approach, however, does not include a full likelihood specification, making it ill-suited for Bayesian implementation. 
% See \cite{Zeger1992, Gardiner2009, Zhang2012} for discussion on the differences between GLMM and GEE approaches. 
Since characterizing the marginal/population-level effects of predictors within a coherent Bayesian framework is often of interest, neither GLMM or GEE  are satisfactory approaches here.

As an alternative to GLMMs inducing correlation through random effects, we consider copula-based dependence modeling  to construct a joint structure for  multivariate modeling. 
See \cite{Kolev2009} for a  survey of copula-based regression models. 
% The basic  idea is to choose univariate models specific to each margin and then imbed this in a copula to form multivariate dependence \citep{Sklar1959}. 
% The representation of a multivariate distribution as a composition of a copula and its univariate margins was first introduced by \cite{Sklar1959}. 
This modeling framework is arguably more flexible than GLMM in the sense that the parameters associated with the dependence between observations are separate from the parameters associated with the mean response, whereas in the GLMM the random effects impact both.
% that modeling the univariate margins can be made arbitrarily complex, typically each margin having its own unique set of parameters describing the mean structure; while the dependence across the margins can be modeled separately with its own set of parameters. 
% The copula margins are specified , distinct from those involved in the specification of the the dependence structure.
% \org{[unique in the sense that each margin different or unique from the dependence parameters?]}
% while the specification of the dependence structure involves a completely different parameter set.
Here, a Gaussian copula  is employed, where each repeated measurement from an individual corresponds to a copula margin, specified as a Negative Binomial (NB) hurdle model to account for over-dispersion. 
A distinctive feature of our model compared to most copula-based regression models is that the marginal distributions are connected through sharing a common set of parameters.
% We choose a hurdle  distribution for the response variable to account for over-dispersed counts.
% , and we use global-local shrinkage priors on the regression parameters to perform variable selection. 
Flexible and interpretable dependence within the copula is determined by  a Simultaneous Autoregessive model \citep[SAR;][]{Banerjee2004} to account for multiple types of adjacency relationships.

% Consequently, we choose to employ a copula-based approach to obtain marginal population level inference while determining dependence through a Gaussian copula.
% that can incorporate many features of count data encountered in practice. 

To perform Bayesian inference for models involving copula-based dependence structures, different Markov chain Monte Carlo (MCMC) algorithms have been suggested. 
% \org{[if this is first use of MCMC, it must be defined]}
% \cite{Min2010, Min2011} implemented MCMC methods for pair-copula constructions with continuous and discrete time data. 
\cite{Pitt2006} proposed a data augmentation MCMC scheme 
% based on a data augmentation approach 
for Gaussian copulas, which was extended by \cite{Smith2012} for non-elliptical copulas. 
% Despite the efforts, implementation of MCMC for an arbitrarily complex copula structure still remains a challenge.
% \org{[potentially highlight that marginal parameters are not specific to each margin which is why some of these methods fail for us]}
Since our modeling involves discrete count data and shares parameters across margins, these algorithms are not applicable,
% having fairly complex dependence structure, 
% computing the likelihood for sampling is not feasible,
and data augmentation MCMC does not mix effectively.
% computing the likelihood turns out to be fairly difficult making a fully Bayesian estimation approach very challenging.
Thus, we employ 
% a likelihood-free
a Approximate Bayesian Computation \citep[ABC;][]{Sisson2018} for posterior inference.
% \org{[this paragraph still needs a bit of work]}

% \notes{Layout the organization for the rest of the paper.}

The rest of this article is arranged as follows. 
In Section \ref{sec:motivation}, we describe the motivating data that necessitate our modeling setup. 
The proposed model is presented in Section \ref{sec:model}, followed by a description of the posterior computation scheme in Section \ref{sec:computation}. 
We then employ our method to analyze the IFS data in Section \ref{sec:IFS_study}. 
Section \ref{sec:simulation} presents a thorough simulation analysis. 
The main manuscript ends with a discussion in Section \ref{sec:discussion} that briefly summarizes the applicability of our approach, potential extensions, and future directions.
Additional details on our proposed approach, along with further empirical details, can be found in the supplementary materials document.
% associated with this main manuscript.

\section{Iowa Fluoride Study  Data} \label{sec:motivation}

This work is motivated by the Iowa Fluoride Study (IFS). 
% Loss of tooth substance, a disease known as dental caries, occurs as the tooth-adherent bacteria produce acid by metabolizing sugarsthat over time demineralizes the tooth structure and causes cavities.
One of the main objectives of this cohort study is to investigate the associations between the outcome variable of dental caries and different risk 
% (e.g., consumption of sugary beverages) 
and protective factors, including tooth-brushing, fluoride ingestion, and consumption of sugary beverages
% Earlier analyses of the IFS data by 
\citep{Levy2003, Broffitt2013}.
% have shown that high intake of sugary beverages is detrimental to dental health, while 
% water and milk consumption, 
% frequent tooth-brushing and fluoride in toothpaste have protective effects.
The  cohort was born between 1992 and 1995. 
A caries experience score for each observable tooth surface was recorded when each participant received dental examinations at ages 5, 9, 13, 17 and 23. 
This caries status of each tooth surface was determined by trained and calibrated dentists. 
Sound and filled surfaces that had non-cavitated (incipient) caries were scored as 1, and surfaces with definitive, cavitated (frank) caries or missing due to caries were scored 2. 
The integer-valued caries experience score for each tooth was found by summing these scores across all tooth surfaces, with higher scores indicating more caries experience.
%The cavitated-level caries experience scores for each observable tooth \citep[page 157]{MERCHAN2021154}
%%The caries experience scores for each observable tooth 
%were recorded when each participant received a dental examination at ages 5, 9, 13, 17 and 23. 
%This caries status was determined by calibrated dentists, and
%% on each dental surface zone of each tooth.
%% Based on these, 
%an integer-valued caries experience score was assigned to each tooth, with higher scores indicating more caries experience. 
% Based on the caries status determined by a calibrated dentist, each tooth is assigned an integer-valued caries experience score, with higher scores indicating more damage. 
The number of patients observed varies across ages from 696 at age 5 to 342 at age 23. 
The total number of available tooth-specific caries scores in the dataset is 64,926.

%and by measurement occasion the numbers are 
%with 13,751, 14,623, 14,523, 12,667 and 9362 teeth observed at ages 5, 9, 13, 17 and 23, respectively.
%\anish{(cut out age wise teeth information from here)}.
% \org{[commas for 5+ digit numbers]}

% We comment on  few key characteristics of the dataset. 
Since all  teeth of an individual share similar environmental factors, caries scores are expected to exhibit some form of dependence. 
We believe caries scores to be temporally correlated, and the correlation among scores within same time point/dental observation are expected to be related to the distance between the corresponding teeth and/or other features of dental anatomy.
% Caries scores are temporally correlated as we expect  scores for the same tooth  to be similar across time. 
% Within the same time point/dental observation, teeth are also expected to be correlated, but the form of such dependence may be related to distance between teeth or other features of periodontal anatomy.
% The spatially adjacent teeth are also expected to be more highly correlated than teeth further away. 
Therefore, we desire a flexible specification of the  complex correlation structure across the longitudinally- and spatially-related scores. 
Additionally, 
the set of observed teeth changes across  ages
% there are not measurements on every tooth of an individual at all  Figure \ref{fig:appn_dentition}s 
due to the mixed dentation seen in late childhood/early adolescence, as primary teeth are replaced by permanent teeth. 
% While primary teeth usually fall out by age thirteen, permanent teeth do not start erupting before nine.
% Since primary teeth usually fall out by age thirteen, one cannot characterize  caries for  primary teeth after age thirteen. 
% Similarly, no caries data is recorded on permanent teeth before nine, since these teeth have yet to erupt. 
% Another feature of the data is zero inflation as depicted in Figure \ref{fig:zero-inflation}. 
% It is evident that the distribution of the non-zero caries scores is dominated by the high proportion of zeros.
% \begin{figure}[tb]
%     \centering
%     \includegraphics[width=0.65\textwidth]{pics/IFS_overview.png}
%     \caption{The figure in the left panel shows the overall distribution of data demonstrating clear zero inflation, and the figure on the right shows distribution of the non-zero data indicating potential over-dispersion.
%     }
%     \label{fig:zero-inflation}
% \end{figure}
A key feature of the caries scores is the high proportions of zero counts across all ages with 93\% of teeth having no caries overall (see Figure \ref{fig:zero-inflation} and Table \ref{tab:appn_data_summary} in Appendix \ref{appn:IFS_setup}).
% The overall proportion of zero counts in the data is 93\%,
% (see Figure \ref{fig:zero-inflation})
% while that at ages 5, 9, 13, 17, and 23 are approximately 95\%, 95\%, 96\%, 88\%, 92\% respectively, indicating zero-inflation overall and at all the time points.
Moreover, the distribution of the positive counts 
% appear to be highly positively skewed with a longer tail 
indicates potential over-dispersion.

% It is now convenient to mention that, the copula framework introduced earlier is rich enough to accommodate all these features. To summarize, 
% In light of the above data features, 
Therefore, analysis of the IFS data requires statistical methodology that can accomodate:
% the following desiderata:
(1) zero-inflation; 
(2) structural missingness associated with mixed dentation;
% in the data arising due to the nature of the study must be allowed in the model, 
(3) a flexible  dependence structure that accounts for the  multiple adjacency relations relevant to dental anatomy; and 
(4)  population-level interpretations of the predictor effects to assist clinicians in selecting measures to improve dental health. 
In the next section we propose a modeling strategy that is able to achieve all of these.
% In the following section, we  present our proposed model that possesses all these features.
% , and then describe the computational scheme to do posterior inference.

\section{Hurdle Count Model with Latent Copula Structure} \label{sec:model}

\subsection{Marginal Response Model}

We let $Y_{ij}$ denote the $j$-th measurement of the non-negative integer-valued outcome for the $i$-th individual ($i=1,\ldots,n$; $j=1,\ldots,J$).  We let $\mcJ=\{1,\ldots,J\}$ denote the full set of potential measurements, which are aligned in the sense that the $j$-th response for the $i$-th and $i'$-th patients correspond to the caries scores of the same tooth measured at the same time point. Here, $j$ is indexing both the spatial location of the tooth inside the mouth, as well as the longitudinal time point.  When convenient, we consider the pair $(l_j,t_j)$ to designate the time point $t_j\in\mathcal{T}=\{1,\ldots,T\}$ and the location/tooth $l_j\in\mathcal{L}=\{1,\ldots,L\}$.  
% Each $j\in\mcJ$ corresponds to a unique combination of $(l_j,t_j)$.

Let $\boldsymbol{x}_{ij}$ be the $(d+1)$-dimensional vector of fixed effects predictors with $x_{ij,0} = 1$ for a  model intercept.  
Since a zero count indicates a healthy tooth and implies  a single source of zeros in our data, we opt for a hurdle model. 
Let $Z_{ij} = \mathbb{1}(Y_{ij} > 0)$.
% be an indicator  for a non-zero score. 
% A hurdle model is defined in two parts: (a) the presence  model for whether an outcome is zero or positive in terms of the binary variable $Z_{ij}$ and (b) the severity  model for the positive outcomes, which are defined in terms of a count distribution with a zero-truncated support.
In the presence part of the model, 
% the probability 
$\pi_{ij} = \mathrm{P}(Y_{ij} >0)= \mathrm{P}(Z_{ij} = 1)$ is specified through 
% considered by 
a logistic regression as
\begin{equation} \label{eq:presence}
\mathrm{logit}(\pi_{ij}) = \boldsymbol{x}_{ij}'\balpha.
\end{equation}
In the severity model, a typical approach is to model the count distribution truncated at zero. A truncated PMF will take the form $f(y)/[1-f(0)]$ for $y = 1, 2, \ldots$, where $f(y)$ represents the PMF of the untruncated count distribution (such as Poisson or Negative Binomial). 
However, this can be numerically unstable if $f(0)$ is near one.
% This ratio of  two probabilities  may have both components very close to zero if $f(0)$ is near one, resulting in numerical instability.
% in the  estimation of the parameters. 
% Note that, depending on $\bsx_{ij}$, $\bbeta$ and $\phi$, it might not be unlikely to have value of $f(0)$ close to one. 
In order to alleviate this issue, we shift---rather than truncate---the NB distribution \citep{Kang2021}.
The shifted counts $Y_{ij}^* = Y_{ij} - 1$ are assumed to follow NB distribution as 
\begin{equation} \label{eq:severity}
% \begin{array}{rcl}
Y_{ij}^*\mid Z_{ij}=1 \sim \mathrm{NB} \left( \mu=\mu_{ij}, \phi \right), \quad \log(\mu_{ij}) = \bsx'_{ij} \bbeta,
% \end{array}
\end{equation}
where $\mu_{ij}$ and $\phi$ represent the mean and size parameters.
Here, $\E(Y_{ij} | Z_{ij}=1) = 
% 1 + \E(Y_{ij}^* | Z_{ij} = 1) = 
1 + \mu_{ij}$ and $\textrm{V}(Y_{ij} | Z_{ij}=1) = 
% \textrm{V}(Y_{ij}^* | Z_{ij}=1) = 
\mu_{ij} (1+\mu_{ij}/\phi)$. 
This formulation implies that 
  $\phi$ is inversely related to the dispersion. 
% with the dispersion increasing to infinity as $\phi$ goes to zero. 
% $\textrm{V}(Y_{ij}) \uparrow \infty$ as $\phi \downarrow 0$.
% We note that the Hurdle model requires a zero-truncated count distribution for the severity model. \org{[remove?]}
Note that this distribution for $Y_{ij}$ is defined marginally, without accounting for the dependence across $j$, so we refer to this as the marginal response model and denote its full set of parameters by $\boldsymbol{\theta}_M = (\balpha', {\bbeta}', \phi)'$.

% We now specify the prior choices for the parameters $\balpha$, $\bbeta$ and $\phi$. 
% It is often of interest to regularize the effects of the predictors in accordance with their relevance in determining the outcome.
% To that end,  
To regularize the effects of the predictors, we consider a normal-gamma \citep[NG;][]{Griffin2010} shrinkage prior for $\balpha$ and $\bbeta$.
Let $\sigma_{\alpha_k}^2$, $\sigma_{\beta_k}^2$ represent the local variance parameters for the $k$-th coefficient in the presence and severity models, with global variance parameters $\tau_{\balpha}^2$ and $\tau_{\bbeta}^2$. We denote these hyper-parameters as $\btheta_H = (\tau_{\balpha}^2, \tau_{\bbeta}^2)$.
% denote the global variance parameter inducing shrinkage. 
We assume 
% the following hierarchy for the regression coefficients:
\begin{equation*}
\begin{array}{rcccl}
\alpha_0 \sim \mathrm{N}(0, c_{\alpha}^2), & \quad & (\alpha_k | \sigma_{\alpha_k}^2) \stackrel{\mathrm{iid}}{\sim} \mathrm{N}(0, \sigma_{\alpha_k}^2), & \quad & (\sigma^2_{\alpha_k} | \lambda_{\balpha}, \tau_{\balpha}^2 ) \stackrel{\textrm{iid}}{\sim} \text{Ga}(\lambda_{\balpha}, \lambda_{\balpha}/\tau^2_{\balpha}), \\
\beta_0 \sim \mathrm{N}(0, c_{\beta}^2), & \quad & (\beta_k | \sigma_{\beta_k}^2) \stackrel{\mathrm{iid}}{\sim} \mathrm{N}(0, \sigma_{\beta_k}^2), & \quad & (\sigma^2_{\beta_k} | \lambda_{\bbeta}, \tau_{\bbeta}^2 ) \stackrel{\textrm{iid}}{\sim} \text{Ga}(\lambda_{\bbeta}, \lambda_{\bbeta}/\tau^2_{\bbeta}),
\end{array}
\end{equation*}
% $\alpha_0 \sim \mathrm{N}(0, c_{\alpha}^2)$, $(\alpha_k | \sigma_{\alpha_k}^2) \stackrel{\mathrm{iid}}{\sim} \mathrm{N}(0, \sigma_{\alpha_k}^2)$, $(\sigma^2_{\alpha_k} | \lambda_{\balpha}, \tau_{\balpha}^2 ) \stackrel{\textrm{iid}}{\sim} \text{Ga}(\lambda_{\balpha}, \lambda_{\balpha}/\tau^2_{\balpha})$, and
% $\beta_0 \sim \mathrm{N}(0, c_{\beta}^2)$, $(\beta_k | \sigma_{\beta_k}^2) \stackrel{\mathrm{iid}}{\sim} \mathrm{N}(0, \sigma_{\beta_k}^2)$, $(\sigma^2_{\beta_k} | \lambda_{\bbeta}, \tau_{\bbeta}^2 ) \stackrel{\textrm{iid}}{\sim} \text{Ga}(\lambda_{\bbeta}, \lambda_{\bbeta}/\tau^2_{\bbeta})$,
for $k = 1, \ldots, d$. 
% The values of $c_{\alpha}^2$ and $c_{\beta}^2$ are chosen to provide a disperse prior for the intercepts $\alpha_0$ and $\beta_0$, respectively. 
We use $c_{\alpha} = c_{\beta} = 2$ for a reasonably disperse prior on the intercepts $\alpha_0$ and $\beta_0$. 
% \org{[Maybe instead use $c_\alpha$ or $c_{\alpha 0}$]}
One  feature of this prior choice is that $\sigma_{\alpha_k}^2$ and $\sigma_{\beta_k}^2$ can be analytically marginalized out,
% over $\sigma_{\alpha_k}^2$ and $\sigma_{\beta_k}^2$,
% \org{[REMOVED THE DISTS AFTER MARGINALIZATION. ADD THIS BACK INTO A.2 AS A COMMENT ABOUT COMPUTING THE PRIOR DISTRIBUTION]}
% An advantage of this marginalization is 
such
that the prior densities  can be computed without the values of the local parameters, in contrast to the horseshoe prior \citep{Carvalho2010}; this  is beneficial in our sampling algorithm.
% \todo{kurtosis} 
For simplicity and stability, we generally set the hyperparameters $\lambda_{\balpha} = \lambda_{\bbeta} = 1$,  
% yielding a prior
as in the Bayesian lasso \citep{Park2008}. 
%To that end, we choose $\lambda_{\balpha}, \lambda_{\bbeta} \sim \text{Exp}(1)$ to have a prior centered around the Bayesian lasso.  
%d\org{[Is $\lambda$ random?]}
We also assume $\tau_{\balpha}^2, \tau_{\bbeta}^2 \sim \mathrm{IG}(l_1,l_2)$, and choose $l_1 = l_2 = 1$ to obtain a reasonably non-informative prior. 
% We denote these hyper-parameters together by $\btheta_H = (\tau_{\balpha}^2, \tau_{\bbeta}^2)$.
% A well-known alternative to the normal-gamma prior is the horseshoe prior \citep{Carvalho2010}. 
% Although it has attractive shrinkage properties, the prior distribution marginalized over the scale parameters, unlike normal-gamma prior, does not have a closed form. 
% This advantage of using normal-gamma will be utilized in a later section.
For the remaining parameter $\phi$, we assume a moderately disperse prior  $\log \phi \sim \textrm{N}(0,c_{\phi}^2)$ with $c_{\phi} = 2$.
Note that due to the large sample size of the IFS data, sensitivity analysis showed that posterior inference is not sensitive to the choices of $c_{\alpha}$, $c_{\beta}$ and $c_{\phi}$ here.
%, when we experimented with alternative choices for them ranging between 0.5 and 30.
% to make the prior reasonably non-informative.

\subsection{Modeling Dependence through a Gaussian Copula}
\label{sec:model_copula}

We now discuss our proposed copula-based approach to  model the correlation across the margins via a latent Gaussian structure. 
The marginal CDF of $Y_{ij}$, given by $F_{ij}(y | \btheta_M) = \mathrm{P}(Y_{ij} \leq y)$, depends on  $\btheta_M$ and the predictors $\boldsymbol{x}_{ij}$;  for simplicity, it is simply  denoted by $F_{ij}(\cdot)$ in the following unless we need to emphasize the role of $\btheta_M$. 
For each $(i,j)$, $F_{ij}(\cdot)$ marginalized over $Z_{ij}$ is given by
\begin{equation} \label{eq:cdf}
% \begin{array}{rcl}
  F_{ij}(y) = 
  % P(Y_{ij} \leq y) = 
  \frac{1}{1 + e^{\bsx_{ij}' \balpha}} + \left\{ \frac{e^{\boldsymbol{x}_{ij}'\balpha}}{1 + e^{\bsx_{ij}' \balpha}} \sum_{u=1}^{y} \mathrm{NB} \left(u-1 \mid \mu = e^{\bsx_{ij}' \bbeta}, \phi \right) \right\} \mathbb{I}(y > 0),
% \end{array}
\end{equation}
for $y \in \mathbb{Z}_0^{+}=\{0,1,2,\ldots\}$, where $\mathrm{NB}(\cdot|\mu, \phi)$ represents the mass of the Negative Binomial distribution with mean $\mu$ and size $\phi$. 
% \org{[I would use $y$ in the function instead of $y_{ij}$ since the point is the definition of the function, not in the observed value.  That will also be consistent with the followign sentence.]}
Taking a pseudo-inverse of  the CDF yields the associated quantile function as $Q_{ij}(p | \btheta_M) = \inf\limits_{y \in \mathbb{Z}_0^+} \{ y : p \leq F_{ij}(y | \btheta_M) \}$. We now define the multivariate CDF for $\bsy_i = (y_{i1}, \ldots, y_{iJ})'$, represented by $F_i(y_{i1}, \ldots, y_{iJ})$, by formulating the dependence structure across the margins $F_{ij}(\cdot)$ in terms of a Gaussian copula. 

% In the copula framework a joint distribution of a set of random variables
% % , for example 
% $(Y_{ij})_{j=1}^J$ is obtained by matching the quantiles of each margin with the quantiles from a correlated multivariate distribution.
% Here the correlation structure among the marginal hurdle models, $(F_{ij}(\cdot))_{j=1}^{J}$, is inherited from a Gaussian distribution.
% To specify the distribution of $\bsY_i$ in terms of this Gaussian copula framework, we first let
Let $\bsV_{i} = (V_{i1}, \ldots, V_{iJ})' \sim \text{MVN}_{J}(\boldsymbol{0}, \bR(\btheta_D))$, where $\bR(\btheta_D)$ (or more simply, $\bR$) represents a correlation matrix parameterized by the dependence parameters $\btheta_D$. 
% Henceforth, we will simply use $\bR$ to represent the correlation matrix, and will only specify $\btheta_D$ when necessary. 
The latent Gaussian vector $\bsV_i$ maps to $\bsY_i$ through $Y_{ij} = h_{ij}(V_{ij})$ for each $j \in \mathcal{J}$, where $h_{ij}(v | \btheta_M) = Q_{ij}(\Phi(v) | \btheta_M)$ and $\Phi(\cdot)$ is the CDF of N(0,1).  
Importantly, the role of the psuedo-inverse $Q_{ij}(\cdot)$ leads the mapping $h(v)$ to be many-to-one.
In particular, $h(\cdot)$ maps every value of $v$ in the range $\left(h_{ij}^{-1}(y-1), h_{ij}^{-1}(y) \right]$ to the same $y\in \mathbb{Z}_0^{+}$, where $h_{ij}^{-1}(y) = \Phi^{-1}(F_{ij}(y))$ with $h_{ij}^{-1}(-1) = -\infty$.
The joint distribution of $\bsY_i$ can, therefore, be 
obtained from
% expressed in terms of that of
the latent $\bsV_i$ as
\begin{equation} \label{eq:lhd}
\begin{array}{rcl}
f_i(\bsy_i) & = & \mathrm{Pr}(Y_{i1} = y_{i1}, \, \ldots,\,  Y_{iJ} = y_{iJ}) \\
% &=& \mathrm{Pr}(y_{i1} - 1 < Y_{i1} \leq y_{i1}, \, \ldots,\, y_{iJ} - 1 < Y_{iJ} \leq y_{iJ}) \\
& = & \mathrm{Pr}\left(h_{i1}^{-1}(y_{i1}-1) < V_{ij} \leq h^{-1}(y_{i1}), \, \ldots,\, h_{iJ}^{-1}(y_{iJ}-1) < V_{iJ} \leq h^{-1}(y_{iJ})\right) \\
% & = & \int\limits_{h_{i1}^{-1}(y_{i1}-1)}^{h_{i1}^{-1}(y_{i1})} \ldots \int\limits_{h_{iJ}^{-1}(y_{iJ}-1)}^{h_{iJ}^{-1}(y_{iJ})} \bphi(v_{i1}, \ldots, v_{iJ} | \bR) \,dv_{i1} \ldots dv_{iJ} \\
& = & \int_{h_{i1}^{-1}(y_{i1}-1)}^{h_{i1}^{-1}(y_{i1})} \cdots \int_{h_{iJ}^{-1}(y_{iJ}-1)}^{h_{iJ}^{-1}(y_{iJ})} \bphi(v_{i1}, \ldots, v_{iJ} | \bR) \,dv_{i1} \cdots dv_{iJ},
%& = & \sum\limits_{c_1 = 0}^1 \cdots \sum\limits_{c_J = 0}^1 (-1)^{\sum_{i=1}^J c_i} F(y_{i1} - c_1, \ldots, y_{iJ} - c_J) \\
%& = & \sum\limits_{c_1 = 0}^1 \cdots \sum\limits_{c_J = 0}^1 (-1)^{\sum_{i=1}^J c_i} \mathbb{C}(F_{i1}(y_{i1}-c_1), \ldots, F_{iJ}(y_{iJ}-c_J))
\end{array}
\end{equation}
where $\bphi(\cdot | \bR)$ denotes multivariate normal density with mean $\boldsymbol{0}$ and covariance matrix $\bR$. 
However, despite the lack of a closed-form likelihood, the copula structure provides an accessible strategy for data generation under a given set of the parameters $\btheta_M$ and $\btheta_D$ which can be leveraged to perform inference using Approximate Bayesian Computing.

% We now consider how our use  of the Gaussian copula to induce dependence accommodates missingness in the data.
% Recall that 
We recall here that the set of observations $\mcJ$ is defined using all pairs of time and tooth, and this will be used to facilitate the parameterization of $\bR$.  
However, all of these pairs necessarily cannot be observed, as
% as discussed in Section \ref{sec:motivation}.  
primary teeth will not be observed past age 13, permanent teeth are not observed at age 5, and the primary and permanent teeth at the same anatomical location in the mouth cannot be observed simultaneously.
Let $\mathcal{J}_i \subset \mathcal{J}$ denote the set of measurements observed for the individual $i$, and
 $\mathcal{D}$ be the set of  all $(i,j)$ for which $y_{ij}$ is recorded in the dataset.
Throughout, we assume all missing data are structural and/or ignorable, and we return to this  in the Discussion (Section \ref{sec:discussion}).

A distinctive advantage of using a Gaussian copula is that ignorable missing data can be dealt with naturally.
Let $\widetilde{\bsV}_i$ be the sub-vector of $\bsV_i$ corresponding to the observed data $\mathcal{J}_i$, and $\bR_i$ represent the sub-matrix of $\bR$ formed with rows and columns indexed by $\mathcal{J}_i$. 
Since the Gaussian distribution is closed under marginalization, $\widetilde{\bsV}_i \sim \text{MVN}_{|\mcJ_i|} (\boldsymbol{0}, \bR_i)$. Therefore, the joint distribution of the observed $\bsy_i$ is
determined through the (sub)matrix $\bR_i$.  
From this perspective, it is important that we use a multivariate distribution family that is closed under marginalization to accommodate ignorable missing data.

The other key benefit to using a Gaussian copula is that the zeroes in the inverse of $\bR$ impose conditional independence relationships among the elements of the latent $\bsV$, and these conditional independences  are then inherited by the response variables $\bsY$.  This, in particular, is in contrast to other elliptical distributions such as multivariate $t$ or Laplace whose concentration matrices do not encode independence.  
% In the following section, we will discuss parameterization of the correlation matrix $\bR$ that will capture complex spatio-temporal relations.

\subsection{Simultaneous Autoregressive Correlation Structure} \label{sec:corr_structure}

As noted above, the dependence structure across the observations $\bsY_i$ is determined by the  correlation matrix $\bR=\bR(\btheta_D)$ describing the relationships across the latent Gaussian variables $\bsV_i$.
% defined by the correlation matrix $\bR=\bR(\btheta_D)$.  
% It is worth emphasizing 
As the index $j$  represents a tooth-time pair $(l_j, t_j)$,
% This allows us to define
the elements in $\bR$ represent the correlations between any two tooth-time pairs, facilitating the  design of complex spatio-temporal correlation structures.
To that end, we consider the Simultaneous Autoregressive model (SAR).
% \citep[SAR;][]{Whittle1954}.
% which specifies the joint structure of spatially auto-correlated variables $\bsV_i$ explicitly. 
In the classical SAR model, the adjacency of two data points $v_{ij}$ and $v_{ij'}$ is determined by a (single) proximity relation encoded in a binary adjacency matrix $\bW = [w_{jj'}]_{j,j' \in \mathcal{J}}$
with zero elements on the diagonal.
% , such as $w_{jj'} = 1$ if data points at $v_{ij}$ and $v_{ij'}$ are adjacent and 0 otherwise.
% For example, a temporal adjacency can be characterized by the proximity relation $w_{jj'} = \mathbb{1}(|t_{j}-t_{j'}| = 1)$, where $t_j$ represents the time point associated with $v_{ij}$.
% Note that the diagonal $\bW$ must be zero.
% A standard SAR model for $\bsV_i$ would be defined through a model where each $V_{ij}$ is  
% The model defines a structure with 
Each element of $\bsV_i$
is  regressed on its neighbors through the model  $\bsV_{i} = \rho \bW \bsV_i + \beps_i$, where 
% $\bW$ denotes the binary adjacency matrix,
% with its $j$-th row representing the proximity of the $j$-th tooth-time pair with the other pairs, 
$\rho$ represents the spatial weight and $\beps_i$ is a Gaussian error vector.
% typically following normal distribution, such that each $V_{ij}$ will be regressed on its neighbors.
\cite{Pace2000}; 
% extended this setup, by expressing a spatio-temporal proximity relation in terms of a weighted linear combination of spatial-only and temporal-only components along with their interactions.
% % , capturing the lag structure in time, space and their interaction separately.
% Also see 
\cite{Badinger2011, Debarsy2022} all consider various extensions of SAR models with multiple proximity relations, which is the general strategy that we choose.
% in the context of spatial econometrics.

% In this work, we will
To that end, we consider a composite of proximity relations allowing $K$ different adjacency types, with $\bW^{(k)}$ characterizing the $k$-th proximity relation. 
The resulting composite adjacency matrix $\bB = [b_{jj'}]_{j,j' \in \mathcal{J}}$ is a  weighted linear combination  $b_{jj'} = \sum_{k=1}^K \rho_k w^{(k)}_{jj'}$, where $\rho_k$ is the autoregressive parameter associated with the $k$-th association type.  
Hence, the resulting SAR model for the latent Gaussian variables is
\begin{equation} \label{eq:SAR}
V_{ij} = \sum_{j' \in \mathcal{J}} b_{jj'} V_{ij'} + \epsilon_{ij} = \sum_{j' \in \mathcal{J}} \left( \sum_{k=1}^K \rho_k w^{(k)}_{jj'} \right) V_{ij'} + \epsilon_{ij},
\end{equation}
where $\beps_{i} = (\epsilon_{i1}, \ldots, \epsilon_{iJ})' \sim \textrm{MVN}_{J}(\boldsymbol{0}, \bGamma)$ with $\bGamma = \text{diag}(\gamma_1^2, \ldots, \gamma_J^2)$.
% , and $\rho_k$ is the autoregression parameter associated with the $k$th association type. 
% It is evident from the above formulation that $V_{ij}$ is conditionally dependent on $V_{ij'}$ only if $w^{(k)}_{jj'} = 1$ for at least one $k$.
This implies that $V_{ij}$ is conditionally independent of $V_{ij'}$ if  $w^{(k)}_{jj'} = 0$ for all $k$, that is, if they share no adjacency relationships.
An equivalent vectorized representation of the model \citep[][Chapter~4]{Banerjee2004} is  $\bsV_i = \bB\bsV_i + \beps_i$, which implies
% \begin{equation} \label{eq:latent_model}
$\bsV_i \sim \text{MVN}_{J} \left(\boldsymbol{0}, \bR \right)$,
% \end{equation}
where $\bR = (\bI - \bB)^{-1} \bGamma (\bI - \bB)^{-1}$. To ensure that $\bR$ is a correlation matrix 
% with unit diagonals 
as required for the copula model, the residual variance parameters in $\bGamma$ are constrained to be $\gamma_j^2 = \sum_{j' \in \mathcal{J}} \widetilde{b}_{jj'}$, where $\widetilde{b}_{jj'}$ is the $(j,j')$-th element of the matrix $\left[(\bI - \bB)^{-1} \circ (\bI - \bB)^{-1}\right]^{-1}$, with $\circ$ representing an element-wise product.
Throughout, we require that
% , under the assumption that 
$(\bI - \bB)$ and $(\bI - \bB)^{-1} \circ (\bI - \bB)^{-1}$ are both invertible 
so that all necessary quantities are well-defined.
% with $\circ$ representing the element-wise product.
Importantly, this representation of $\bR$ implies that the vector of autoregressive coefficients $\btheta_D = (\rho_1, \ldots, \rho_K)'$ are the set of free parameters that determine the correlation matrix $\bR$.
The support for $\btheta_D$, denoted by $\bTheta_D \subset \mathbb{R}^K$, must consist of vectors such that the invertability restrictions hold for the resulting $\bB$, yielding positive definite $\bR(\btheta_D)$ \citep{Elhorst2012}. 
Note that, for any reasonably chosen $\{\bW^{(k)}\}$, $\bTheta_D$ is  non-empty since there is at least a small neighborhood around $\btheta_D=\bzero$ in which the relevant matrices can be inverted.
We choose the prior for $\btheta_D$ to be uniform on $\bTheta_D$.
% However, for multivariate $\btheta_D$, the boundary of $\bTheta_D$ 
% depends on the corresponding adjacency matrices and is generally difficult to find as it is a complex function of the $\bW^{(k)}$s \citep{Elhorst2012}.

% We refer the reader to \cite{Elhorst2012} and the references therein, for a discussion on this.
% An alternative parameterization for  composite SAR models that requires a positivity assumption on all $\rho_k$ has been proposed by \cite{Debarsy2022}.
% Our posterior sampling scheme for the marginal as well as the dependence structure parameters is elaborated in the next section.

A common alternative to SAR for modeling spatially correlated variables relies on conditional specification of their distributions, as in Conditional Autoregressive (CAR) models \citep{Besag1974}. 
% See \citet[Chapter~4]{Banerjee2004} for a  review on the CAR model and its relation to the SAR model.
Since further restrictions are required for CAR to yield a joint correlation structure \citep[][Chapter~4]{Banerjee2004}, we do not consider CAR here.
% Briefly, the CAR model would instead model $\bsV_i$ 
% by specifying the conditional distribution
% $V_{ij} | \bsv_{i(-j)} \sim \mathrm{N}(\bsb_j' \bsv_{i}, \gamma_j^2)$, where $\bsb_j'$ is the $j$-th row of the adjacency matrix $\bB$.
% % , consisting of the neighborhood information for $V_i$. 
% One challenge to using CAR is that specifying the model through conditional distributions often fails to produce a corresponding joint model,
% % For the conditionals to have a valid joint structure, $b_{jj'}/\gamma_j^2 = b_{j'j}/\gamma_{j'}^2$ must be satisfied for all $j, j'$, 
% as discussed in \cite{Besag1974}. 
% Consequently, we focus on using the SAR model to define the correlation matrix of our copula models.

\section{Posterior Computation and Inference with ABC}
\label{sec:computation}

% It is important to first emphasize 
% Recall that our model specification involves a NB  hurdle model as the marginal model for each $Y_{ij}$, so all margins within our copula have a common set of parameters $\btheta_M$.  We additionally have the parameters $\btheta_D$ which determine the  dependence structure through the  Gaussian copula. 
% The resulting 
% As noted previously,  
Our NB hurdle model with Gaussian copula  dependence results in a 
likelihood  (\ref{eq:lhd}) that lacks a closed form, and 
% . 
% Furthermore, the support for $\btheta_D$ is a subset of $\mathbb{R}^K$ following certain restrictions discussed in the previous section, that makes the specification more complex.
% The 
the posterior distribution for $\btheta$ is given by
\begin{equation} \label{eq:posterior}
\pi(\btheta\mid\bsy_\obs) \propto \pi(\btheta) p(\bsy_\obs\mid \btheta) = \pi(\btheta) \prod_{i=1}^n \int \left[ \prod_{j \in \mathcal{J}} \mathbb{I}\left\{ y_{ij}=h_{ij}(v_{ij} | \btheta_M)
% , \ j\in\mcJ 
\right\} \right] \phi(\bsv_i | \btheta_D)\,d\bsv_i.
\end{equation}
Most standard MCMC algorithms for estimating 
% $\boldsymbol{\theta}=(\btheta_M,\btheta_D)$ 
the parameters in a copula model consider a data augmentation approach that generates the latent vectors $\bsV_i$ and then samples $\btheta=(\btheta_M,\btheta_D)$  conditional on the latent vectors $\bsV_i$. 
However, it is usually challenging and inefficient to sample  $\bsV_i$ under  the restriction $h_{ij}^{-1}(y_{ij}-1) < V_{ij} \leq h^{-1}(y_{ij})$, as argued in \cite{Pitt2006}.
% They develop an MCMC scheme that  marginalizes out the latent variables for the $j$-th response to efficiently update the parameters that define the $j$-th marginal distribution.
% The authors considered a setup, where the margin $F_{ij}$ is parameterized by a distinct $\btheta_j$ when deriving an efficient MCMC scheme was possible. 
Furthermore, our parameter $\btheta_M$ is shared across all the margins, making the proposal by \cite{Pitt2006} inapplicable. 
Here, we have opted for an Approximate Bayesian Computation (ABC) strategy for posterior inference.
% , which employs a data generative scheme to implicitly incorporate the likelihood, instead of evaluating it directly.

\subsection{ABC Background}

% We briefly provide some  background on 
Approximate Bayesian Computing \citep[ABC,][]{Sisson2018} is a data-generation based approach, which 
% For a more detailed review,  see the collection from \cite{Sisson2018}.
is most useful in settings where generating datasets is fairly easy, even if the likelihood is not directly defined or is difficult to compute analytically.
% Approximate Bayesian computing 
% It is a 
% likelihood-free 
% data-generation based approach, where 
In its most basic form, a large number of parameter values and  corresponding datasets are generated, and posterior inference is done based only on those parameter values for which the simulated datasets are similar to the observed data.
Consider an observed dataset $\boldsymbol{y}_{\text{obs}}$ from the density $p(\boldsymbol{y} | \boldsymbol{\theta})$ 
% $\pi(\boldsymbol{y}_{\text{obs}} | \boldsymbol{\theta})$ 
with parameter  $\boldsymbol{\theta}$.
% Let $\mathcal{Y}$ denote the support for the data.
Under ABC, the approximate posterior distribution is defined to be 
$\pi_{\ABC}(\btheta | \bsy_{\obs}) \propto \pi(\btheta) \int
% _{\mathcal{Y}} 
K_h(\Delta(\bsy, \bsy_\obs)) \, p(\bsy | \btheta) \, d\bsy$,
where $K_h(\cdot)$ is a kernel function with bandwidth parameter $h$ and $\Delta(\cdot, \bsy_\obs)$ is a suitably chosen distance function so that $\Delta(\bsy, \bsy_\obs)$ (or denoted simply by $\Delta(\bsy)$) measures how dissimilar a generated dataset $\bsy$ is from the observed data $\bsy_\obs$.
% Since $\bsy_\obs$ is fixed, we will often write $\Delta(\bsy)$ to represent the distance to the generated data $\bsy$.
%denoted as $\{(\boldsymbol{\theta}_g, \boldsymbol{y}_g)\}_{g=1}^G$ by (a) first sampling $\boldsymbol{\theta}_g$ from a proposal distribution $q(\boldsymbol{\theta})$, (b) generating a dataset $\boldsymbol{y}_g$ conditional on $\boldsymbol{\theta}_g$ and then (c) accepting $(\boldsymbol{\theta}_g, \boldsymbol{y}_g)$ only when $\boldsymbol{y}_g$ is sufficiently close to $\boldsymbol{y}_\text{obs}$ in terms of some suitably defined kernel. 
Often the data are high-dimensional, and
% as is often the case in real world problems, 
$\Delta(\bsy)$ is formulated in terms of a low-dimensional vector of summary statistics  $\bss = \bss(\bsy)$, chosen to capture the relevant information in  the data. 
% To simplify notation, let $\bss = \bss(\bsy)$ and $\bss_\obs = \bss(\bsy_\obs)$.
Note that, unless the low-dimensional summaries are sufficient for $\btheta$, the target partial ABC posterior becomes
\begin{equation} \label{eq:posterior_ABC}
    \pi_{\ABC}(\btheta | \bss_{\obs}) \propto \pi(\btheta) \int
    % _{\mathcal{S}} 
    K_{h}(\Delta(\bss, \bss_{\obs})) p(\bss | \btheta) d\bss,
\end{equation} 
where $\bss_\obs = \bss(\bsy_\obs)$ and $p(\bss|\btheta) \propto \int
% _{\mathcal{Y}} 
\mathbb{I}(\bss(\bsy) = \bss) \, p(\bsy|\btheta) \, d\bsy$.
% \org{[SHOULD $p(\bss\mid\btheta)$ be $\propto$ or $=$???]}
When the summary statistics are sufficient for $\btheta$, $\pi_{\ABC}(\btheta | \bss_{\obs})$ will be same as $\pi_{\ABC}(\btheta | \bsy_{\obs})$ \citep[][Chapter~1]{Sisson2018}.
% \org{[You haven't defined $\pi_{\ABC}(\btheta, \bss | \bss_{\obs})$ only $\pi_{\ABC}(\btheta| \bss_{\obs})$.  Presumably the statement is still true if you only use the post dist]}
While there are many extensions and adaptations of this general ABC strategy 
that should be tailored to the particular context,
% for posterior inference in the literature. 
% We refer the reader to \cite{Sisson2018} for further details.
the crucial elements of an ABC algorithm are the following:
% (1) a proposal distribution for the model parameters $\btheta$ close enough to the target along with reasonable initial value to start the ABC-MCMC algorithm with, 
(a) a set of summary statistics that captures the essential features of the high dimensional data, (b) a suitable choice for the kernel, and (c) an approach to obtain parameter samples from the ABC posterior in (\ref{eq:posterior_ABC}). 

\subsection{Summary Statistics and Kernel Choice} \label{sec:ABC-summary-statistics}

% Choosing a set of summary statistics that is low-dimensional while being informative about all relevant features of the model is an active research  problem in ABC. 
One standard approach to selecting a low-dimensional set of summary statistics
is to find a tractable auxiliary model that provides a reasonable approximation to the target model and derive summary statistics based on the auxiliary model \citep{Drovandi2011}.
% There have been three main strategies proposed in the literature for 
% % optimal 
% selection of summary statistics  \citep{Beaumont2019}.
% % which can be categorized into three groups.  
% First are subset selection methods, where the idea is to select a subset from a large collection of possible summary statistics based on some optimality criterion \citep{Joyce2008, Nunes2010}.  
% Secondly, researchers have considered projection methods, that aim to find informative low-dimensional projections from a vector of data features
% % , have been considered 
% \citep{Wegmann2009, Fearnhead2012}.  
% Thirdly, auxiliary likelihood methods define summary statistics from a more tractable approximate model \citep{Drovandi2011}. 
% % We refer the reader to \cite{Beaumont2019} for further details on different methods. 
To that end, we  derive the statistics for summarizing the marginal parameters $\btheta_M$ by considering 
% an auxiliary likelihood of the data, that is based on 
our target likelihood at $\btheta_D = \bzero$ given by $\prod_{(i,j) \in \mathcal{D}} (1-\pi_{ij})^{z_{ij}} [\pi_{ij}p(y_{ij}-1 | 
% \exp(\bx_{ij}'\bbeta), 
\mu_{ij},
\phi)]^{1-z_{ij}}$.
% Motivated by the third approach,  we first choose summary statistics providing information about  the parameters $\btheta_M$ by considering an auxiliary likelihood of the data, based on our target likelihood of $\btheta_M$ at $\btheta_D=\mathbf{0}$. This choice assume independence across all observations in the auxiliary likelihood
% \begin{equation} \label{eq:ind_lhd}
% % \begin{array}{rcl}
% L_{\mathrm{ind}}(\btheta_M | \bsy) = L(\btheta_M, \btheta_D=\mathbf{0} | \bsy) = \prod\limits_{(i,j) \in \mathcal{D}} \left\{ (1-\pi_{ij}) \mathbb{I}(y_{ij}=0) + \pi_{ij} p(y_{ij}-1 | \exp(\bx_{ij}'\bbeta), \phi) \mathbb{I}(y_{ij} > 0) \right\}.
% % \end{array}
% \end{equation}
We denote this auxiliary model, obtained under an independence misspecification of our target model, by $\mathcal{M}_0$. 
% \anish{(Note: We need this notation to refer back to in later sections and also in appendices.)}
We define a set of summary statistics $\bs_M(\bsY)$ for $\btheta_M$ to be
% We choose 
the maximum likelihood estimates of $\balpha$, $\bbeta$ and $\phi$ under $\model{0}$.  
This $\bs_M(\bsY)$ will be a $(2d+1)$-dimensional vector with the same dimension as  $\btheta_M$.
% to be the summary statistics for $(2d+1)$-dimensional parameter $\btheta_M$. 
The estimate for $\balpha$ is obtained from the standard logistic regression model with 
% $\bsZ$ being 
the response $Z_{ij} = \mathbb{I}(y_{ij} > 0)$, $(i,j) \in \mathcal{D}$;
% , while 
estimates of $\bbeta$ and $\phi$ are obtained by fitting a Negative Binomial regression model to the count data $y^*_{ij} = y_{ij} - 1$, for all $(i,j) \in \mathcal{D}^*
= \{ (i,j) \in \mathcal{D} | y_{ij} > 0 \}$.

To define summary statistics that inform about $\btheta_D$, we draw connection to the regression model (\ref{eq:SAR}) that defined the SAR correlation structure.
Due to the many-to-one relationship between $V_{ij}$ and $Y_{ij}$, the true $V_{ij}$ are not available from an observed dataset.
For each margin $j$, we 
% derive an empirical CDF from observed $y_{ij}$s, and 
estimate $V_{ij}$ 
 by $\hat{v}_{ij}=\Phi^{-1}(\tilde{F}_{j}(y_{ij}))$ where $\tilde{F}_{j}(\cdot)$ is an estimate of the empirical  CDF for tooth $j$.
 We obtain estimates of  the regression coefficients from \eqref{eq:SAR} using  these plug-in values $\hat{v}_{ij}$; the resulting $\hat{\rho}_1,\ldots,\hat{\rho}_K$ serve as the summary statistics $\bss_D(\bsY)$ for $\btheta_D$.
We provide further details on this step  in Appendix \ref{appn:initial_estimation}.

We represent  the full vector of summary statistics by $\bss(\bsY) = (\bss_{M}(\bsY), \bss_{D}(\bsY))$. 
It is worth acknowledging that there are many alternative strategies and choices that could  instead be used to define the summary statistics for this model. 
While we make no claims that these are an optimal set of summary statistics, we do find that they work well in our examples, as will be demonstrated 
in the simulation studies (Section \ref{sec:simulation}).
% via extensive simulation analysis.

Another crucial component of any ABC method is the choice of a kernel function $K_h(\cdot)$ and its corresponding bandwidth parameter $h$. 
Rather than a uniform kernel, we use a Gaussian kernel, as algorithmic performance was less sensitive to tuning choices
%made achieving good MCMC performance relatively easy 
in our experiments
(see Appendix \ref{appn:IFS_computation}).
The Gaussian kernel is defined as $K_h(\Delta) = \exp(-\frac{1}{h} \Delta)$, where $\Delta(\bsy, \bsy_{\obs})=(\boldsymbol{s} - \boldsymbol{s}_{\text{obs}})' \mathbf{A} (\boldsymbol{s} - \boldsymbol{s}_{\text{obs}})$ for some choice of $\mathbf{A}$ (see \cite{Beaumont2019} for further discussion).
% measuring the closeness between the simulated and the observed data in terms of summaries $\bss$ and $\bss_\obs$. 
Since $\Delta(\bsy, \bsy_{\obs})$ measures the disagreement between the simulated and observed summary statistics $\bss$ and $\bss_{\obs}$, we will denote it as $\Delta(\bss, \bss_{\obs})$ or $\Delta(\bss)$.
Note that $h$ controls the global closeness between $\bss$ and $\bss_\obs$, while the scaling matrix $\mathbf{A}$ %=\bSigma_s^{-1}$ 
in $\Delta(\bss)$ controls the relative deviations allowed in the individual components of $\bss$.
We provide more  discussion on the bandwidth selection strategy in the context of IFS data analysis in Section \ref{sec:IFS_model_setup} and with simulation analysis in Section \ref{sec:simulation}.
We use a diagonal $\bA$,  which is estimated prior to ABC-MCMC sampling (further details in Appendix \ref{appn:initial_estimation}).

\subsection{ABC-MCMC Sampling} \label{sec:ABC-MCMC-algo}

% The choice of summary statistic and kernel defines the target distribution in (\ref{eq:posterior_ABC})
% \org{[ref eqn 10}
% that we base inference on.  
% Now, we must determine how to obtain samples of $\btheta$ from this.
% \cite{Fan2018} provides a brief overview of the different computational strategies 
% % available in the literature to efficiently 
% to sample from the approximate posterior distribution. 
Rejection and importance sampling methods and their variants have been the cornerstone of inference in the ABC framework \citep{Fan2018}.
% adapted to work in the ABC framework. 
However, when the parameter vector is moderate- to high-dimensional, as in our case, generating  posterior samples concentrated in a narrow region of the support is often infeasible with rejection algorithms. 
Importance sampling performs poorly as well, yielding  highly unbalanced weights,
% and  small effective sample sizes
making the posterior inference unreliable. 
% More sophisticated Sequential Monte Carlo based ABC algorithms \citep{Beaumont2009} also struggled to achieve reasonable effective sample sizes in our investigations.
% In our experiments, none of these classic ABC algorithms were able to yield reasonable inference for this setting in a time efficient manner.

In contrast, \cite{Marjoram2003} proposed an ABC version of the classic MCMC algorithm, and  we adopt this strategy. 
Their approach is based on using the Metropolis-Hasting (MH) algorithm to sample a pair $(\btheta, \bsy)$ from a data augmentation ABC posterior $\pi_{\ABC} (\btheta,\bsy | \bss(\bsy_{\obs})) \propto \pi(\btheta)  K_h(\Delta(\bss(\bsy), \bss(\bsy_\obs))) \, p(\bsy | \btheta)$
% \begin{equation}
% \label{eq:posterior_ABC_full}
% \pi_{\ABC} (\btheta,\bsy | \bss(\bsy_{\obs})) \propto \pi(\btheta)  K_h(\Delta(\bss(\bsy), \bss(\bsy_\obs))) \, p(\bsy | \btheta) ,
% \end{equation}
and perform inference using the retained $\btheta$ samples.
% To that end, each iteration of the ABC-MCMC algorithm starts with drawing a candidate sample of $\btheta'$ from a proposal distribution.
% One generates the data augmentation dataset $\bsy'\sim p(\cdot | \btheta')$, 
% and the pair $(\btheta',\bsy')$ is accepted or rejected using the Metropolis-Hastings (MH) acceptance probability.
The key is that the intractable  likelihoods $p(\bsy|\btheta)$ and $p(\bsy'|\btheta')$ appear in both the posterior and proposal distributions, canceling out one another.  Thus, the MH transition probability can  be computed numerically.
Here, we choose a multivariate normal as the random walk proposal distribution for the parameter vector $\btheta=(\btheta_M,\btheta_D)$ and 
 consider  adaptive Metropolis by updating 
% We update 
the proposal covariance with a vanishing adaptation scheme \citep{Andrieu2008}.
Let the proposal density in iteration $g$ be denoted by $q_{g}(\cdot | \btheta^{(g-1)})$.
% , where the subscript $g$ emphasizes the adaptation.
% To simplify notation for the following discussion, we denote it by $q$ without the subscript.
% $\btheta^{(g-1)}$ representing the current value of the parameter vector.
% and $\bSigma_{g-1}$ being the corresponding covariance matrix
% Note that, we do not consider the covariance matrix $\bSigma_g$ as a parameter, the subscript here indicates the fact that it will be updated in every iteration of the MCMC algorithm.
See Appendix \ref{appn:initial_estimation} and \ref{appn:algo_details} for details on initialization steps and covariance adaptation, respectively.
% The details for covariance adaptation are included in Appendix \ref{appn:algo_details}, along with discussion of 
% choosing the initial value $\btheta^{(0)}$ and the initial proposal covariance matrix in Appendix \ref{appn:initial_estimation}. 
The main steps at iteration $g$ are the following:

% We choose a multivariate normal proposal density, with the covariance matrix adapted in each MCMC step. Initial value for $\btheta_M$ is obtained from the posterior samples drawn under independence miss-specification, and that for $\btheta_D$ is obtained empirically using a data generative approach. 
% We choose the initial covariance matrix for the MH proposal to be block diagonal, with covariance for $\btheta_M$ estimated from the posterior samples under independence and that for $\btheta_D$ estimated empirically.

% Iteration $g$ of the ABC-MCMC algorithm starts with drawing a sample of $\btheta^{(g)}$ from the proposal distribution and generate data $\bsy^{(g)}$ conditional on $\btheta^{(g)}$, followed by computation of the corresponding summary statistic $\bss^{(g)}$. 
% The new parameter value is then accepted or rejected based on an MH acceptance probability. 
% A step to update the MH covariance matrices $\bSigma_M$ and $\bSigma_D$ will follow. 
% In addition to updating $\btheta$, we also update the normal-gamma hyper-parameters $\btheta_H$ completing the iteration. 
% Since $\btheta_H$ does not depend on $\bsy^{(g)}$ conditional on $\btheta$, or in particular $\btheta_M$, it can be updated outside the ABC-MCMC framework, using simply a classical MH step. A complete iteration $g$ of our full MCMC algorithm is presented as follows,

\begin{enumerate}
  \item \textit{Update} $(\btheta, \bsy)$: Generate a candidate parameter vector $\btheta' \sim q_g(\btheta'|\btheta^{(g-1)}) $, generate $\bsy'$ from the copula data model given $\btheta'$, and compute the summary statistics $\bss' = \bss(\bsy')$. We accept $(\btheta', \bsy')$ with probability $A\left( (\btheta',\bsy'), \, (\btheta^{(g-1)}, \bsy^{(g-1)})\right)$ given by
    % \begin{equation} \label{eq:acc_prob}
    %  \min \left\{ 1, \frac{ K_h(\bss') }{ K_h(\bss^{(g-1)}) } \frac{\pi(\btheta') \, q(\btheta' | \btheta^{(g-1)})}{ \pi(\btheta^{(g-1)}) \, q(\btheta^{(g-1)} | \btheta')} \right\},
    % \end{equation}
    \begin{equation*} \label{eq:acc_prob}
    \begin{array}{rcl}
     & \min \left\{ 1, \frac{ \pi(\btheta') \,K_h(\bss') p(\bsy'|\btheta') }{ \pi(\btheta^{(g-1)}) \, K_h(\bss^{(g-1)}) p(\bsy^{(g-1)} | \btheta^{(g-1)}) } \frac{ q_g(\btheta^{(g-1)} | \btheta') p(\bsy^{(g-1)} | \btheta^{(g-1)}) }{ q_g(\btheta' | \btheta^{(g-1)}) p(\bsy' | \btheta') } \right\} 
     =  \min \left\{ 1, \frac{ K_h(\bss') }{ K_h(\bss^{(g-1)}) } \frac{\pi(\btheta') 
     % \, q_g(\btheta^{(g-1)} | \btheta') 
     }{ \pi(\btheta^{(g-1)}) 
     % \, q_g(\btheta' | \btheta^{(g-1)})
     } \right\} & 
     % = & \min \left\{ 1, \frac{ K_h(\bss') }{ K_h(\bss^{(g-1)}) } \frac{\pi(\btheta') 
     % % \, q_g(\btheta^{(g-1)} | \btheta') 
     % }{ \pi(\btheta^{(g-1)}) 
     % % \, q_g(\btheta' | \btheta^{(g-1)})
     % } \right\}&
    \end{array}
    \end{equation*}
    and set $(\btheta^{(g)}, \bsy^{(g)}) = (\btheta', \bsy')$.  Otherwise, reject $(\btheta', \bsy')$ and set $(\btheta^{(g)}, \bsy^{(g)}) = (\btheta^{(g-1)}, \bsy^{(g-1)})$.
    % Note that the likelihood terms from the numerator and denominator get cancelled resulting in a likelihood-free acceptance probability.
    % We also update the MH proposal covariance matrix using the proposed $\btheta'$.
    We also update the MH proposal distribution $q_g(\cdot|\cdot)$ (Appendix \ref{appn:algo_details}).
  % We generate candidate vector $\btheta' \sim q(\btheta'|\btheta^{(g-1)}, \bSigma_{g-1}) $, generate $\bsy'$ from our copula based data model given $\btheta'$ and compute the summary statistics $\bss' = \bss(\bsy')$. We accept $(\btheta', \bsy')$ with probability
  %   \begin{equation} \label{eq:acc_prob}
  %    \min \left\{ 1, \frac{ K_h(\bss') }{ K_h(\bss^{(g-1)}) } \frac{\pi(\btheta') \, q(\btheta' | \btheta^{(g-1)}, \bSigma^{(g-1)})}{ \pi(\btheta^{(g-1)}) \, q(\btheta^{(g-1)} | \btheta', \bSigma^{(g-1)})} \right\},
  %   \end{equation}
  %   and set $(\btheta^{(g)}, \bss^{(g)}) = (\btheta', \bss')$, otherwise reject $(\btheta', \bsy')$ and set $(\btheta^{(g)}, \bss^{(g)}) = (\btheta^{(g-1)}, \bss^{(g-1)})$.
  %   Note that we also update the MH proposal covariance matrix using the proposed $\btheta'$.
    % according to the vanishing adaptation rule, using the algorithm 4 suggested in \cite{Andrieu2008}.
  
%  \item \textit{Update} $\btheta_H$: We use classical MH  to update $\btheta_H$ using two random walk steps for $\log(\tau^2_\alpha)$ and  $\log(\tau^2_\beta)$.
 \item \textit{Update} $\btheta_H$: We use two random walk MH steps for $\log(\tau^2_\alpha)$ and  $\log(\tau^2_\beta)$.

\end{enumerate}

Multiple ABC-MCMC chains are run in parallel based on  different initial $\btheta$ values, 
generating
% and this algorithm results in the generation of
a sequence of  samples $((\btheta^{(g)}, \bsy^{(g)}))_{g=1}^G$. 
The corresponding sequence of summary statistics is $(\bss^{(g)})_{g=1}^G$.
Let $\vartheta$ denote an estimand of interest, which may simply be $\btheta$, a subset of its components or may consist of more complicated functionals  of $\btheta$.
Posterior inference for $\vartheta$ with respect to the ABC posterior (\ref{eq:posterior_ABC}) can be performed based on the corresponding samples $\vartheta^{(g)}$ obtained from $\btheta^{(g)}$.
% Here an additional regression adjustment step (discussed in the following section) is performed on $\vartheta^{(g)}$ and the final samples are used for inference.
We remove the first few iterations as burn-in until the chains stabilize, and the remaining samples are further processed with regression-adjustment followed by thinning.
% Adjusted samples are finally thinned to reduce correlation in the retained set of approximate posterior samples.
We investigate the mixing and convergence of the pre-adjustment ABC-MCMC chains by visual inspection, as well as with Gelman-Rubin $\hat{R}$ proposed by \cite{Gelman1992}
(Appendix \ref{appn:IFS_computation}).
% \org{[Maybe some more discussion about how you investigate mixing/convergence/etc.  Describe it consistent with what you are showing in B2]}
% Since the algorithm accepts proposed parameters using a data generative approach, this added stochasticity encourages the choice of a smaller target acceptance probability than would be typical for Metropolis-Hastings based MCMC algorithms.
% \org{[dont follow this sentence]}

\subsection{Regression Adjustments} \label{sec:regression_adjustment}

% % We will discuss the regression adjustment strategy in the case of a univariate estimand $\dot{\theta}$. 
% Recall that as the kernel bandwidth $h$ decreases, 
% % the ABC-based samples $\dot{\theta}^{(g)}$ would better approximate its true posterior. 
% % \org{[I don't think the samples better approximate]}
% the ABC posterior (\ref{eq:posterior_ABC}) better approximates the true posterior distribution of interest (\ref{eq:posterior}).  
% However, running ABC-MCMC with very small $h$
% % How small $h$ could be, however, will be dictated by the computational limitations, in that small $h$ 
% will often result in very low MCMC acceptance rates and poor mixing. 
% In contrast, when $h$ is taken to be large, 
% we may be able to obtain more posterior samples, but these will come from
% accepted datasets $\bsy^{(g)}$ that may differ substantly from $\bsy_{\obs}$ with respect to the chosen summary statistics, leading to bias and/or high uncertainty in inference.
% % With $h$ not small enough, the discrepancies between $\bss^{(g)}$ and $\bss_{\obs}$ will have a negative impact on the associated ABC samples $f(\btheta^{(g)})$.

To reduce the effect of the ABC bandwidth, regression-adjustment is often considered as a post-processing step \citep{Beaumont2002}.
The basic strategy  is to build a model 
%$\vartheta=\E(\vartheta|\bss)+\epsilon$ 
for the estimand $\vartheta$ as a function of $\bss$ using the ABC samples of $(\vartheta^{(g)}, \bss^{(g)})$.
From this model, the estimated residual $\hat{\epsilon}^{(g)} = \vartheta^{(g)}-\widehat{\E}(\vartheta|\bss=\bss^{(g)})$ is combined with the predicted parameter value under the observed data to obtain the regression-adjusted sample 
% $\ddot{\theta}^{(g)}$, given by 
$\ddot{\vartheta}^{(g)} = \widehat{\E}
(\vartheta|\bss=\bss_\obs)+\hat{\epsilon}^{(g)}$, which is used for inference.
We use the \textsf{R} \textit{abc} package \citep{Csillery2012} for local linear regression adjustment with conditional heteroscedasticity \citep{BlumFrancois2010}.

We individually regression-adjust each component of $\btheta=(\btheta_M,\btheta_D)$, as well as  
elements from $\bR(\btheta_D)$.
Appendix \ref{appn:algo_details} contains further details.
As the MH step in ABC-MCMC results in many repetitions among the retained posterior samples $(\btheta^{(g)}, \bss^{(g)})$, we consider only the unique samples for regression adjustment and reweigh the adjusted samples in accordance with their sampling frequencies during ABC-MCMC.
Therefore, it is important that ABC-MCMC is run for long enough and/or with a bandwidth not too small to ensure a large number of unique samples.
See Appendix \ref{appn:sim_compare_regression_strategies} for comparison to an alternative adjustment strategy that adds noise to avoid repeated samples.
% with similar performance.

% An alternative strategy based on an addition of random noise to the summary statistics is discussed in Appendix \ref{appn:sim_compare_regression_strategies}, where we compare it against our proposed approach.

% As we noted in section \ref{sec:corr_structure}, $\bR$ is a function of $\btheta$, we adjust
% the dependence parameters $\btheta_D$ have bounded support $\bTheta_D$. 
% Considering the range for each component of $\btheta_D$ to be $(-1,1)$, a modified $z$-transformation is applied coordinate-wise on each $\btheta_D^{(g)}$ to obtain $\Breve{\btheta}_D^{(g)}$, where the $k$-th component is given by $\Breve{\theta}^{(g)}_{D;k} = \log \left(( \theta^{(g)}_{D;k} + 1) / (\theta^{(g)}_{D;k} - 1) \right)$. Let $\Breve{\btheta}^{(g)} = (\btheta_M^{(g)}, \Breve{\btheta}^{(g)}_D)$ denote the transformed parameter vector. 
% The transformed set of sample parameters are then adjusted and then inverse-transformed to yield the final version of regression adjusted parameters, which we still denote by $\btheta^{(g)}$ for simplicity. 
% It is worth mentioning that, since the $\btheta_D$ parameters have non-rectangular support, the component-wise transformation will not always work, in that some of the $\btheta^{(g)}_D$ samples resulting from the aforementioned regression adjustement may not lie within $\bTheta_D$. 

\subsection{Model Assessment} \label{sec:model_validation}

% Using our ABC-MCMC algorithm along with regression adjustment provides the framework for performing posterior inference of our model.  
From these parameter samples, we next  consider methods to validate  model fit and compare across competing models using 
% One commonly used Bayesian approach to validate the fit of a model is 
posterior predictive checks \citep{Rubin1984, Gelman1996}.
Let $t(\bsY)$ denote a test statistic capturing a feature of interest from the data. 
To test the model fit with respect to this feature, the observed value of the test statistic $t(\bsy_{\obs})$ is compared against the posterior predictive distribution of $t(\bsY)$. 
% To that end, a collection of replicated datasets $\bsy^{\mathrm{rep}}$ is generated from the (regression-adjusted and thinned) posterior samples of $\btheta$, and the distribution of $t(\bsY^{\rep})$ is characterized (typically by violin plot or histogram) for various choices of the test statistic $t(\cdot)$.
% a histogram is drawn based on the corresponding test statistic values $t(\bsy^{\text{rep}})$ that depicts the posterior predictive distribution of $t(\bsY)$. 
If  $t(\bsy_{\obs})$ is located in  the tail of this distribution, the model may be viewed as inadequately  explaining this  feature.
To numerically summarize the posterior predictive accuracy for $t(\cdot)$, 
% particularly when considering a large number of potential test statistics, 
we consider the two-sided posterior predictive p-value $ppp = 2 \min\{ \mathrm{Pr}(t(\bsY^{\mathrm{rep}}) \geq t(\bsy_\obs)), \mathrm{Pr}(t(\bsY^{\mathrm{rep}}) \leq t(\bsy_\obs)) \}$,
% \[
% ppp = ppp(\bsy_\obs) = 2 \min\{ \mathrm{Pr}(t(\bsY^{\mathrm{rep}}) \geq t(\bsy_\obs)), \mathrm{Pr}(t(\bsY^{\mathrm{rep}}) \leq t(\bsy_\obs)) \},
% \]
where the probability is with respect to $p(\bsy^{\mathrm{rep}}|\bsy_\obs)= \int p(\bsy^{\mathrm{rep}}|\btheta) \pi_{\ABC}(\btheta|\bss_{\obs}) d\btheta$.
Small $ppp$ (such as $ppp < 0.05$) suggests poor fit to the feature in the test statistic $t(\bsY)$.
However,  $ppp$ is criticized for double-counting the data, and the proposed calibration schemes \cite[e.g.,][and references]{
Hjort2006} are too computationally intensive to be practically useful.
% \cite{Nott2018} proposed an approximate way to perform the calibration utilitizing ABC regression adjustment procedure to bypass the computational overload.
% % However, this is possible with appropriate choice for the summary statistics and involves drawing parameters from the prior, which is not ideal for a high-dimensional parameter space as the most of the drawn values will lie far from the area of posterior concentration.
% \org{However, this is requires drawing parameters directly from the prior, so it is not applicable in cases with moderate-dimensional parameter space and disperse priors, such as ours, as most generated parameters values will lie far from the area of posterior concentration.
% }
Since our goal is to assess the model fit, not interpreting the exact value of $ppp$, we do not view the calibration issue to be particularly important in our applications.

% \org{[should start with chosing t based on abc summary stats, then discuss features outside of s.]}
%To assess the modeling of the mean structure, we choose $\bss_M(\bsY)$ defined in Section \ref{sec:ABC-summary-statistics} as test statistics for posterior predictive checks; we note that these were used as the kernel summary statistics in the ABC-MCMC algorithm, and therefore we anticipate good fit for these summaries.
To assess the modeling of mean structure, the test statistics $\bst_M$ include the mean and variance of counts, the proportion and average of the non-zero counts, along with some statistics characterizing the associations between the covariates and the outcome (Section \ref{appn:IFS_modeling}).
% It is worth observing that the summary statistics we accepted during the ABC-MCMC sampling can not be used for posterior predictive checks.
% The marginal distribution of the accepted summary statistics is $p_{\ABC}(\bss|\bss_{\obs}) = \int \pi_{\ABC}(\btheta, \bss | \bss_{\obs}) d\btheta $, which is different from our target posterior predictive distribution for $\bss$:  $p(\bss | \bss_{\obs}) = \int \pi_{\ABC}(\btheta|\bss_{\obs}) p(\bss|\btheta) d\btheta$. 
% % \org{[i wonder if we should be using $\bss_\obs$ instead of $\bsy_\obs$ as the conditioning statement here (and elsewhere)]}
% In particular, due to kernel based weighting of the likelihood for the summary statistics drawn during ABC, $p_{\ABC}(\bss | \bss_{\obs})$ will contract more towards $\bss_{\obs}$ compared to $p(\bss| \bss_{\obs})$.
For assessing the dependence model fit, a small set of representative pairwise correlation coefficients is selected to capture a collection of temporal and/or spatial correlations that we aim to investigate, denoted as $\btheta_\bR$ (Table \ref{tbl:theta_R} lists $\btheta_\bR$ for the IFS analysis).
We define the corresponding test statistics, denoted by $\bst_{\bR}(\bsy)$ or simply $\bst_{\bR}$, to be the corresponding Spearman correlation coefficient
between $Y_{ij}$ and $Y_{ij'}$ for each pair $(j,j')$ in the collection.  
Note that, unlike the $\bss_D(\bsy)$ used during ABC-MCMC based on the particular choice of the SAR adjacencies, $\bst_\bR$ does not depend on the choice of SAR model and are therefore well-suited to comparing across various SAR specifications.

\section{Iowa Fluoride Study Data Analysis} \label{sec:IFS_study}

\subsection{Data and Modeling Details} \label{sec:IFS_model_setup}

% We provided a brief overview of the IFS data 
Extending the discussion in Section \ref{sec:motivation}, 
 we now consider further details of the IFS data in the context of fitting our models.
% which we will now explore from the modeling perspective. 
Recall that our outcome variable of interest is the caries score representing the extent of the dental caries experience for each tooth.
% on a given tooth.
% at different surfaces of the tooth.
We consider a set of 
behavioral predictors, including dental visit frequency and brushing frequency during past 6 months, daily total fluoride consumption and total sugary beverage consumption. Additionally, we consider indicators for tooth type (molar, premolar, canine, incisor), dentation (primary vs.\ permanent), and age
% teeth indicator as well as time (age) indicators to characterize the effect of different teeth and time points 
(see Appendix \ref{appn:IFS_setup} for more details).
% More details on the predictor variables for this analysis are described in Appendix \ref{appn:IFS_setup}. 
% Table \ref{tbl:predictors}.
% , and their effects along with the intercept constitute the marginal parameter $\btheta_M$.  
Values of the behavioral predictors are obtained from semi-annual surveys, and
% The predictor values were recorded using semi-yearly surveys.
the value used at time $t$ comes from
an imputation employing an AUC trapezoidal method from the recorded survey values of that variable since the previous measurement occasion \citep{Choo-Wosoba2018, Kang2021}.
% \org{[include refs for using this approach. Hyoyoung, Amanda's papers.]}
All continuous variables are normalized to have the same scale.
% We propose negative-binomial hurdle model to describe the marginal structure for the IFS data.
% \org{[For Brushing variable, is this categorical? what are you doing with the don't know category?]}

%We have also considered four indicator variables for age considering age 9 as the baseline, so that a variable is 1 when a caries score corresponds to its related time point, or is 0 otherwise. 
% Further, note that teeth can be categorized into four types: incisors, canines, premolars and molars. 
% We therefore choose to have three indicator variables, and consider molar type as the baseline. 
% Since the target ages encompass childhood, adolescence as well as young ages of an individual, we would expect primary teeth to fall out eventually and be replaced by permanent teeth. 
% To that end, we also consider another indicator variable for distinguishing between the primary and permanent teeth, the variable being 1 in the case of a primary teeth and 0 otherwise. 
% Our choices for the baselines are guided by the fact that only the participants at age 9 have both primary as well as permanent teeth, age lower than that only had primary teeth while higher ages only had permanent teeth recorded in the data. 
% Our goal was to determine the effects of the indicators as independently as possible. 

To build the dependence structures for SAR,
% across the caries scores of an individual, 
we use different combinations of the proximity relations.
% \begin{figure}
%     \centering
%     \includegraphics[width=0.7\textwidth]{pics/Dental structure.png}
%     \caption{Dentition \anish{(Collected from the internet, probably we will change the figure, keeping it as a quick reference for now).}
%     \org{[If you can, add a citation and reference here in the caption.]}}
%     \label{fig:dentition}
% \end{figure}
% Figure \ref{fig:dentition} provides the standard naming and placement of all the primary and permanent teeth inside the mouth.
Throughout, we consider the standard naming convention for primary and permanent teeth (see Figure \ref{fig:appn_dentition}).
To capture temporal proximity, the adjacency matrix $\bW^{(t)}$ includes $w_{jj'}^{(t)} = 1$ when $t_j$, $t_{j'}$ are adjacent time points and $l_j = l_{j'}$ are the same tooth; otherwise, $w_{jj'}^{(t)} = 0$.
Horizontal proximity is encoded by $\bW^{(h)}$, where $w_{jj'}^{(h)} = 1$ if $t_j = t_{j'}$ and $l_j$ and $l_{j'}$ are horizontally adjacent (such as teeth 8 and 9 in Figure \ref{fig:appn_dentition}). 
% Note that, no tooth from one jaw is horizontally adjacent to a tooth from the other jaw.
Additionally, we consider a vertical proximity $\bW^{(v)}$ connecting the nearest teeth from the upper and lower jaws (e.g., teeth 8 and 25).
We also have a primary-permanent proximity $\bW^{(pp)}$ to connect the primary tooth to the permanent tooth that replaces it at the next measurement occasion (e.g., tooth E at age 5 and tooth 8 at age 9). 
We further consider structures that provide equal connections within time $\bW^{(ct)}$, meaning that scores from all teeth at the same time are adjacent, and equal-connection everywhere $\bW^{(ce)}$, where all teeth and all time points are considered adjacent (similar to equicorrelation).
% Note that, we consider two teeth, which are located in opposite jaws at similar places, to be vertically adjacent. For example, primary teeth 'A' and 'E' are vertically adjacent to tooth 'K', 'P' respectively, as are permanent teeth 1 and 8 to 32 and 25 respectively.
% The corresponding adjacency matrices along with the associated parameters are 
These are summarized in Table \ref{tbl:dependence_structures}.
A  list of all connections for each $\bW^{(k)}$ is in 
% Table \ref{tbl:adjacency} in 
Appendix \ref{appn:IFS_setup}.
% \org{[table link doesn't work.  Possibly multiply defined?]}
% \org{[Probably include full list of connections in each proximity relationship in the appendix.]}
% \org{[I don't think the $w=1$ part of the table is informative, at least for the first four.  It is just adding notation that we don't really use.]}
% \org{[maybe don't want to call it equicorrelation, bc in m4 we don't actually get equal correlation. maybe equal connection or all (as in all teeth are adjacent).}  
% \org{[also, change ri and rt subscripts to a notation that matches the discussion]}

\begin{table}[!tb]
\centering
\footnotesize
\begin{tabular}{ccl}
\hline \hline
Parameters  & Adjacency matrix & {Interpretation} \\ \hline
$\rho_t$    & $\bW^{(t)}$      & Temporal adjacency         \\
$\rho_h$    & $\bW^{(h)}$      & Horizontal teeth adjacency        \\
$\rho_v$    & $\bW^{(v)}$      & Vertical teeth adjacency      \\
$\rho_{pp}$ & $\bW^{(pp)}$     & Primary/Permanent adjacency     \\
$\rho_{ct}$ & $\bW^{(ct)}$     & Equal connection within time         \\
$\rho_{ce}$ & $\bW^{(ce)}$     & Equal connection everywhere (across time and location)      \\
\hline \\ \hline \hline
Parameters & Model & Dependence structure \\ \hline
- & $\model{0}$ & Independent \\
$(\rho_t, \rho_h)$ & $\mathcal{M}_1$ & $\bB_1 = \rho_t \bW^{(t)} + \rho_h \bW^{(h)}$ \\
$(\rho_t, \rho_{h}, \rho_{pp})$ & $\mathcal{M}_2$ & $\bB_2 = \rho_t \bW^{(t)} + \rho_{h} \bW^{(h)} + \rho_{pp} \bW^{(pp)}$ \\
$(\rho_t, \rho_h, \rho_v, \rho_{pp})$ & $\mathcal{M}_3$ & $\bB_3 = \rho_t \bW^{(t)} + \rho_h \bW^{(h)} + \rho_v \bW^{(v)} + \rho_{pp} \bW^{(pp)}$ \\
$(\rho_t, \rho_{ct})$ & $\mathcal{M}_4$ & $\bB_4 = \rho_{ct} \bW^{(ct)} + \rho_t \bW^{(t)}$ \\
$(\rho_{ct}, \rho_t, \rho_{h})$ & $\mathcal{M}_5$ & $\bB_5 = \rho_{ct} \bW^{(ct)} + \rho_t \bW^{(t)} + \rho_{h} \bW^{(h)}$ \\
$(\rho_{ce})$ & $\mathcal{M}_6$ & $\bB_6 = \rho_{ce} \bW^{(ce)}$ \\
$(\rho_{ce}, \rho_{t})$ & $\mathcal{M}_7$ & $\bB_7 = \rho_{ce} \bW^{(ce)} + \rho_t \bW^{(t)}$ \\
$(\rho_{ce}, \rho_t, \rho_h)$ & $\mathcal{M}_8$ & $\bB_8 = \rho_{ce} \bW^{(ce)} + \rho_t \bW^{(t)} + \rho_h \bW^{(h)}$ \\
\hline
\end{tabular}
\caption{The top portion  describes the adjacency matrices considered, along with their interpretations and corresponding parameters.
Full details about the elements of these can be found in Appendix \ref{appn:IFS_setup}.
The lower portion provides model choices with different dependency structures considered in the analysis.}
\label{tbl:dependence_structures}
\end{table}

% \org{[Update to match Table 2]}
These proximity relationships are used in various combinations as described in Table \ref{tbl:dependence_structures}, and each model is fit to the IFS data.
% We consider the various models with different dependence structures as described in the bottom part of Table \ref{tbl:dependence_structures}.
Model $\model{1}$, including proximity across time and  horizontally adjacent teeth,
% However, this two parameter model
is overly simplistic as it induces independence between the upper and lower jaws, as well as between all primary and permanent teeth. 
Model $\model{2}$ adds the connection between primary and permanent teeth.
% which are located at similar locations.
More realistic is $\model{3}$ which includes four types of proximities within the SAR framework such that all teeth are (marginally) correlated.
% , where the form of the correlation is structured by the various adjacency relationships.  
Rather than considering the correlations that depend solely on the spatial/anatomical structure, we also consider models that treat the dependence as exchangeable across and within time points. 
% $\model{4}$ assumes all teeth within a time point share the same connection coefficient which is combined with a temporal adjacency relationship.
% $\model{5}$ adds more complexity to the dependence structure by allowing for horizontal proximity.
$\model{4}$ and $\model{5}$ assume temporal adjacency and that all teeth within a time point share the same connection coefficient; $\model{5}$ also includes horizontal adjacency.
In $\model{6}$, we assume all teeth to be equally correlated across all time points (i.e., equicorrelation), while $\model{7}$ and $\model{8}$ extend $\model{6}$ by providing more complex correlation structures.
% It is worth observing that teeth located at different jaws are not connected in $\bB_1$, and same is the case for the primary and permanent teeth.
% The resulting correlation matrix for model $\model{1}$ would have a block-diagonal form, essentially providing an incomplete dependence structure for the caries scores.
% This model is therefore not preferable, and can only be considered as a baseline to be compared against.
% $\model{2}-\model{4}$ would allow non-zero correlation for any two tooth-time pairs.
% We consider four composite dependence structures given as, (1) $\bB_1 = \rho_t \bW^{(t)} + \rho_h \bW^{(h)}$, (2) $\bB_2 = \rho_t \bW^{(t)} + \rho_h \bW^{(h)} + \rho_v \bW^{(v)} + \rho_{pp} \bW^{(pp)}$, (3) $\bB_3 = \rho_{ce} \bW^{(ce)}$, and (4) $\bB_4 = \rho_t \bW^{(t)} + \rho_{ct} \bW^{(ct)}$. 
% \org{[should include this in a bottom portion of Table 2]}
% The $\rho$ parameters characterizing these $\bB$ matrices define the set of dependence parameters $\btheta_D$. 
The simplest model $\model{0}$ is obtained assuming $\bR = \bI$, that is, different caries scores of an individual are independent.

We employ  Gibbs sampling to estimate the independence model $\model{0}$ as discussed in Appendix \ref{appn:initial_estimation}. 
For the other models, we fit the IFS data using the proposed ABC-MCMC algorithm with 
bandwidths of $h = 1, 10, 30$, and $100$. 
For each model and each $h$, we generated 3 chains, each with 185,000 posterior samples. 
We removed the first 5000 samples as burn-in from each chain and used the remaining samples for regression adjustment.
The adjusted samples are then thinned to obtain a final sample size of 3000 per chain.
% \org{[Are we using thinning or just counting the unique samples?]}
Bandwidth selection 
% was done using an informal approach, based on the number of distinct accepted samples the ABC-MCMC algorithm generated while also taking into account the mixing of the MCMC chains.
% Bandwidth selection in the context of IFS data is 
is described  in Appendix \ref{appn:IFS_computation},  and we choose $h = 10$ as it is the smallest $h$  with good ABC-MCMC mixing. 
 
% Note that we performed thinning on the adjusted samples instead of the original ABC samples to have a larger sample size for regression adjustment.
% \org{[Is this right?  It sounds like parameter estimation is based on 9000 samples.  Or should we say:]  Posterior prediction results are based on a thinned sample of the adjusted parameters that retained 3000 samples per chain.}

% All the following IFS data analysis results are based on this choice for the bandwidth.
% We investigate the quality of the posterior estimation in terms of mixing and convergence of the ABC-MCMC chains in Appendix \ref{appn:IFS_computation}.

% In the following subsections, we will first interpret the estimated model and then validate it based on posterior predictive checks. 
% To that end, we choose a set of test statistics shown in Table , that we believe highlight important features of the dataset that a potential model must able to replicate.

\subsection{Model Comparison and Validation} \label{sec:IFS_model_comparison}

We use posterior predictive checks based on $\bst_M$ (Figure \ref{fig:pp_M_IFS}) and $\bst_\bR$ (Table \ref{tbl:theta_R}) as discussed in Section \ref{sec:model_validation} for validation of the marginal and dependence structure of our model.
% For model validation based on posterior predictive checks, we consider two sets of test statistics. 
% To assess the models in terms of the marginal parameter estimation, the summary statistics $\bss_M$ are considered as discussed in Section \ref{sec:model_validation}. 
% To assess the estimation of the dependence structure, we consider our test statistics $\bss_{\bR}$ to consist of the Spearman correlation coefficients between tooth-time pair $j$ and $j'$, corresponding to the representative set of elements 
% from the correlation matrix $\bR$; the selected set is presented in Table \ref{tbl:theta_R} with pairs denoted by $\left((l_j,t_j), (l_{j'},t_{j'})\right)$ with $l_j$ and $t_j$ representing the spatial location and the age at measurement.
% \org{[Section 2 indicates that we use $(l_j,t_j)$, so you need to swap one of these places for consistency.]}
The chosen correlation coefficients are associated with
% measure the (marginal) associated of the 
the same types of relationships that motivated our choices: $\bW^{(t)}$, $\bW^{(h)}$, $\bW^{(v)}$, and $\bW^{(pp)}$.
% evaluate dependencies that may arise between two tooth-time pairs due to temporal as well as spatial (horizontal, vertical) proximity in the IFS dataset. 
% To assess any correlation between the caries scores for primary and permanent teeth, we have included correlation coefficients between pairs of primary and permanent teeth located roughly at the same position.
% Note that we use the same set of predictive statistics for the marginal model, while the dependence statistics are separate from those used in model estimation.
%Note that the test statistics for the marginal model are the same statistics that were used for estimation, while the dependence parameters represent features that were not directly considered during ABC-MCMC.

% \begin{figure}[!bt]
%   \centering
%   \includegraphics[width=\textwidth]{pp_M1_plots_IFS.png}
%   \caption{Comparing posterior predictive plots for $\btheta_M$ from different models based on IFS data. The title of each plot shows the $ppp$ values for the corresponding summary statistic obtained from $\model{0}$ -- $\model{8}$ consecutively and the horizontal line indicates the observed summary statistics.}
%   \label{fig:pp_M1_IFS}
% \end{figure}

\begin{figure}[!bt]
  \centering
  \includegraphics[width=\textwidth]{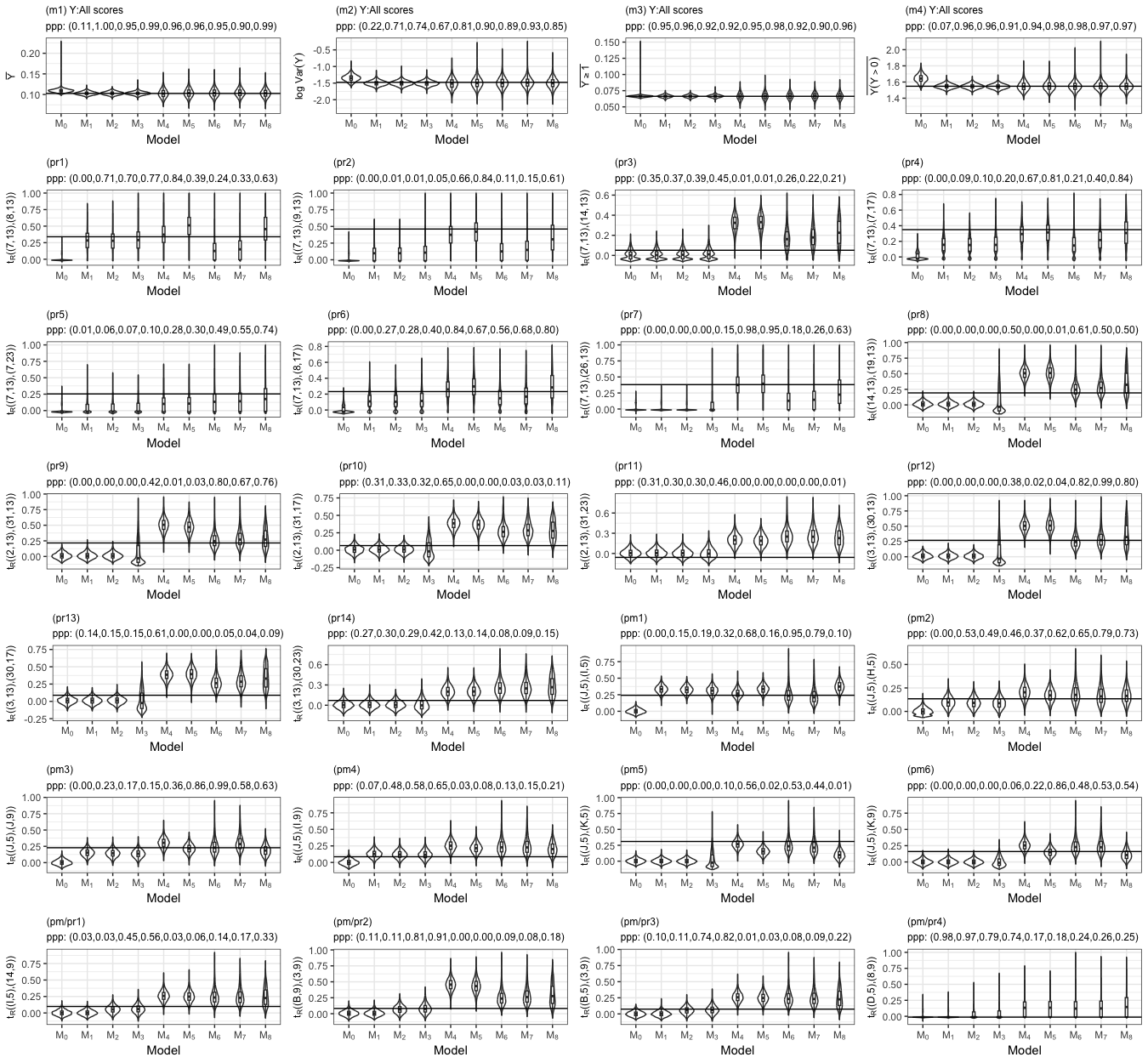}
  \caption{Comparisons of the posterior predictive plots for a subset of $\bst_M$ and $\bst_\bR$ elements based on IFS data. The title of each panel shows the $ppp$ values ($ppp = 0.00$ indicates $ppp < 0.01$) for the corresponding summary statistic obtained from $\model{0}$--$\model{8}$. The horizontal line indicates the observed summary statistics.}
  \label{fig:pp_M_R_IFS}
\end{figure}

We plot the posterior predictive distributions of the marginal and correlation test statistics for different models using box/violin plots.
Figure \ref{fig:pp_M_R_IFS}(m1-m4) shows a subset of the summary statistics chosen to assess the marginal model fit.
All the SAR models $\model{1}$--$\model{8}$ have comparable predictive performance while $\model{0}$, with independence misspecification, shows inferior performance in capturing the overall mean and the mean of non-zero scores, implying some sensitivity to the dependence structure.
Note that the predictive distributions are considerably wider in $\model{4}$--$\model{8}$ compared to those in $\model{1}$--$\model{3}$.
The dependence structure for $\model{4}$--$\model{8}$ includes the overall connectivity relationships $\bW^{(ct)}$ and $\bW^{(ce)}$ which assumes higher levels of correlation and less information from the data than the sparser $\model{1}$--$\model{3}$.
%Figure \ref{fig:pp_M_R_IFS}(a0,a7,a10) presents the posterior predictive distributions for a subset of presence model parameters: $\alpha_0$, $\alpha_7$ and $\alpha_{10}$, indicating very good coverage for the  observed test statistics (denoted by the horizontal line) lying near the center of the distributions. 
%However, the second row panels (b0, b7, b10) show some sensitivity to the dependence model with respect to the MLEs of the severity parameters $\beta_0$, $\beta_7$ and $\beta_{10}$; 
% of the predicted dataset under independence $\model{0}$, the observed statistics appear to be farther away from the center of their respective predictive distributions.
%However, all of the SAR models $\model{1}$--$\model{8}$ are comparable and centered at the IFS data value.
We  show the full set of posterior predictive plots regarding the marginal fit and provide a thorough discussion in Appendix \ref{appn:IFS_modeling}.
A comparison of alternative marginal model specifications is provided in Appendix \ref{appn:IFS_compare_marginal_models}.
% By comparing the position of the test statistics obtained from the IFS data as denoted by the horizontal line against the posterior predictive distributions, we can assess the goodness of fit for different models.
% The $ppp$ value provides a numeric summary and is reported above the figure.
% 

% \org{[FIX THE LABELS IN FIGURE 2]}

% We now turn to the
% The posterior predictive plots for the correlation parameters are presented in Figure \ref{fig:pp_R_IFS}. 
Figure \ref{fig:pp_M_R_IFS}(pr1-pr14) illustrates the predictive distribution of a representative set of correlations between pairs of permanent teeth, 
% at different temporal and spatial (horizontal/vertical) distances apart,
while (pm1-pm6) consist of those for primary teeth pairs, and (pm/pr1-pm/pr4) describe the primary-permanent correlation structure.
The distributions for all the correlation parameters under model $\model{0}$ are  centered around zero, inconsistent with the observed data for many of the considered correlations.
% in the case of model $\model{0}$. 
% Therefore, only those test statistics which have values close to zero, have good coverage \anish{(will include the panel numbers)}.
% This is further supported from the $ppp$ values.
Model $\model{1}$, accounting only for temporal and horizontal adjacency, provides a sparse correlation structure as evident from the underestimation in many of predictive distributions.
$\model{2}$ accounts for primary-permanent correlation and naturally performs better in (pm/pr1-pm/pr3).
Although $\model{3}$ connects all tooth-time pairs, its predictve plots turn out to be very similar to those for $\model{2}$, except having a wider coverage (particularly for vertically adjacent pairs such as pr8 and pr9).
This is because $\rho_v$ in $\model{3}$ is estimated to be close to zero, leading to similar point estimates of $\bR$, but with greater variability.

While $\model{3}$ provides the model with the most structure, it has poorer fit than $\model{4}$--$\model{8}$, which connect tooth-time pairs across space and time more directly through $\bW^{(ct)}$ and $\bW^{(ce)}$. 
However, we acknowledge that these models provide poor coverage for the IFS data value in a few Spearman correlations (pr11, pr13), where the observed correlations are close to zero, but these cases may not be representative of the full set of correlations for the type of adjacency represented. Overall, the equal connection models $\model{6}$--$\model{8}$ appear to perform the best, and we may say that $\model{8}$ is slightly better based on this set of $\bst_\bR$.

Appendix \ref{appn:IFS_modeling} provides an overall assessment of the model fit for the dependence structure, considering $ppp$ values from all the tooth-time pairs. Combined with the above, this leads us to take $\model{8}$ as the preferred model for the IFS data.
While we anticipated that the SAR model combining different spatial relationships based on anatomical consideration ($\model{3}$) would best fit the data, it turned out that the dependence model must include an equicorrelation component.  
As saliva allows food particles and oral bacteria associated with the development of caries to easily travel within the mouth, it is reasonable to conclude that a level of risk is shared across all teeth beyond the associations  among adjacent teeth encoded in $\bW^{(h)}$, and this perspective is consistent with the $\bW^{(ce)}$ structure in $\model{8}$.
% \org{[LEVY: Does this explanation seem plausible?]}
% It is important to point out that our preference for a particular model is based on both model assessment metrics and is grounded on intuitive arguments.
% One needs to employ rigorous model selection strategies if the aim is to formally validate model choice.

% Appendix \ref{appn:IFS_modeling} reports further investigations regarding different model choices with the IFS data analysis.  We report on the sensitivity of parameter estimation across these different model choices.
% % marginal models on dependence model specification, comparing uncertainty in estimation of the correlation parameters across different models. 
% We also compare the our Negative Binomial hurdle model against alternative marginal models including a Poisson-hurdle and Negative Binomial model (without zero inflation) in Appendix \ref{appn:IFS_compare_marginal_models}.

% \org{[what about some of the other model comparisons we've discussed?  hurdle poisson? NB (no hurdle)? can include this in the appendix but should make a statement/quick reference to them here]}

\subsection{IFS Interpretation under the Best Model} \label{sec:IFS_best_interpretation}

% \org{[If we are using equal tail intervals in the simulations, we should use tailed intervals in these results also.]}

We now interpret the best-fitting model $\model{8}$.
Coefficient estimates and 95\% credible intervals (CI) are shown in Table \ref{tbl:theta_M_D}.
% Recall that the presence model parameter $\alpha_k$ in equation (\ref{eq:presence}) represents  the change in the  log-odds of a non-zero caries score at the population level, when the value of the $k$-th predictor increases by one unit (i.e., one standard deviation).
It is evident from the presence model (\ref{eq:presence}) that the odds of a non-zero caries score decreases as the number of dental visits and the frequency of brushing increases, while the risk of caries increases with higher consumption of sugary beverages.
Total daily fluoride ingested does not appear to be  associated with the presence of dental caries in this analysis.
In the severity model (\ref{eq:severity}), $\beta_k$ represents the change in the log-transformed mean caries score for a unit change (one standard deviation) in the $k$-th predictor variable. 
Table \ref{tbl:theta_M_D} shows that the directions of the behavioral predictor effects are similar to what was seen in the presence model, except for a null effect for brushing.
% with predictors, except the total fluoride ingested. 
% However, total fluoride ingested and frequency in brushing appear to not have significant effects in decreasing the expected caries score.

\begin{table}[!tb]
\centering
\footnotesize
\begin{tabular}{lcccccc}
\hline
\multicolumn{1}{l}{Predictor}             &  & \multicolumn{2}{c}{Presence Model} &  & \multicolumn{2}{c}{Severity Model}      \\ 
\cline{3-4} \cline{6-7}
&  & $\hat{\alpha}$  & 95\% CI          &  & $\hat{\beta}$        & 95\% CI          \\ \hline
Intercept                              &  & -3.056                             & (-3.178, -2.934)            &  & -0.879                                   & (-1.015, -0.725)                   \\
Dental Visit (past 6 months)                          &  & -0.223                             & (-0.326, -0.115)            &  & -0.186                                   & (-0.296, -0.069)                   \\
Daily total fluoride ingested (mgF)          &  & 0.007                              & (-0.079, 0.087)             &  & -0.079                                   & (-0.172, 0.016)                    \\
Frequency of brushing (past 6 months)                 &  & -0.135                             & (-0.244, -0.032)            &  & -0.035                                   & (-0.133, 0.063)                    \\
Daily total sugar beverage (oz)          &  & 0.226                              & (0.145, 0.301)              &  & 0.210                                     & (0.124, 0.287)                     \\ \hline
Tooth Type                             &  & \multicolumn{1}{c}{}               & \multicolumn{1}{c}{}        &  & \multicolumn{1}{c}{}                     & \multicolumn{1}{c}{}               \\
\hspace{1em}Molar     &  & \multicolumn{1}{c}{Ref}            & \multicolumn{1}{c}{--}      &  & \multicolumn{1}{c}{--}                   & \multicolumn{1}{c}{--}             \\
\hspace{1em}Premolar &  & -0.905                             & (-0.971, -0.844)            &  & 0.138                                    & (0.042, 0.233)                     \\
\hspace{1em}Canine    &  & -0.497                             & (-0.539, -0.457)            &  & -0.293                                   & (-0.383, -0.213)                   \\
\hspace{1em}Incisor   &  & -0.512                             & (-0.558, -0.470)            &  & -0.154                                   & (-0.235, -0.073)                   \\
Primary                                &  & -0.068                             & (-0.147, 0.017)             &  & 0.268                                    & (0.151, 0.384)                     \\ \hline
Observation Time                       &  & \multicolumn{1}{c}{}               & \multicolumn{1}{c}{}        &  & \multicolumn{1}{c}{}                     & \multicolumn{1}{c}{}               \\
\hspace{1em} Age 5    &  & -0.107                             & (-0.195, -0.022)            &  & -0.094                                   & (-0.209, 0.012)                    \\
\hspace{1em} Age 9    &  & \multicolumn{1}{c}{Ref}            & \multicolumn{1}{c}{--}      &  & \multicolumn{1}{c}{--}                   & \multicolumn{1}{c}{--}               \\
\hspace{1em} Age 13   &  & -0.029                             & (-0.092, 0.033)             &  & -0.252                                   & (-0.365, -0.135)                   \\
\hspace{1em} Age 17   &  & 0.404                              & (0.341, 0.467)              &  & -0.006                                   & (-0.094, 0.088)                    \\
\hspace{1em} Age 23   &  & 0.154                              & (0.091, 0.222)              &  & 0.071                                    & (-0.030, 0.168)                    \\ \hline
NB size $\phi$                         &  & \multicolumn{1}{c}{}               & \multicolumn{1}{c}{}        &  & \multicolumn{1}{c}{$\hat{\phi}$ = 0.853} & \multicolumn{1}{c}{(0.722, 1.015)} \\ \hline
\\ \hline
SAR Parameters & & Mean & 95\% CI     &  & \multicolumn{2}{l}{}
\\ \hline
Temporal adjacency ($\rho_{h}$) & & 0.080 & (0.048, 0.103)  & & \multicolumn{2}{l}{}\\
Horizontal teeth adjacency ($\rho_{t}$)  & & 0.256 & (0.232, 0.278)  & & \multicolumn{2}{l}{}\\
Equal connection everywhere ($\rho_{ce}$)  & & 0.0013 & (0.0010, 0.0016)  & & \multicolumn{2}{l}{} \\ 
% & & & & & \multicolumn{2}{l}{(across time and location)} \\ 
\hline
\end{tabular}
\caption{Posterior parameter estimates of $\btheta_M$ and $\btheta_D$ for the IFS data under the best-fitting model $\model{8}$.
}
\label{tbl:theta_M_D}
\end{table}

We further observe that, compared to molar teeth, premolar, canine and incisor teeth are less likely to have a positive caries score, as $\hat{\alpha}_k$  are all negative. 
In the severity model, 
among those teeth that have a positive caries score,
% Looking into the $\hat{\beta}_k$ values shows that the 
molar teeth also have a higher mean  score compared to the incisors and canines.
The premolar teeth appear to have significantly higher mean caries score
(conditional on having caries) compared to molar teeth, even though they have much lower prevalence.
% compared with molars.
The odds of a non-zero caries score for permanent teeth does not appear to be significantly higher than that for the primary teeth, while for those with positive scores, primary teeth tend to have higher severity.
Further, we observe generally increasing odds of caries, with the smallest odds at age 5 and the highest odds at age 17, followed by a small reduction at age 23.
% significant differences between the odds of non-zero scores at age 9 and at other time points except at age 13.
% The odds appear to increase with age until age 23, where we observe a slight reduction.
% A large jump in $\hat{\alpha}$ from age 13 to 17 is observed though, indicating the odds of non-zero score at age 17 is much higher compared to the previous time points. 
% The pattern does not continue at age 23, nor is it seen in the mean score among the non-zeros.
% The mean caries scores shows an increasing trend with age starting from age 13, however the increment from age 9 is only significant at age 13.
Among teeth with caries, the mean score is smaller at 13 than age 9, but there are no other trends among measurement occasions.
% Compared to the caries score at age 9, age 13 has significantly smaller caries score, age 17 appears to be similar, while age 23 has a significantly higher expected score.

The estimation of the SAR model parameters, presented at the bottom of Table \ref{tbl:theta_M_D}, shows that both temporal and horizontal adjacency have a statistically significant impact on the dependence structure of the data, and unlike models $\model{1}$--$\model{3}$, a significant $\rho_{ce}$ helps maintain non-zero correlations between tooth scores farther away in time or horizontal distance.
Estimates of the correlation coefficients $\btheta_\bR$ are reported in Table \ref{tbl:theta_R}.

\section{Simulation Studies} \label{sec:simulation}

We further validate our approach by considering simulation studies 
similar to  the analysis of our IFS data.
% under different model specifications. 
% We have considered the IFS data str ucture to setup the simulation analysis. 
We imitate the same data structure as the IFS data by generating datasets with the same set of individuals having the same values for the predictors.
We consider a Negative Binomial hurdle model for the marginal distributions and  
the dependence is modeled using the Gaussian copula specified by the SAR model
$\model{4}$. 
This choice assumes all teeth at a given measurement occasion are connected through the adjacency matrix $\bW^{(ct)}$ and  imposes autoregressive correlation across time through   $\bW^{(t)}$.
% While $\model{8}$ is preferred to best describe the IFS data, we found model $\model{4}$ to be a close competitor and the model makes sense intuitively in the context of any dental caries dataset.
% Moreover compared to $\model{4}$, model $\model{5}$ additionally has the horizontal teeth connectivity. Therefore it would be interesting to generate simulated datasets from model $\model{4}$ and analyze their $\model{5}$ model fit from the perspective of model assessment/selection.
% To that end we have chosen $\model{4}$ to be the true data generating model for the simulation setting.
The data-generating values of all the parameters are similar to  the  point estimates from the IFS analysis (see Appendix \ref{appn:simulation_setup}).
% \org{[update to justify choice of M4.]}
We generate 100 datasets accordingly, and for each, run three ABC-MCMC chains for 60,000 iterations  with $h=0.1, 1, 10, 30$.
After regression adjustment, burn-in and thinning, we obtain 9000 posterior samples for each dataset.
% to generate 3 chains of 60,000 posterior samples followed by regression adjustment, removal of burn-in and thining to obtain a final sample size of 3,000 per chain.
% We also compare inference with and without the regression adjustments.

We compare  estimation of our  ABC-MCMC algorithm against ABC rejection and importance samplers
% We employed the standard rejection ABC algorithm 
\citep[][Chapter~1]{Sisson2018}, without and with regression adjustment.
See Appendix \ref{appn:rej_imp_algo} for specifics  on our implementation of these approaches.
% The samples obtained from each are regression adjusted to yield the final set of samples to be used for comparison.
% We briefly mention here that b
% Both 
% the rejection and importance 
These algorithms
% are run as long as required for drawing 
draw 250,000 samples from  the prior and the proposal distributions, respectively,  
which requires  similar computational time as the ABC-MCMC sampler.
% generating 185,000 total samples.
% Further note that, 
In our implementation of the rejection algorithm, only 0.1\% of the samples closest to $\bss_{\obs}$ are accepted, which is equivalent to having a uniform kernel with data-determined bandwidth.
% determined by the maximum of the smallest 0.1\% distances.
% (see Appendix \ref{appn:rej_imp_algo}).
% In our simulated datasets, $h^*$ turned out to have pretty large values more than 5,000.
% This is much larger than the maximum $h$ considered for the Gaussian kernels in MCMC and importance samplers.
% \org{[REMOVE LAST TWO SENTENCES???]}

We assess  estimation performance using a range of metrics.
% Let $\theta_{M;k}$ represent the $k$-th component of the marginal parameter vector $\btheta_M$. 
% When assessing estimation accuracy for the dependence structure, we consider estimation of the collection of correlation from $\bR(\theta_D)$ presented in Table \ref{tbl:theta_R}; $\btheta_\bR$ represents this 24-vector, and $\theta_{\bR;k}$ the $k$-th element.
Let $\theta_{\cdot;k}$ denote either $\theta_{M;k}$ or $\theta_{\bR;k}$, the $k$-th component of $\btheta_M$ or $\btheta_\bR$, respectively.
Let $\hat{\theta}_{\cdot;k,b}$ represent the estimate obtained from the $b$-th simulated data, and $\theta_{\cdot;k,0}$ denote the corresponding true parameter value.
% \org{[don't use $h$ here since that represents bandwidths.  Maybe $b$??]}
Average bias and root mean squared error (RMSE) for each parameter 
% across all the simulated datasets 
are given by $\frac{1}{B} \sum_{b=1}^B (\hat{\theta}_{\cdot;k,b} - \theta_{\cdot;k,0})$ and $\sqrt{\frac{1}{B} \sum_{b=1}^B (\hat{\theta}_{\cdot;k,b} - \theta_{\cdot;k,0})^2}$, respectively.
To investigate 
% whether the posterior samples for parameters cover their respective true values, we compare their empirical coverage rates across multiple  significance levels using 
posterior coverage, we use
a similar strategy as in \citet{Uddin2023}.
Letting $(\hat{\theta}_{\cdot;k,b}^{(l_\zeta)}, \hat{\theta}_{\cdot;k,b}^{(u_\zeta)})$ be a $100\zeta\%$ 
equal-tailed
credible interval,
% for parameter $\theta_{\cdot;k}$ obtained for $b$-th dataset.
% with $\hat{\theta}_{\cdot;k,h}^{(u_\zeta)}$ and $\hat{\theta}_{\cdot;k,h}^{(l_\zeta)}$ representing the corresponding $100\zeta\%$ upper and lower bounds.
we define the empirical coverage rate (ECR) 
% of the $100\zeta\%$ interval 
for parameter $\theta_{\cdot;k}$ as $\ECR(\zeta) = \frac{1}{B} \sum_{b=1}^{B} \bI \left(\theta_{\cdot;k,0} \in (\hat{\theta}_{\cdot;k,b}^{(l_\zeta)}, \hat{\theta}_{\cdot;k,b}^{(u_\zeta)})\right)$.
To summarize this coverage,
% for the parameter $\theta_{\cdot;k}$,
we compute $\ECR(\zeta)$ for each $\zeta = 0.025, 0.050, \ldots, 0.975$ and obtain the overall empirical coverage score ($\OECS$) as the area under the $\ECR$ curve (Appendix \ref{appn:sim_results}).
An $\OECS = 0.5$ 
indicates 
% optimal coverage rate,
that the empirical coverage matches the target coverage rate on average,
% $100\zeta\%$ at all $\zeta$; 
while greater or less than 0.5 indicates over- or under-coverage, respectively.
% We  graphically depict these empirical coverage curves in the Appendix \ref{appn:sim_results}.
% Thus, for every parameter in $\btheta_M$ and selected elements of $\bR$ we obtain the corresponding $\OECS$, and the closer it is to 0.5 the better is the overall coverage for the parameter.
Additionally, to characterize the level of concentration in the estimated posterior distribution, we report the average
width of the 80\% credible intervals for each parameter.
% We also consider the width of the 80\% credible interval to be another important metric, as it captures the uncertainty around the point estimate.
% A narrow credible interval will indicate higher precision in parameter estimation.
% It is worth noting, however, that an MCMC chain, stuck in a local mode, may also report high concentration around the mode, and such behavior does not  necessarily mean better estimation.
% Therefore, we have run multiple MCMC chains and investigated their mixing and convergence.
% We compute the effective sample size for all the parameters.

%We compute the estimation accuracy metrics for each of the 27 marginal parameters in $\btheta_M$ and 24 correlation parameters in $\btheta_\bR$, averaged over all the simulated datasets.
%To compare the sampling strategies based on these metrics we use a rank aggregation approach proposed by \cite{Pihur2007}.
%\anish{(Brief description how this works. The way to do it that we have in mind, is first we will rank the samplers based on individual metrics aggregated over marginal parameters, correlation parameters and combined (across marginal and correlation). Once we have the combined ranks for each metric, we aggregate them to have overall ranks. The related table is Table \ref{tbl:rank_aggregation}.)}

We use box/violin plots to show the performance on each criterion across all the marginal $(\btheta_M)$ and correlation $(\btheta_\bR)$ parameters.
% distribution across parameters. 
% We compare boxplots/violinplots obtained for different sampling strategies 
% in Figures \ref{fig:model_fit_M_compare_sim} and \ref{fig:model_fit_R_compare_sim}.
% \org{[we used violin plots in the IFS section.  how do the violin plots compare to the boxplots?]}
As in the IFS analysis, we include model $\model{0}$ with the independence misspecification and use a Gibbs sampler to fit it.
% It is worth noting here that the model $\model{0}$, with independence misspecification, has same marginal structure as the true model and marginal parameters are estimated using a Gibbs sampler.
Hence, we treat $\model{0}$ as a baseline and compare the performances of the ABC strategies targeting the true model against it.
We only show the results for the importance sampler at bandwidth $h=10$,
since this is the $h$ with the best performance of this sampler and of our method.
% and therefore are only included in the figures.
Performance 
% of the importance sampler 
at other bandwidths can be found in Appendix \ref{appn:sim_results}.
Also note that the number of unique samples for ABC-MCMC with $h=0.1$ was typically too low to perform regression adjustment, and hence this combination is excluded.

% \org{[REWORKING TO SHORTEN FOR MAIN MANUSCRIPT.  MOVE THE EXTENDED DISCUSSION TO C.1.]}

% The table also provides sampler ranks based on the width of 80\% credible interval.
% Note that, $\model{0}$ does not have a rank for $\btheta_\bR$. Due to its independence misspecification the correlation parameters, instead of getting estimated, are set to be zero. 
% Hence the credible interval widths for $\btheta_\bR$ are not defined for $\model{0}$.
% Additionally, since credible interval width is not usually considered as a metric for measuring estimation performance, the corresponding ranks are not used to find the overall rank of the samplers.

% \org{[need to add some initial direction for Shoumi about what to add here and possibly include a skeleton table for her to fill out.]}

To further summarize differences across the metrics, we compute a ranked order of the accuracy (lowest absolute bias, lowest RMSE, absolute difference between OECS and 0.5) across all methods for each parameter.  
These ranks are aggregated across the 27 parameters in $\btheta_M$, the 24 parameters in $\btheta_\bR$, and the 51 parameters in $(\btheta_M,\btheta_\bR)$ to obtain a consensus ranking of the methods for each metric.
Rank aggregation is performed by minimizing the sum of Spearman footrule distances across all parameters \citep{Pihur2007, pihur2008finding}. 
Additionally, an overall aggregated ranking is found by combining the ranks for all three metrics across parameters ($\btheta_M$, $\btheta_D$, or combined).
% A ranking of the CI widths is also found but not included in the overall ranking as narrower CIs do not necessarily imply more appropriate inference. 
Further details of the rank aggregation approach can be found in Appendix \ref{appn:simulation_setup}.
For the independence model $\model{0}$, we consider all estimates of $\bR(\btheta_D)$ to be zero, compute OECS based on their credible intervals being $[0,0]$, and exclude consideration of the width of the 80\% CI.

\begin{figure}[!tb]
    \centering
    \includegraphics[width=\textwidth]{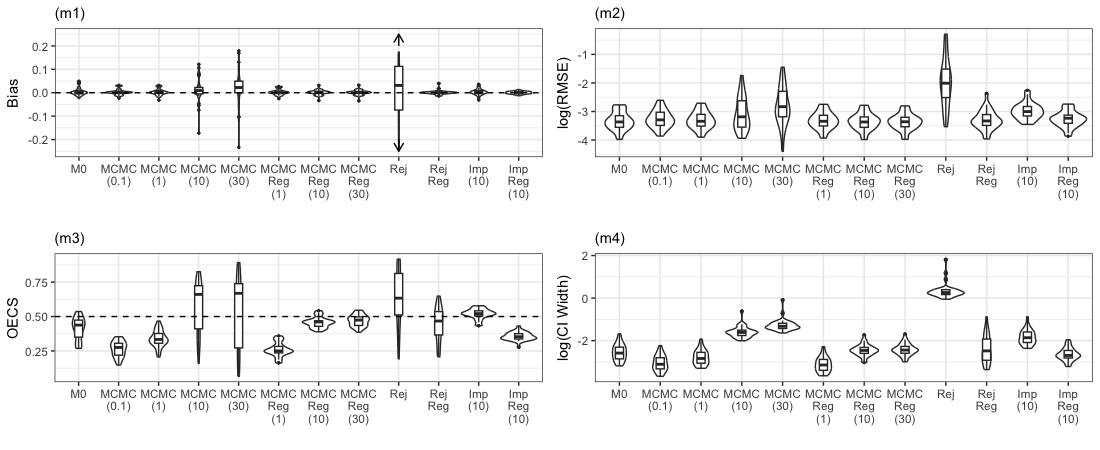}
    \caption{Estimation performance for the marginal parameters $\btheta_M$ from different posterior sampling strategies. In panel (m1), arrows in the boxplot for the ABC rejection indicate truncation, as its tails range from -0.30 to 0.75. All methods except $\model{0}$ are ABC.}
    \label{fig:model_fit_M_compare_sim}
\end{figure}

In Figure \ref{fig:model_fit_M_compare_sim}(m1), we observe that 
$\model{0}$ performs fairly well in terms of bias for the marginal parameters.
% the bias for the marginal parameters under the misspecified model $\model{0}$ is centered around zero, with a small variability across parameters. 
The bias for ABC-MCMC increases with bandwidth.
% , as higher bandwidth results in larger approximation error. 
% In particular for bandwidth 10 and 30, some parameters 
% % appear to 
% have large bias.
% that are considered as outliers.
However,  post-sampling regression adjustment significantly improves the bias across all the bandwidths and provides competitive performance when compared with $\model{0}$. 
The ABC  rejection sampler performs the worst, with a wide range of bias among the marginal parameters.
This is unsurprising as the 0.1\% rejection rate selects a (uniform kernel) bandwidth of $h^*>20000$, yielding samples of $\Delta(\bss)$ which are orders of magnitude larger than those from the alternative methods.  
% \org{[IS THIS TRUE???]}
% By using samples that are widely different from $\bss_\obs$, biased parameter estimation can result,
% since the corresponding uniform kernel has very large bandwidth $h^*$, accepting parameters even when the generated summary statistics are widely different from $\bss_{\obs}$, which in turn results in unreliable point estimates.
Noticeable improvement is achieved with regression adjustment.
The ABC  importance sampler is comparable to that of ABC-MCMC sampler for $h=10$.
% , due to a much stricter Gaussian kernel compared to the rejection sampler.
The first row of 
Table \ref{tbl:rank_aggregation} indicates that ABC importance sampler with regression adjustment performs best,
although Figure \ref{fig:model_fit_M_compare_sim}(m1) indicates only slight differences amongst all methods for the criteria (except for rejection and MCMC $h=10,30$).
% It is worth noting that, while the other samplers have comparable bias except rejection, MCMC with $h=10$ and 30 we assign them different ranks which may overemphasize their differences.
% In particular MCMC ($h=10$) with regression adjustment performs fairly well while ranking 8 among the 12 sampling strategies we considered.
% \org{[For conciseness, we should just call it ABC rejection, not ABC rejection]}

\begin{table}[!tb]
	\centering
	\footnotesize
	\begin{tabular}{lccccclccclcccc}
		\hline
		& $\model{0}$          & \multicolumn{4}{c}{MCMC}                                                                  &                      & \multicolumn{3}{c}{MCMC+Reg}                                       &                      & Rej                  & Rej+Reg              & Imp                  & Imp+Reg              \\ \cline{3-6} \cline{8-10} \cline{14-15} 
		$h$                                      &                      & 0.1                  & 1                    & 10                   & 30                   &                      & 1                    & 10                   & 30                   &                      &                      &                      & 10                   & 10                   \\ \hline
		Bias                                     &                      &                      &                      &                      &                      &                      &                      &                      &                      &                      &                      &                      &                      &                      \\
		\hspace{1em}$\btheta_M$ & 6                    & 3                    & 7                    & 10                   & 11                   &                      & 5                    & 8                    & 4                    &                      & 12                   & 2                    & 9                    & 1                    \\
		\hspace{1em}$\btheta_\bR$ & 10                   & 6                    & 4                    & 7                    & 9                    &                      & 1                    & 2                    & 5                    &                      & 12                   & 11                   & 8                    & 3                    \\
		\hspace{1em}Combined    & 10                   & 4                    & 5                    & 7                    & 9                    &                      & 2                    & 1                    & 6                    &                      & 12                   & 11                   & 8                    & 3                    \\ \hline
		RMSE                                     &                      &                      &                      &                      &                      &                      &                      &                      &                      &                      &                      &                      &                      &                      \\
		\hspace{1em}$\btheta_M$ & 2                    & 7                    & 5                    & 8                    & 11                   &                      & 6                    & 3                    & 1                    &                      & 12                   & 4                    & 10                   & 9                    \\
		\hspace{1em}$\btheta_\bR$ & 10                   & 4                    & 3                    & 6                    & 9                    &                      & 1                    & 2                    & 7                    &                      & 12                   & 11                   & 8                    & 5                    \\
		\hspace{1em}Combined    & 9                    & 6                    & 3                    & 7                    & 11                   &                      & 1                    & 2                    & 4                    &                      & 12                   & 10                   & 8                    & 5                    \\ \hline
		OECS                                     &                      &                      &                      &                      &                      &                      &                      &                      &                      &                      &                      &                      &                      &                      \\
		\hspace{1em}$\btheta_M$ & 4                    & 10                   & 7                    & 8                    & 11                   &                      & 12                   & 3                    & 2                    &                      & 9                    & 5                    & 1                    & 6                    \\
		\hspace{1em}$\btheta_\bR$ & 12                   & 5                    & 1                    & 6                    & 9                    &                      & 8                    & 2                    & 7                    &                      & 11                   & 10                   & 4                    & 3                    \\
		\hspace{1em}Combined    & 12                   & 8                    & 5                    & 6                    & 9                    &                      & 10                   & 2                    & 3                    &                      & 11                   & 7                    & 1                    & 4                    \\ \hline
		% WCI                                      & \multicolumn{1}{l}{} & \multicolumn{1}{l}{} & \multicolumn{1}{l}{} & \multicolumn{1}{l}{} & \multicolumn{1}{l}{} &                      & \multicolumn{1}{l}{} & \multicolumn{1}{l}{} & \multicolumn{1}{l}{} &                      & \multicolumn{1}{l}{} & \multicolumn{1}{l}{} & \multicolumn{1}{l}{} & \multicolumn{1}{l}{} \\
		% \hspace{1em}$\btheta_M$ & 5                    & 2                    & 3                    & 10                   & 11                   & \multicolumn{1}{c}{} & 1                    & 6                    & 7                    & \multicolumn{1}{c}{} & 12                   & 8                    & 9                    & 4                    \\
		% \hspace{1em}$\btheta_\bR$ & -                    & 5                    & 7                    & 10                   & 11                   & \multicolumn{1}{c}{} & 1                    & 6                    & 8                    & \multicolumn{1}{c}{} & 2                    & 4                    & 9                    & 3                    \\ \hline
		Overall                                  &                      &                      &                      &                      &                      &                      &                      &                      &                      &                      &                      &                      &                      &                      \\
		\hspace{1em}$\btheta_M$ & 4                    & 10                   & 7                    & 8                    & 11                   &                      & 5                    & 3                    & 1                    &                      & 12                   & 2                    & 9                    & 6                    \\
		\hspace{1em}$\btheta_\bR$ & 10                   & 4                    & 1                    & 6                    & 8                    &                      & 5                    & 2                    & 9                    &                      & 12                   & 11                   & 7                    & 3                    \\
		\hspace{1em}Combined    & 10                   & 5                    & 4                    & 7                    & 9                    &                      & 6                    & 1                    & 2                    &                      & 12                   & 11                   & 8                    & 3                    \\ \hline
		
	\end{tabular}
	\caption{Aggregated ranks of the different sampling strategies based on their estimation performance across marginal $(\btheta_M)$ and correlation $(\btheta_\bR)$ parameters. All methods except $\model{0}$ are ABC.}
	\label{tbl:rank_aggregation}
\end{table}

Similar increases in RMSE as $h$ increases are  evident in Figure \ref{fig:model_fit_M_compare_sim}(m2).
% with regard to change to bandwidth.
% For smaller bandwidths 0.1 and 1, RMSE of the ABC-MCMC samples are much smaller compared to that obtained from higher bandwidths.
% Regression adjustment helps improve the samples from larger bandwidths to have comparable RMSE.
% \org{[Don't say effective sample size.  Also, if you haven't adjusted these samples, you shouldn't include them in the figure.]}
The rejection sampler again performs the worst.
% with largest RMSE, 
% although regression adjustment yields competitive estimation.
% which is improved when the samples are regression adjusted.
Regression adjustment is helpful throughout, and is particularly noticable for the rejection sampler.
% Although MCMC + Reg ($h=10$) appears to be very similar to Imp + Reg ($h=10$), Table \ref{tbl:rank_aggregation} gives a lower rank to the latter strategy.
Parameter estimates from the importance sampler have slightly higher RMSE before regression adjustment (compared to the corresponding MCMC($h=10$)), and after regression adjustment  performance is somewhat comparable to the adjusted MCMC (although the aggregate rank is worse).
However, the distribution of the importance weights is highly imbalanced; on average, fewer than 15  of 250,000 samples account for more than 50\% of the posterior distribution, and hence Imp and Imp+Reg estimation  effectively relies on a very small number of posterior samples.
MCMC($h=30$) with regression adjustment is the best ranking method  in terms of RMSE, 
followed by $\model{0}$ and MCMC($h=10$) with regression (Table \ref{tbl:rank_aggregation}).
% better than $\model{0}$ ranking second and MCMC ($h=10$) being at the third place.

Turning to Figure \ref{fig:model_fit_M_compare_sim}(m3), we recall that the optimal value for the area under the ECR curve is 0.5. 
% For each sampling strategy, $\OECS$ is measured for every parameter from its ECR curve, and a boxplot is generated from these scores.
$\model{0}$ frequently undercovers the true parameter values.
ABC-MCMC with $h=0.1, 1$ yield good point estimates but consistently undercovers, while larger bandwidths ($h=10, 30$) overcover.
This is further evidenced by Figure \ref{fig:model_fit_M_compare_sim}(m4) with the narrowest CIs for smaller bandwidths and very wide intervals when the bandwidth is larger.
This is because ABC-MCMC algorithm with small $h$ is unable to effectively explore the full posterior, since the smaller bandwidth leads to low acceptance rates and small steps. 
The regression adjustment provides improvement, except under $h=1$ where it
% fails to solve this, and potentially 
exacerbates the issue by further concentrating the posterior. 
% In contrast, for the larger bandwidths ($h=10,30$), the MCMC posterior is overly disperse and over-covers, but the regression adjustment effectively corrects the tails of the ABC posterior to achieve approximately correct coverage.
The rejection sampler, with a much wider kernel, 
% allowing generated data to lie further away from the observed data, 
results in overcoverage with widest credible intervals.
While the unadjusted importance sampler (with poorer point estimation) provided good coverage, Imp+Reg (which had good estimation) undercovers on average.
This may be due to the small number of corrected samples getting most of the weights, causing over-shrinkage of the CIs.
Overall, Imp($h=10$) and MCMC+Reg($h=10, 30$)
show  acceptable coverage across $\btheta_M$.
% has the best ranking for marginal parameter coverage, with MCMC+Reg ($h=10, 30$) performing similarly.

% The widths of the credible intervals (Figure \ref{fig:model_fit_M_compare_sim}(d)) are sensitive to the bandwidth parameter and remain sensitive to a lesser degree even after regression adjustment. 
% These results complement the OECS conclusions. 
% Bandwidths $h=0.1$ and 1 yield the most narrow intervals, but OECS verifies that these are narrower than appropriate. 
% ABC-MCMC (h=10,30) have the widest intervals, which are substantially shrunk after applying regression adjustment.
% The credible intervals are widest in the case of rejection samplers since the resulting bandwidth yields the most dispersed target distribution.

\begin{figure}[!tb]
    \centering
        \includegraphics[width=\textwidth]{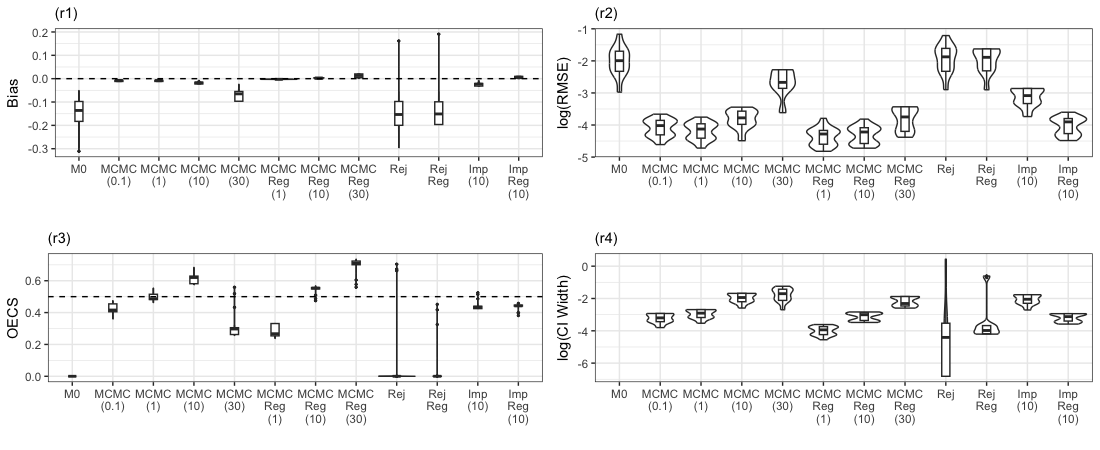}
    \caption{Estimation performance for the correlation parameters $\btheta_\bR$ from different posterior sampling strategies. All methods except $\model{0}$ are ABC.}
    \label{fig:model_fit_R_compare_sim}
\end{figure}

Similar plots for the selected correlations  are shown in Figure \ref{fig:model_fit_R_compare_sim}. 
$\model{0}$ assumes $\bR = \bI$ and has high bias and poor performance in all  metrics.
The rejection sampler also has severe bias and estimation error, showing complete failure to learn the dependence structure;  
regression adjustment provides little help here.
This is also reflected by the consistently poor ranks in Table \ref{tbl:rank_aggregation}.
For the ABC-MCMC samplers, the estimation performance deteriorates  across all  metrics as the bandwidth increases and is more sensitive to $h$ than the marginal parameters.
% with bandwidth 30 being noticeably worse than the rest, compared to what we observed in the case of marginal parameters in Figure \ref{fig:model_fit_M_compare_sim}. 
The regression adjustment of the ABC-MCMC samples improves the metrics, with the sampler having the best overall performance at $h = 10$. 
% It appears from Table \ref{tbl:rank_aggregation} that MCMC ($h=10$) with regression adjustment ranks near the top of the list across all performance metrics.
% The unadjusted ABC sampler at $h=1$ also performs fairly well (not as much for the marginal parameters), while regression adjustment made the coverage worse. 
% This may be due to having not enough variability in the posterior samples that regression adjustment was not reliable and ended up in over-correction.
% Rejection sampler performs the worst as expected, similar to what was observed for the marginal parameters in Figure \ref{fig:model_fit_M_compare_sim}.
% Regression adjustment does not noticeably improve the performance metrics except shrinking the credible intervals.
% The performance of the importance sampler  is worse for the correlation parameters compared to the marginal parameters we noted earlier, and ranks lower than the regression adjusted MCMC samplers in bias and RMSE.

% \org{[HOW IS OECS FOR M0 NOT RANK 12?  I DON'T SEE HOW REJ CAN BE BETTER.]}

The last three rows of Table \ref{tbl:rank_aggregation} provides the sampler ranks taking all the metrics together.
MCMC+Reg($h=30$) is the best with regard to $\btheta_M$ fit, while Rej+Reg and MCMC+Reg($h=10$) are in second and third place.
For $\btheta_\bR$, MCMC($h=1$) performed best combining all the metrics, with MCMC+Reg($h=10$) second best and Imp+Reg($h=10$) in third place.
Combing all the performance metrics across all the parameters, MCMC+Reg($h=10$) is found to be the best.

Tables \ref{tbl:sim_compare_M} and \ref{tbl:sim_compare_R} in Appendix \ref{appn:sim_results} include further details on the bias, RMSE, and coverage  for each parameter.
In addition to the aforementioned analyses,
we conduct some further investigations about the sensitivity of the importance sampler to the bandwidth choice, an alternative regression adjustment approach, and estimation performance under misspecification of the SAR model
in Appendices \ref{appn:sim_results}, \ref{appn:sim_compare_regression_strategies} and \ref{appn:sim_model_misspecification}, respectively.
% While the Figures \ref{fig:model_fit_M_compare_sim} and \ref{fig:model_fit_R_compare_sim} show overall sampler performances across all the parameters, the estimation accuracy for the individual parameters are reported in Table \ref{tbl:sim_compare_M} and \ref{tbl:sim_compare_R} in Section \ref{appn:sim_results}.

% along with a discussion on sensitivity to bandwidth choice.
% Alternative strategies for regression adjustment are explored in model misspecification.
% % A discussion on sensitivity of the importance sampler to bandwidth choice and model misspecification are also included.
% We further investigate features of our regression adjustment approach in Appendix \ref{appn:sim_compare_regression_strategies}, and we compare estimation under misspecfication of the SAR model in Appendix \ref{appn:sim_model_misspecification}.

% We add some final comments on the simulation analyses.
% % based on the observations noted above.
% While MCMC with small bandwidths theoretically leads to better approximation of the true posterior, we often fail to attain good mixing/convergence in the MCMC sampling and estimation is negatively impacted.
% Larger bandwidth  achieves better convergence by allowing chains to move to samples farther away from observed data at the cost of introducing more bias in the point e100stimates, but as long as $h$ is not too large, regression adjustment can effectively shift parameter sample back $\bss_\obs.$
% Bandwidth $h=10$ appears to strike a balance between the two in our experiments.
We make some final comments regarding the simulation analysis.
In this setting, a bandwidth of $h=10$ appears to strike a balance between good MCMC mixing, while also reducing approximation error associated with the difference between the true posterior \eqref{eq:posterior} and the ABC posterior \eqref{eq:posterior_ABC}, after regression post-processing.
% Regression adjustment leads to even smaller bias in the point estimation.
%The ranks provided in Table \ref{tbl:rank_aggregation} shows that regression adjusted MCMC $(h=10)$ is consistently better in estimating $\btheta_M$ as well as $\btheta_\bR$ across all performance metrics and overall (combining $\btheta_M$ and $\btheta_\bR$ across all metrics) is the best model fitting strategy.
In terms of the aggregated ranks, MCMC+Reg($h=10$) was the only method with good performance (low ranks) for both point estimation and uncertainty quantification.
The data structure considered here is similar to that of the IFS data, in terms of the number of observations and the proportion of zero inflation, so these results further validate the choice of $h=10$ in Section \ref{sec:IFS_study}.  
For other data settings, the user will need to run ABC-MCMC for multiple bandwidths and determine which $h$ should be used based on similar considerations to those made here.

\section{Discussion} \label{sec:discussion}

In this work, we have proposed a zero-inflated spatio-temporaly correlated count data model.
% , where the population-level effects of the predictors can be estimated. 
A Negative Binomial hurdle model defines each marginal distribution, facilitating population-level interpretations.
% while their dependence structure is modeled using a Gaussian copula. 
A   Gaussian copula specifies dependence using a Simultaneous Autoregressive model with 
% , where the correlations across the margins are expressed through 
multiple adjacency relationships.
Standard Bayesian inference is unavailable since the resulting likelihood is intractable, and
% making standard Bayesian inference challenging. 
we employed ABC to estimate an approximate posterior distribution.
The posterior samples from ABC-MCMC are further processed using a regression adjustment, and the final adjusted samples are used for inference and validated with posterior predictive checks.

We have employed this  model to analyze the Iowa Fluoride Study data. 
% The population-level effects of the predictors including dental appointments, total fluoride ingested, frequency of brushing, amount of sugar beverages on the presence of caries and its severity have been determined. 
The  estimated effects are consistent with conclusions from elsewhere in the dental literature; namely, that low dental visit frequency, low brushing frequency, high soda intake, increased age, and molar tooth type are all risk factors for the appearance and/or severity of caries.
Moreover, our approach finds predictor effects at the population-level,  unlike \cite{Choo-Wosoba2018, Kang2021, Kang2023}, where the effects only had individual-level interpretations.
The flexibility of our model in specifying the dependence enabled us to fit and compare a variety of potential spatio-temporal correlation structures. 
The structures based on equal connection choices turned out to be more consistent with the IFS data than structures assuming conditional independencies through spatial structural relationships.
% \org{[adjusted to make this more concrete about the conclusions drawn from the analysis, instead of just describing the methodology]}
% considered in this work, that are conveniently expressed in terms of temporal proximity, horizontal and vertical teeth proximity as well as primary-permanent proximity. 
% Model fitting with different correlation structures was assessed using posterior predictive checks.
% \anish{Need to include discussion on model selection.} 

% We have demonstrated through both simulations and the IFS analysis that employing classical ABC rejection and importance sampling algorithms 
% struggle to yield a stable inference.
% While not included  here, additional experiments using a 
% sequential Monte Carlo (SMC)  algorithm and its extensions for ABC \citep{Sisson2007, bonassi2015} also failed to achieve  balanced weights in our experiments.
% By obtaining posterior samples through (correlated) Markov chains, we found  success, but
% our proposed ABC-MCMC algorithm also had several challenges.
% The bandwidth was treated as a tuning parameter, and a set of potential values was chosen based on the convergence and mixing of the Markov chains.
% More structured strategies for bandwidth selection and/or improving MCMC mixing with small bandwidths could further improve estimation performance.
% One such approach could incorporate parallel tempering with chains  run in parallel at different values of $h$ with state transitions between the chains \citep{Baragatti2013}.

Choosing summary statistics is  a critical step  in ABC since the model learns  only through $\bss(\bsy)$ and not from the likelihood of $\bsy$.
Our choice of statistics has been motivated through the  auxiliary likelihood approach \citep{Drovandi2011}, where we consider 
estimates from a simpler model to assess data fit.
% summarize the marginal and correlation structure parameters.
% a hurdle model under independence misspecification to learn  the marginal parameters $\btheta_M$ and use a naive estimate of the normal quantiles to fit the SAR model defining the correlation structure.
% while an approximate SAR model was used to describe the correlation structure in terms of the dependence parameters ($\btheta_D$).
% Summary statistics for each set of parameters were MLE of their respective model.
Posterior predictive checks indicate that our model may struggle with some features of the dependence, requiring more flexible $\bR(\btheta_D)$ and/or additional parameters in $\btheta_D$.  However,  this may also be due in part to inadequency of $\bss_D(\bsY)$ for learning $\btheta_D$.
Rather than approximating the SAR model,
% model estimates based on 
% imputed quantiles, 
an alternative could  directly use the pairwise correlation estimates, similar to how these $\bst_{\bR}(\bsy)$ were used for posterior predictive assessment.
There are  more than 8,000 such correlations in the IFS data, so some level of dimension reduction  would be  required.
% so some level of 
% dimension reduction.
% would be needed.
% A priori selection of relevant pairs (as in the set from Table \ref{tbl:theta_R}) could be considered, as well as 
% More sophisticated strategies could be considered
This could potentially be done using
 entropy-minimizing subset selection
 % under information theoretical considerations 
 \citep{Nunes2010}, projection methods \citep{Fearnhead2012}, or from an a priori chosen collection.
% While a small number of dependence parameters is enough to define the full correlation matrix when SAR model correctly specifies the dependence model, the corresponding number of summary statistics may not be enough to account for the large number of correlation coefficients they are accounting for.
% A much larger set of statistics summarizing a range of representative correlation coefficients would likely be more informative.
% A potential approach in this direction would be to start with the set of summary statistics we used for posterior predictive checks, or an even larger set, and use an information theoretic subset selection method \citep{Nunes2010} to reduce the dimension.
% Another less computationally intensive approach could be using a projection method \citep{Fearnhead2012} that, instead of selecting a subset from a large group of statistics, considers different linear projections.

We have performed model comparison simultaneously with  model validation by considering the predictive distributions.
Standard model selection for ABC typically involves generating parameters and their corresponding data from the prior under each model and applying a rejection algorithm using  summary statistics shared across all models \citep{Fagundes2007, 
% Cornuet2010, 
Pudlo2016}.
However, as demonstrated in the simulation and IFS examples, 
% algorithms drawing  parameters independently 
% from the prior (or another proposal distribution) 
approaches like these
fail to learn the regions of highest posterior mass and, therefore, will not provide trustworthy estimates of the posterior model probabilities in our context.  
Other standard Bayesian model selection methods such as DIC 
% and WAIC 
\citep{Spiegelhalter2002} 
 are not useful here since the likelihood is intractable.
% Another difficulty in ABC based inference, connected with the choice of summary statistics, is model selection.
% Since the likelihood is not directly available, implementation of standard DIC, WAIC \citep{Gelfand1994, Gelman2014} based model selection strategies is not feasible.
% Any indirect model selection strategy treats the model as a parameter with an associated prior distribution and formulates the model selection problem as estimating the model parameter in an ABC framework. 
% Many samples each consisting of a model, parameter and a corresponding data triplet are generated and a probability for each model conditional on the observed data are estimated \citep{Fagundes2007, Cornuet2010, Pudlo2016}.
% The summary statistics required in this approach would ideally be informative enough for all the models while also being low-dimensional, both of which are practically difficult to achieve \citep{Robert2011}.
% Moreover, finding region of posterior concentration by drawing parameters from a non-informative prior turned out to be very inefficient in our experiments.
Hence, we have opted for model comparison   based on intuitive arguments guided by our posterior predictive checks, but further work on ABC model selection methods would be useful in this context.
As noted in Section \ref{sec:model_copula}, the Gaussian copula facilitates marginalization over the missing data, and ABC generates data $\bsy$ with the same missing data pattern as in $\bsy_\obs$.  Thus, we have not required  special consideration to account for missingness.  In the IFS data, we can consider  two   sources of missingness.  The first is in the spirit of structural missingness
% [ref https://www.nature.com/articles/s42256-022-00596-z]
\citep{Mitra2023}
and arises due to the timing of tooth eruption.  
Unlike analyses specific to high-risk populations
% [ref doi: 10.1214/16-AOAS917]
\citep[e.g.,][]{Jin2016},
there is no information to be gained by considering which teeth are observed given that the patient is seen at time $t$.  Whether an individual tooth is observed is primarily a function of patient age and the transition from primary to permanent teeth, not underlying dental health.  
More relevant is missingness from whether patients are observed at time $t$.  
% Under an assumption of ignorability
% % [ref little, rubin book]
% \citep{Little2019},
% the missing data mechanism need not be specifically modeled or included in the analysis.  That is, we must believe that whether or not a patient is seen for their dental examination depends only on their previously observed caries scores but not on their current dental health.  A more plausable assumption would be auxiliary variable missing at random \citep{Daniels2008}, when missingness depends not on the unobserved caries score but on an observed predictor such as frequency of attending dental examinations.  Under this reasonable assumption all of our inferences remain valid.
An assumption of auxiliary variable missing at random \citep{Daniels2008} is potentially reasonable, as we believe  missingness depends not on the unobserved caries score, but on  observed predictors such as frequency of  dental visits.  In that case the missing data mechanism need not be modeled, and  our analyses remain valid.
We note that our model structure could be easily extended to account for non-ignorable missingness
with 
% .  The most obvious would be to incorporate
a selection model approach 
\citep{Little2019}.
% that models whether or not a patient attends their examination given their (potentially unobserved) caries scores and other covariates.  
% This choice, or others the incorporate the missingness indicators and caries scores, 
% Such a modeling choice would be incorporated into 
The data generation model $p(\bsy\mid\btheta)$ would then include the missing data mechanism, and $\bss(\bsy)$ must be  adjusted   to include statistics related to the missing data indicators.  
% Other adjustments to $(\bss_M,\bss_D)$ may be necessary to ensure that they are useful and informative across all of the missing data patterns generated from the model.  
Implementation and investigation of such strategies is beyond the scope of this project.

\section*{Acknowledgements}

This research was funded in part by NIH/NIDCR grant R03 DE030502.

% Finally missing data is an important aspect that must be taken care of.
% There could be two primary sources of missingness in the recorded caries scores: the structural missingness due to the time of tooth irruption and patients not showing up for dental examination.
% Since the structural missingness is not believed to be associated with caries scores, the inference based on our model remains valid.
% Under the assumption that the the missingness in the caries scores are associated with the corresponding values of the known predictor variables, instead of the level of deterioration in dental health, we can treat the missing data as auxiliary variable missing at random \citep{Daniels2008}.
% Our model will also be a good fit in this scenario.
% When the missingness is non-ignorable, that is, missingness is associated with the scores themselves, the analysis becomes more difficult and one possible approach to address this would be based on the pattern mixture model presented by \cite{Kaciroti2006}.

\spacingset{1.5}
{\small \bibliography{references.bib}}
% { \bibliography{references.bib}}
\bibliographystyle{apalike}

\newpage

\appendix
\appendixpage
\renewcommand{\theequation}{\thesection.\arabic{equation}}
\setcounter{equation}{0}
\renewcommand{\thefigure}{\thesection.\arabic{figure}}
\renewcommand{\theHfigure}{\thefigure}
\setcounter{figure}{0}
\renewcommand{\thetable}{\thesection.\arabic{table}}
\renewcommand{\theHtable}{\thetable}
\setcounter{table}{0}

\section{Further Algorithmic Details}

\begin{figure}[!tb]
	\centering
	\begin{tikzpicture}[node distance = 2.7cm]
		\centering
		\node (data) [input_narrow] {\scriptsize Data};
		\node (est_theta_M) [proc, right of=data] {\scriptsize Initial estimation of marginal parameters };
		\node (init_theta_M) [param, right of=est_theta_M] {\scriptsize $\widetilde{\btheta}_M, \widetilde{\bSigma}_M$};
		\node (est_theta_D) [proc, above of=est_theta_M] {\scriptsize Initial estimation of Dependence Structure through simulation};
		\node (init_theta_D) [param, right of=est_theta_D] {\scriptsize $\widetilde{\btheta}_D, \widetilde{\bSigma}_D$};
		\node (est_kern) [proc, right of=init_theta_D] {\scriptsize Kernel Estimation through simulation};
		\node (tuning_h) [input_wide, above of=est_kern, yshift=-0.5cm] {\scriptsize Bandwidth $(h)$};
		\node (kern_params) [param, right of=est_kern] {\scriptsize $K_h(\Delta | \bA)$};
		\node (ABC) [proc, right of=init_theta_M] {\scriptsize ABC-MCMC};
		\node (theta) [param, right of=ABC] {\scriptsize $(\btheta_M, \btheta_D)$};
		\node (reg) [proc, right of=kern_params] {\scriptsize Regression Adjustment};
		\node (final_theta) [output, right of=theta] {\scriptsize $\{(\btheta_M^{(g)}, \btheta_D^{(g)})\}$, $g = 1, \ldots, G$};
		
		\draw [arrow] (data) -- (est_theta_M);
		\draw [arrow] (est_theta_M) -- (init_theta_M);
		\draw [arrow] (init_theta_M) -- (est_theta_D);
		\draw [arrow] (est_theta_D) -- (init_theta_D);
		\draw [arrow] (tuning_h) -- (est_kern);
		\draw [arrow] (est_kern) -- (kern_params);
		\draw [arrow] (init_theta_M) -- (est_kern);
		\draw [arrow] (init_theta_D) -- (est_kern);
		\draw [arrow] (init_theta_M) -- (ABC);
		\draw [arrow] (init_theta_D) -- (ABC);
		\draw [arrow] (kern_params) -- (ABC);
		\draw [arrow] (ABC) -- (theta);
		\draw [arrow] (theta) -- (reg);
		\draw [arrow] (reg) -- (final_theta);
	\end{tikzpicture}
	\caption{Flow chart of the ABC algorithm. The trapazoid-shaped boxes represent input to the algorithm, rounded-corner boxes are procedures, circles indicate intermediate estimates of parameters and the rightmost box with a double border indicates the final output of the algorithm.}
	\label{fig:ABC_algo}
\end{figure}
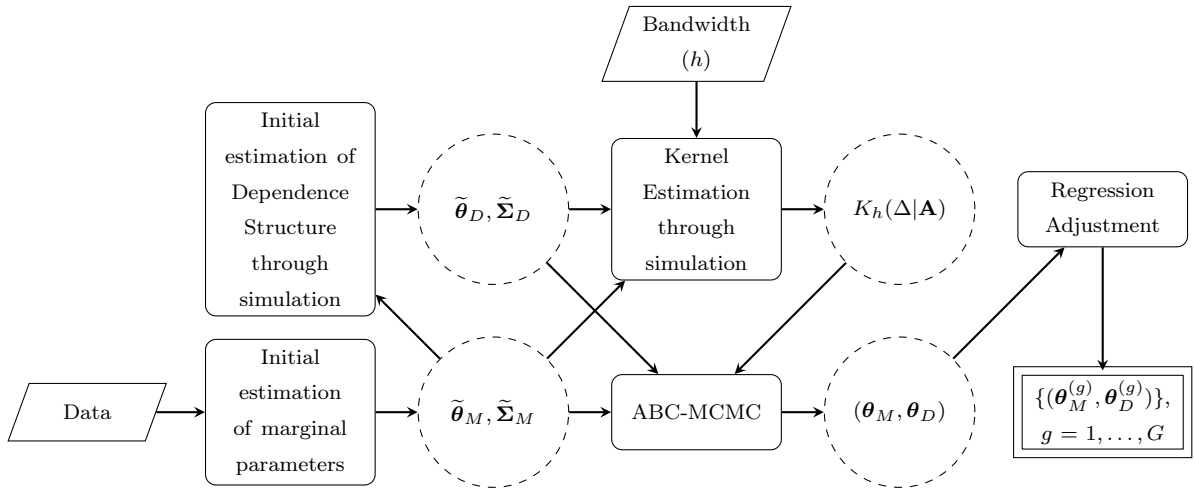

In this section, we provide further details of our algorithm and computational strategy that were not included in Section \ref{sec:computation} of the main manuscript. 
Figure \ref{fig:ABC_algo} summarizes the full sequence of steps we take to perform parameter estimation. 
We will first describe our approach to find initial estimates of $\btheta_M$ and $\btheta_D$, along with that for derivation of summary statistics $\bss_{D}(\bsY)$ and the estimation of kernel scaling matrix $\bA$.
% \org{[is this the name?  we didn't really call it that in 4.3.]}
The following subsections will present a discussion on the specifics of ABC-MCMC algorithm, including covariance adaptation and hyperparameter $\btheta_H$ sampling. 
Section \ref{appn:regression_adjustment} will provide additional details on post-ABC regression adjustment.
The final subsection includes details on the ABC rejection and importance sampling algorithms.
% \org{[not the last thing, right?]}

\subsection{Initial Estimation} \label{appn:initial_estimation}

As noted in the main text, our proposed model simplifies to a more tractable model under the assumption $\btheta_D = \bzero$. 
In addition to using this auxiliary likelihood to compute the ABC summary statistics regarding $\btheta_M$, we also use the simplified model to find initial values for our ABC-MCMC algorithm.
To that end, we first devised an algorithm for initializing our ABC-MCMC based on estimating $\btheta_M$ conditional on $\btheta_D = \bzero$, and then  estimating $\btheta_D$ conditional on the estimated $\btheta_M$. The estimation of the initial proposal covariance matrix is also discussed.

As suggested by \cite{Pillow2012}, we reparameterize the Negative Binomial distribution such that $\phi$, the size parameter, is included in the mean term.
Note this reparameterization is only for this step of finding an initial value for $\btheta_M$.
The likelihood under $\btheta_D = \bzero$ reduces to
\begin{equation}
	L^*_{\mathrm{ind}}(\btheta_M | \bsy) = \prod\limits_{(i,j) \in \mathcal{D}} \left\{ \pi_{ij} \mathbb{I}(y_{ij}=0) + (1-\pi_{ij}) p(y_{ij}-1 | \phi \exp(\bx_{ij}'\bbeta), \phi) \mathbb{I}(y_{ij} > 0) \right\}.
\end{equation}
% where $\btheta_M$ represents the marginal parameter vector in the reduced model, which is different from $\btheta_M$.
\cite{Polson2013} and \cite{Pillow2012} have discussed conjugate Gibbs sampling schemes to obtain posterior samples for logistic and Negative Binomial model parameters, respectively.
In both cases, the authors introduced a set of auxiliary variables each following a P\'{o}lya-Gamma distribution and expressed binomial and Negative Binomial likelihoods as a mixture of normal distributions with respect to those latent variables. 
\cite{Neelon2019} has also used P\'{o}lya-Gamma latent variable to implement Gibbs sampling steps for zero-inflated Negative Binomial regression in the context of spatio-temporal analysis. 
Here, we follow their approaches for the estimation of marginal model parameters $\balpha$ and $\bbeta$. 

For those $(i,j) \in \mathcal{D}^*$, we have  $ Y_{ij} - 1 \sim \text{NB}(\mu_{ij} = \phi \exp({\boldsymbol{x}}'_{ij}{\bbeta}), \phi) $ with the density, 
\[ p(y) = \frac{\Gamma(y-1+\phi)}{\Gamma(\phi) (y-1)!} \xi_{ij}^{\phi} (1-\xi_{ij})^{y_{ij}-1}, \ \  \xi_{ij} = \frac{1}{(1+\exp(\bsx_{ij}'\bbeta))}.\] 
To devise a conjugate sampler for $\bbeta$, we introduce an auxiliary parameter $\omega_{ij}^{(\bbeta)}$ for each $(i,j) \in \mathcal{D}^*$ such that $\omega_{ij}^{(\bbeta)}|y_{ij}, \bbeta \sim \text{PG}(y_{ij}-1+\phi, \bsx'_{ij} \bbeta)$.
% \org{[The second parameter in the prior here needs to be $\bx'_{ij}\bbeta$, right?  Otherwise, where does the $e^{-\omega_{ij}(\bx'_{ij}\bbeta)^2/2}$ term come from?]}
% where the NB density function is expressed as 
% \begin{equation} \label{eq:nb}
	% p(y) = \frac{\Gamma(y-1+\phi)}{\Gamma(\phi) (y-1)!} \xi^{\phi} (1-\xi)^{y-1}.
	% \end{equation} 
% We observe that, $\E(Y_{ij}) = 1+\frac{(1-\xi_{ij}) \phi}{\xi_{ij}}$, implying $\text{logit}(1-\xi_{ij}) = {\boldsymbol{x}}'_{ij}{\bbeta}$ or in other words, $(1-\xi_{ij}) = \frac{\exp({\boldsymbol{x}}'_{ij}{\bbeta})}{1+\exp({\boldsymbol{x}}'_{ij}{\bbeta})}$. Substituting $\xi_{ij}$ in terms of ${\bbeta}$ in equation (\ref{eq:nb}) and 
% \org{[introduce the data augmentation parameters here (with distribution conditional on y)]}
Using the integral identity 
% as discussed in 
\citep{Polson2013},
% and conditional on $\phi$ 
we  obtain the likelihood contribution to $\bbeta$ from $y_{ij}$ to be
\[
L^*_{\mathrm{ind}}(\bbeta | y_{ij}) \propto \frac{\exp({\boldsymbol{x}}'_{ij}{\bbeta})^{y_{ij}-1}}{(1 + \exp({\boldsymbol{x}}'_{ij}{\bbeta}))^{y_{ij}-1+\phi}} \propto e^{\kappa_{ij} (\bx'_{ij}\bbeta)} \int_0^\infty e^{-\omega^{(\bbeta)}_{ij}(\bx'_{ij}\bbeta)^2/2} p(\omega^{(\bbeta)}_{ij}) d\omega^{(\bbeta)}_{ij},
\]
where $\kappa_{ij} = \frac{1}{2}(y_{ij}-1-\phi)$ and $p(\omega_{ij}^{(\bbeta)})$ represents the density $\text{PG}(y_{ij}-1+\phi, 0)$. 
We consider $\MVN_{d}(\bzero, \bSigma_{\bbeta})$ as the  prior for $\bbeta$ (conditionally on the NG variance parameters).
% We note that the resulting full conditional distribution for $\bbeta$ is multivariate normal given the data augmentation parameters. 
% The details of the distribution are provided in the following MCMC steps.
% The resulting full conditional distribution for $\bbeta$ is given by $\left( {\bbeta} | \{ \omega^{({\bbeta})}_{ij} \}, \{ y_{ij} \} \right) \sim \text{MVN}({\boldsymbol{\mu}}_{{\bbeta}}^*, {\boldsymbol{\Sigma}}^*_{\bbeta})$, where
%     \[
%     {\boldsymbol{\Sigma}}^*_{\bbeta} = \left(\sum_{\{i,j : z_{ij} = 1 \}} \omega^{({\bbeta})}_{ij} {\boldsymbol{x}}_{ij} {\boldsymbol{x}}_{ij}' + {\boldsymbol{\Sigma}}^{-1}_{\bbeta} \right)^{-1}, \quad {\boldsymbol{\mu}}_{\bbeta} = {\boldsymbol{\Sigma}}^*_{\bbeta} \left[ \sum_{\{i,j : z_{ij} = 1 \}} \left(\frac{y_{ij}-1-\phi}{2}\right) {\boldsymbol{x}}_{ij} \right].
%     \]
% \org{[This $p(\bomega^{(\bbeta)})$ in the line below is not the prior!!!  It is a product of $c=0$ PGs from the inside of the integrals, but it isn't the prior which has $c=x*beta$]}
The resulting posterior distribution for $ \bbeta$, conditional on the $\omega^{(\bbeta)}_{ij}$ augmentation parameters, is proportional to 
\[
e^{-\bbeta' \bSigma_{\bbeta}^{-1} \bbeta/2} \prod\limits_{(i,j) \in \mathcal{D}} e^{\kappa_{ij} (\bx_{ij}'\bbeta) -\omega(\bx_{ij}'\bbeta)^2/2},
\]
% \org{[only way to fix this is to not provide a joint density of $(\bomega^{(\bbeta)}, \bbeta)$.  just introduce the beta prior and give the form of the full conditional for beta in the above eqn.  or change how you talk about $p(\bomega^{(\bbeta)}) $ above]}
% We note that the resulting full conditional distribution for $\bbeta$ is multivariate normal given the data augmentation parameters. 
which is a multivariate normal distribution.  The details of the distribution are provided in the following MCMC steps.
% \org{[I made some slight adjustments here to add back in a few of the details]}

To obtain a conjugate sampler for $\balpha$, the auxiliary variables $\omega_{ij}^{(\balpha)}$ are introduced, where $\omega_{ij}^{(\balpha)} | z_{ij}, \balpha \sim \text{PG}(1,\bsx'_{ij} \balpha)$.
% \org{[x'alpha???]}
Then the logistic regression model with $\text{logit}(\pi_{ij}) = \boldsymbol{x}'_{ij}\balpha$ as specified in equation (\ref{eq:presence}) 
% \org{[should be from $L^*_{ind}$, right?]}
implies that the likelihood contribution to $\balpha$ from $z_{ij}$ would be
\[
L^*_{\mathrm{ind}}(\balpha | z_{ij}) \propto \frac{\exp({\boldsymbol{x}}'_{ij}{\balpha})^{z_{ij}}}{(1 + \exp({\boldsymbol{x}}'_{ij}{\balpha}))} \propto e^{\kappa_{ij} (\bx'_{ij}\balpha)} \int_0^\infty e^{-\omega^{(\balpha)}_{ij}(\bx'_{ij}\balpha)^2/2} p(\omega^{(\balpha)}_{ij}) d\omega^{(\balpha)}_{ij},
\]
where $\kappa_{ij} = z_{ij}-1/2$ and $p(\omega^{(\balpha)}_{ij})$ for each $(i,j) \in \mathcal{D}$ is the PG(1,0) density.
We choose the prior for $\balpha$ to be $\MVN(\bzero, \bSigma_{\balpha})$.
% The joint posterior distribution for $(\bomega^{(\balpha)}, \balpha)$ is proportional to
% \[
% p(\bomega^{(\balpha)}) e^{-\balpha' \bSigma_{\balpha}^{-1} \balpha/2} \prod\limits_{(i,j) \in \mathcal{D}} e^{\kappa_{ij} (\bx_{ij}'\balpha) -\omega(\bx_{ij}'\balpha)^2/2}.
% \] 
Due to the conjugate structure resulting from the augmentation of $\bomega^{(\balpha)}$, we can derive a Gibbs sampling step for sampling $\balpha$ from MVN. 
% With this background, we now discuss the MCMC algorithm.

Under the independence assumption/misspecification in  this auxiliary likelihood,
posterior sampling of the parameters for the presence model ($\balpha, \bsigma_{\balpha}^2, \tau_{\balpha}^2, \lambda_{\balpha}$) depends only on $Z_{ij}$, while the severity model parameters ($\bbeta, \bsigma_{\bbeta}^2, \tau_{\bbeta}^2, \lambda_{\bbeta}$) are only dependent on the non-zero $Y_{ij}$.
Since $Z_{ij}$ are known for given data $Y_{ij}$, updating the presence model parameters using the following steps (1a-1c) can be done separately from the steps (2a-2d) for updating the severity model parameters. As such, we can run (1a-1c) and (2a-2d) in parallel to reduce overall computation time. 
The following sampling steps of the MCMC algorithm are repeated until approximate convergence is achieved.
\begin{enumerate}
	\item[(1a)] \textit{Updating} ${\balpha}$: We update $\balpha$ in two steps: 
	(i) first draw samples for auxiliary variables $\{ \omega^{({\balpha})}_{ij} \}$ for $i = 1, \ldots, n$, $j \in \mathcal{J}_i$, $\omega^{({\balpha})}_{ij} | \balpha \stackrel{\text{iid}}{\sim} \text{PG}(1, {\boldsymbol{x}}'_{ij} {\balpha})$, 
	(ii) then draw $\balpha$ as $\left( {\balpha} | \{ \omega^{({\balpha})}_{ij} \}, \{ y_{ij} \} \right) \sim \text{MVN}({\boldsymbol{\mu}} ^*_{{\balpha}}, {\boldsymbol{\Sigma}}^*_{\balpha})$, where ${\boldsymbol{\Sigma}}^*_{\balpha} = \left(\sum_{i,j} \omega^{({\balpha})}_{ij} {\boldsymbol{x}}_{ij} {\boldsymbol{x}}_{ij}' + {\boldsymbol{\Sigma}}^{-1}_{\balpha} \right)^{-1}$, $\mathbf{\Sigma}_{\balpha} = \text{Diag}(\sigma_{\alpha_0}^2, \sigma_{\alpha_1}^2, \ldots, \sigma_{\alpha_d}^2)$ is the prior covariance of $\balpha$, and ${\boldsymbol{\mu}}^*_{\balpha} = {\boldsymbol{\Sigma}}^*_{\balpha} \left( \sum_{i,j} (z_{ij} - 1/2) {\boldsymbol{x}}_{ij} \right)$. 
	% Note that the prior covariance $\bSigma_{\balpha}$ is determined by $\lambda_{\balpha}$ and $\tau_{\balpha}^2$, since its diagonal entries $\sigma_{\alpha_k}^2, k = 1, \ldots, d$ have gamma prior with parameters $\lambda_{\balpha}$ and $\lambda_{\balpha}/\tau_{\alpha}^2$, which themselves are updated later in steps (1c-1d).
	
	\item[(1b)] \textit{Updating} $\boldsymbol{\sigma}^2_{\balpha}$: The full conditional distribution of $\sigma^2_{\alpha_k}$ for $k = 1, \ldots, d$ is given by the generalized inverse Gaussian (GIG) distribution,  $\textrm{GIG}(\lambda_{\balpha} - 1/2, \alpha_k^2, 2\lambda_{\balpha}/\tau_{\balpha}^2)$.  In this parameterization, the density for $\text{GIG}(x;m,a,b)$ is given by
	\[
	p(x)=\frac{(a/b)^{m/2}}{2K_m(\sqrt{ab})} x^{m-1} \exp \left\{-\frac{1}{2} (ax + b/x) \right\}, \ \ \ x>0,
	\]
	where $K_m(\cdot)$ is the modified Bessel function of the second kind.
	\item[(1c)] \textit{Updating} $\tau_{\balpha}^2$: The full conditional distribution for $\tau_{\balpha}^2$ is given by $\textrm{IG}(1+d\lambda_{\balpha}, 1 + \lambda_{\balpha} \sum_{k=1}^d \sigma_{\alpha_k}^2)$.
	
	% \item[(1d)] \textit{Updating} $\lambda_{\balpha}$: To update $\lambda_{\balpha}$ we need a Metropolis-Hastings (MH) step. The full conditional distribution for $\lambda_{\balpha}$  based on $\boldsymbol{\sigma}_{\balpha}^2$ is proportional to $\pi(\lambda_{\balpha}) \prod_{k=1}^d p(\sigma^2_{\alpha_k} | \lambda_{\balpha}, \tau_{\balpha} ^2)$. The proposal distribution for $\lambda_{\balpha}$ is given as $(\lambda_{\balpha}^* | \lambda_{\balpha}) \sim \text{LN}(\lambda_{\balpha}, \sigma^2_{1})$.
	
	\item[(2a)] \textit{Updating} ${\bbeta}$: The two steps for updating $\bbeta$ are: (i) first draw samples for auxiliary variables $\{ \omega^{({\bbeta})}_{ij} \}_{(i,j) \in \mathcal{D}^*}$, $\omega^{({\bbeta})}_{ij} | \bbeta \stackrel{\text{iid}}{\sim} \text{PG}(y_{ij}-1+\phi, {\boldsymbol{x}}'_{ij} {\bbeta})$, (ii) then draw $\bbeta$ as $\left( {\bbeta} | \{ \omega^{({\bbeta})}_{ij} \}, \{ y_{ij} \} \right) \sim \text{MVN}({\boldsymbol{\mu}}_{{\bbeta}}^*, {\boldsymbol{\Sigma}}^*_{\bbeta})$, where
	\[
	{\boldsymbol{\Sigma}}^*_{\bbeta} = \left(\sum_{\{i,j : z_{ij} = 1 \}} \omega^{({\bbeta})}_{ij} {\boldsymbol{x}}_{ij} {\boldsymbol{x}}_{ij}' + {\boldsymbol{\Sigma}}^{-1}_{\bbeta} \right)^{-1}, \quad {\boldsymbol{\mu}}_{\bbeta} = {\boldsymbol{\Sigma}}^*_{\bbeta} \left[ \sum_{\{i,j : z_{ij} = 1 \}} \left(\frac{y_{ij}-1-\phi}{2}\right) {\boldsymbol{x}}_{ij} \right].
	\]
	Note that the intercept $\beta_0$ is updated in this step along with other $\beta_k$, but it  will also be updated again in the step (2b) jointly with $\phi$. It is, however, worth mentioning that sampling from the subfamily of $\text{PG}(1,\bsx_{ij}'\balpha)$ distributions, as needed for the logistic regression parameters, can be performed very efficiently.
	However, sampling from $\text{PG}(y_{ij}-1+\phi, \bsx_{ij}' \bbeta)$ for Negative Binomial regression is more challenging and less computationally efficient due to the non-integer shape parameter, as discussed in \cite{Polson2013}.

	\item[(2b)] \textit{Updating $\phi$ and $\beta_0$}: We update $\beta_0$ and $\phi$ jointly in a MH step. Note that $\beta_0$ and $\log(\phi)$ are combined to obtain the intercept term in $\log \mu_{ij} = \log \phi + \bsx_{ij}'\bbeta$ of the Negative Binomial distribution, implying that the value of $\phi$ impacts both $\log(\mu)$ as well as the over-dispersion. Hence, we update $\phi$ by jointly proposing new values for $\beta_0$ and $\phi$ in such a way that $\beta_0+\log\phi$ (and hence, $\log(\mu_{ij})$) remains unchanged. We generate $\delta$ from symmetric $\mathrm{Unif}[-l,l]$ for some suitable choice of $l>0$ and propose new values of $\bbeta$ and $\phi$ jointly as $\beta_0^* = \beta_0^{(s-1)} + \delta$, $\log\phi^* = \log\phi^{(s-1)} - \delta $, and the remaining coefficients are unchanged ($\beta_k^* = \beta_k^{(s-1)}$  for $k=1,\ldots,d$). The probability of acceptance for $(\bbeta^*, \phi^*_0)$ is given by 
	\[
	\min \left\{ 1, \frac{p(\bbeta^*) p(\log \phi^*)  \prod\limits_{(i,j) \in \mathcal{D}^*} p_{\mathrm{NB}} (y_{ij} | \phi^* \exp(\bsx_{ij}'\bbeta^*), \phi^*) }{ p(\bbeta^{(s-1)}) p(\log \phi^{(s-1)}) \prod\limits_{(i,j) \in \mathcal{D}^*} p_{\mathrm{NB}} (y_{ij} | \phi^{(s-1)} \exp(\bsx_{ij}'\bbeta^{(s-1)}), \phi^{(s-1)})} \right\},
	\]
	where $p_{\mathrm{NB}}$ represents the NB density.
	
	% The joint proposal distribution turns out to be $q(\beta_0^{(s)}, \log \phi^{(s)} \mid \beta_0^{(s-1)}, \log \phi^{(s-1)}) = \text{N}(\delta = \beta_0^{(s)} - \beta_0^{(s-1)} \mid 0, \sigma_{\delta}^2) \mathbb{I}(\beta_0^{(s)} +  \log \phi^{(s)} = \beta_0^{(s-1)} + \log \phi^{(s-1)})$. 
	
	% We observe that the proposal satisfies the detailed balance condition as follows,
	% \[
	% \begin{array}{rcl}
		% q(\beta_0^{(s)}, \log \phi^{(s)} \mid \beta_0^{(s-1)}, \log \phi^{(s-1)}) & = & \text{N}(\delta = \beta_0^{(s)} - \beta_0^{(s-1)} \mid 0, \sigma_{\delta}^2) \times \\
		% & & \mathbb{I}(\beta_0^{(s)} +  \log \phi^{(s)} = \beta_0^{(s-1)} + \log \phi^{(s-1)}) \\
		% & = & \text{N}(\delta = \beta_0^{(s-1)} - \beta_0^{(s)} \mid 0, \sigma_{\delta}^2) \\
		% & & \mathbb{I}(\beta_0^{(s-1)} +  \log \phi^{(s-1)} = \beta_0^{(s)} + \log \phi^{(s)}) \\
		% & = & q(\beta_0^{(s-1)}, \log \phi^{(s-1)} \mid \beta_0^{(s)}, \log \phi^{(s)}).
		% \end{array}
	% \]

	\item[(2c)] \textit{Updating} $\boldsymbol{\sigma}^2_{\bbeta}$: The full conditional distribution of $\sigma^2_{\beta_k}$ for $k = 1, \ldots, d$ is given as $\textrm{GIG}(\lambda_{\bbeta} - 1/2, \beta_k^2, 2\lambda_{\bbeta}/\tau_{\bbeta}^2)$.
	
	\item[(2d)] \textit{Updating} $\tau_{\bbeta}^2$: The full conditional distribution for $\tau_{\bbeta}^2$ is $\textrm{IG}(1+d\lambda_{\bbeta}, 1 + \lambda_{\bbeta} \sum_{k=1}^d \sigma_{\beta_k}^2)$.
	
	% \item[(2e)] \textit{Updating} $\lambda_{\bbeta}$: To update $\lambda_{\bbeta}$ we need a Metropolis-Hastings (MH) step. The full conditional distribution for $\lambda_{\bbeta}$  based on $\boldsymbol{\sigma}_{\bbeta}^2$ is proportional to $\pi(\lambda_{\bbeta}) \prod_{k=1}^d p(\sigma^2_{\beta_k} | \lambda_{\bbeta}, \tau_{\bbeta} ^2)$. The proposal distribution for $\lambda_{\bbeta}$ is given as $(\lambda_{\bbeta}^* | \lambda_{\bbeta}) \sim \text{LN}(\lambda_{\bbeta}, \sigma^2_{2})$.
	
\end{enumerate}

We calculate the mean from a large number of posterior samples and denote it as $\widetilde{\btheta}_M$.
So that these results correspond to the parameterization in (\ref{eq:severity}), we replace the MCMC sampled $\beta_0$ with $\beta_0+\log\phi$.
This provides our initial estimate for $\btheta_M$. 
We also obtain the posterior covariance matrix and denote it as $\widetilde{\bSigma}_M$. 

To obtain an initial estimate for $\btheta_D$, denoted as $\widetilde{\btheta}_D$, and a corresponding rough estimate for the covariance matrix denoted as $\widetilde{\bSigma}_D$, we take the following approach: 
(a) draw $G=10,000$ samples of $\btheta_D$, represented by $\{ \btheta_{D}^{(g)} \}_{g=1}^G$ from a uniform distribution over the subset of the support where each component of $\btheta_D$ is positive;
(b) for every $ \btheta^{(g)} = (\widetilde{\btheta}_{M}, \btheta_{D}^{(g)}) $ generate a dataset $\bsy^{(g)}$ and a corresponding summary statistic $\bss^{(g)} = \bss(\bsy^{(g)}) = (\bss_M(\bsy^{(g)}), \bss_D(\bsy^{(g)}))$;
(c) select a small set (usually one percent) of the $\btheta_{D}^{(g)}$ samples for which the Euclidean distances between  $\bss_D(\bsy^{(g)})$ and $\bss_{D}(\bsy_{\text{obs}})$ are smallest, and choose $\widetilde{\btheta}_D$ and $\widetilde{\bSigma}_D$ as the mean and the covariance matrix of that set of best performing values. 
The initial estimate of $\btheta$ is then given as $\widetilde{\btheta} = (\widetilde{\btheta}_M, \widetilde{\btheta}_D)$ 
and the initial covariance matrix for the proposal distribution of $\btheta$ in the ABC-MCMC algorithm will be block diagonal with components $\widetilde{\bSigma}_M$ and $\widetilde{\bSigma}_D$.
% and that for the covariance matrix is given as a block diagonal matrix $\widetilde{\bSigma} = ((\widetilde{\bSigma}_M, \bzero), (\bzero', \widetilde{\bSigma}_D)).$

We now elaborate on the construction of the summary statistics for $\btheta_D$.
First, we obtain naive estimates for the latent Gaussian random variables $V_{ij}$ from the observed data $Y_{ij}$, and then following the SAR  model, obtain estimates for the regression coefficients $\btheta_D$ as outlined in Section \ref{sec:ABC-summary-statistics}. 
Let $\mathcal{I}_j$ denote the subset of individuals who have the $j$-th measurement recorded in the data.
We define an empirical CDF for the $j$-th margin, denoted by $\tilde{F}_j$ as $\tilde{F}_j(y_{ij}) = r_{j}(y_{ij})/|\mathcal{I}_j|$, where $r_{j}(y) = \sum\limits_{i \in \mathcal{I}_j} \mathbb{I}(y_{ij} < y) + \frac{1}{2} \sum\limits_{i \in \mathcal{I}_j} \mathbb{I}(y_{ij} = y) $
denotes an estimated rank of $y$ among all the $y_{ij}$ along the $j$-th margin, counting a half  contribution for all observations equal to $y$.
We set $\widehat{v}_{ij} = \Phi^{-1}(\tilde{F}_j(y_{ij}))$.
Note that $r_{j}(\cdot)$ is defined such that for any $y_{ij}$ $\widehat{v}_{ij}$ will be set at the median on the restricted range of values of $V_{ij}$ such that $F_{ij}(y_{ij}-1) < \Phi(V_{ij}) \leq F_{ij}(y_{ij})$.
Additionally, this choice avoids infinite $\hat{v}_{ij}$ for $y=\max_{i \in \mathcal{I}_j} y_{ij}$. 
We further impute $\hat{v}_{ij} = 0$ when $y_{ij}$ is missing. 
The intuition behind this choice is that since a missing $y_{ij}$ can lie anywhere in the support, zero is the median of the corresponding range of $V_{ij}$. 

In line with  the SAR model in (\ref{eq:SAR}), we fit the regression model using these imputed values $\hat{v}_{ij} = \sum_{k=1}^K \rho_k \dbtilde{v}_{ij;k} + \epsilon_{ij}$
for $(i,j) \in \mathcal{D}$,
where $\dbtilde{v}_{ij;k} = \sum_{j' \in \mathcal{J}} w^{(k)}_{jj'} \hat{v}_{ij'}$.
While the variances of $\epsilon_{ij}$ in the SAR model specification differ across $j$, we obtain estimates of $\rho_1, \ldots, \rho_K$ as the regression coefficients in the linear model with an equal variance assumption.
The summary statistic vector $\bs_D(\bY)$  for $\btheta_D$  is given by $(\dbtilde{\bV}' \dbtilde{\bV})^{-1}(\dbtilde{\bV}' \vec{\bsV})$, where $\dbtilde{\bV}_{|\mathcal{D}| \times K} = (\dbtilde{v}_{ij;k})$ is the latent design matrix and $\vec{\bsV}$ is the $|\mathcal{D}|$-dimensional latent outcome vector consisting of $\hat{v}_{ij}$.
Note here that we obtain $\dbtilde{\bV}$ by stacking  the vectors $\dbtilde{\bsv}_{ij} = (\dbtilde{v}_{ij;1}, \ldots, \dbtilde{v}_{ij;K})$ as rows where $(i,j) \in \mathcal{D}$.

We now turn to kernel estimation strategy. 
To estimate the kernel scaling matrix denoted by $\bA$, we generate $G=10,000$ sample datasets denoted by $\{\bsy^{(g)}\}_{g=1}^G$ from $\widetilde{\btheta} = (\widetilde{\btheta}_{M}, \widetilde{\btheta}_{D})$, our initial guess at the parameter vector.
% obtain the summary statistics $\bss^{(g)} = \bss(\bsy^{(g)})$ from each $\bsy^{(g)}$. 
We compute the sample variance of each summary statistics as  $\widetilde{\sigma}^2_k = \frac{1}{G} \sum_{g=1}^G (s^{(g)}_k - \bar{s}_k)^2$, where $s^{(g)}_k$ is the $k$-th summary statistic of the $g$-th sample $\bss^{(g)} = \bss(\bsy^{(g)})$ and $\bar{s}_k = \frac{1}{G} \sum_{g=1}^G s^{(g)}_k$.
% \org{[ARE YOU USING G OR G-1 IN DENOMINATOR?  EITHER IS FINE.]}
We choose $\bA = \textrm{diag}(1/\widetilde{\sigma}^2_1, \ldots, 1/\widetilde{\sigma}^2_K)$, where $K$ is the dimension of the vector $\bss^{(g)}$.
% We define $\widetilde{\bSigma}_{\bss} = \frac{1}{G} \sum_{g=1}^G (\bss^{(g)} - \bar{\bss}) (\bss^{(g)} - \bar{\bss})'$, where $\bss^{(g)} = \bss(\bsy^{(g)})$ and $\bar{\bss} = \frac{1}{G} \sum_{g=1}^{G}\bss^{(g)}$. 
Note, if $\widetilde{\btheta}$ is very close to the true value of $\btheta$, the summary statistics of the generated data $\{ \bss^{(g)} \}$ can be considered as a representative sample capturing the covariance structure if we had generated the data based on true $\btheta$. 
% \org{[need to double check terminology.  Don't think that we previously referred to this as kernel covariance matrix.  Also, don't we diagonalize this?]}
% The bandwidth parameter $h$ regulates the concentration of the kernel.
% and is typically chosen as the $100\epsilon$-th percentile of the $\{\Delta(\bss^{(g)})\}_{g=1}^{G}$, where $\Delta(\bss^{(g)}) = \bss^{(g)} - \bss_{\obs}$ and $\epsilon$ helps control the rate of acceptance for the simulated data.
% \org{[is this what we are doing????]]}

\subsection{ABC-MCMC Algorithm Specifics} \label{appn:algo_details}

Here we elaborate on certain aspects of the ABC-MCMC algorithm that were not included in the main text due to space constraints. 
Description of the proposal covariance adaptation scheme is presented. 
Some details on the posterior sampling of $\btheta_H$ are also provided.

We first discuss the details of the adaptive-MH portion of the ABC-MCMC algorithm, following a strategy from  \citet[Algorithm 4]{Andrieu2008}.
Unlike classical Metropolis-Hastings where the random walk covariance matrix is constant,
adaptive MCMC allows the covariance in the proposal distribution  to change slowly to  improve mixing.
The proposal density in iteration $g$, denoted by $q_{g}(\cdot|\btheta^{(g-1)})$, is multivariate normal with mean $\btheta^{(g-1)}$, the $\btheta$ sample from iteration $g-1$, and with covariance matrix $\bSigma_{g-1}$.
% and the density is denoted as $q(\cdot | \btheta^{(g)}, \bSigma_{g})$.
The aim here is to adapt the proposal covariance for the next iteration $(g+1)$ taking into account the samples of $\btheta$ obtained so far, while making sure that the amount of adaptation vanishes fast enough as $g$ increases.
At the beginning of the chain (iteration $g=1$), the parameter is initalized at $\btheta^{(0)} = \widetilde{\btheta}$ and covariance at
% the proposal mean is $\btheta^{(0)} = \widetilde{\btheta}$, the covariance is chosen as 
$\bSigma_{0} = \widetilde{\bSigma}$, 
as described in Section \ref{appn:initial_estimation}.
Let $\overline{\bmu}_{g}$ and $\overline{\bSigma}_{g}$ denote the weighted sample mean and covariance matrix at iteration $g$, which are also initialized at $\widetilde{\btheta}$ and $\widetilde{\bSigma}$.
% which is obtained at iteration $g$ based on their previous values $\overline{\bmu}_{g-1}$ and $\overline{\bSigma}_{g-1}$ and the currently proposed $\btheta$. 
% The covariance is
Further, let $\eta_g$ be the global covariance scaling factor, initialized at $\eta_0 = 1$.
% We choose $\overline{\bmu}_{0} = \widetilde{\btheta}$ and $\overline{\bSigma}_{0} = \widetilde{\bSigma}$. 

The covariance adaptation steps in iteration $g$ are performed as part of the acceptance/rejection of the pair $(\btheta',\bsy')$  and include the following steps:
% executes the following steps sequentially, 
\begin{itemize}
	\item A value $\btheta' \sim \MVN(\btheta^{(g-1)}, \bSigma_{g-1})$ is proposed
	% \item a value for $\btheta$ is proposed from $q_g$ 
	% \org{[previously used prime, and include the MVN with current theta and cov matrix $\bSigma_g$]}
	and a decision to accept or reject the sample is made based on the acceptance probability 
	% $p_{\text{acc}}(\btheta', \btheta^{(g-1)})$ 
	$A\left( (\btheta', \bsy'), (\btheta^{(g-1)}, \bsy^{(g-1)}) \right)$
	as discussed in Section \ref{sec:ABC-MCMC-algo}. 
	% \org{[use consistent formulas for the MH prob.  Either move this into 4.4 or take the notation I had added and use it here.]}
	Let $\btheta^{(g)}$ denote the final value of the chosen $\btheta$, which will either be $\btheta^{(g-1)}$ or $\btheta'$.
	\item The scaling factor $\eta_g$ is updated as
	\[
	\log\eta_g = \log\eta_{g-1} + \nu_g \left[ A \left( (\btheta', \bsy'), (\btheta^{(g-1)}, \bsy^{(g-1)}) \right) - \pacc \right],
	\]
	where $\pacc$ represents the target acceptance probability. 
	% Note that a higher current acceptance probability compared to the target will yield larger scaling factor compared to the current one, while a current acceptance probability smaller than the target will result in smaller $\eta_g$.
	\item The weighted sample mean and the covariance matrix are updated through
	\[
	\begin{array}{rcl}
		\overline{\bmu}_g & = & (1-\nu_g) \overline{\bmu}_{g-1} + \nu_g \btheta^{(g)}, \\
		\overline{\bSigma}_g & = & (1-\nu_g) \overline{\bSigma}_{g-1} + \nu_g \left\{ (\btheta^{(g)}-\overline{\bmu}_g) (\btheta^{(g)}-\overline{\bmu}_g)' \right\}.
	\end{array}
	\]
	% Observe that the new mean and covariance are obtained by updating their earlier values based on the observed $\btheta$ sample.
	\item The proposal covariance to be used in the next iteration is set to be $\bSigma_g = \eta_g \overline{\bSigma}_g$.
	% , which is used in iteration $g+1$.
\end{itemize}
We choose the vanishing adaptation factor to be $\nu_g = 1/500$ for the first 500 iterations and $1/g$ for iteration $g > 500$.
Note that, our choice of $\nu_g$ is such that $\sum_{g=1}^\infty \nu_g = \infty$ ensuring that all the points in the support of $\btheta$ can be reached, and $\nu_g$ also satisfies $\sum_{g=1}^\infty \nu_g^2 < \pi^2/6 < \infty$ implying that the proposed $\btheta$ samples have bounded fluctuations as discussed in \cite{Andrieu2008}.
%These two conditions together ensure almost sure convergence.
% \org{[??????????????????????????????????????????????]}
% will not prevent convergence.

It is worth making a brief comment on the target acceptance probability $\pacc$.
While it is  common to specify the target  acceptance rate near 0.25, we are generally not able to achieve rates this high since most generated datasets will produce poor kernel values $K_h(\Delta(\bss))$, driving down the MH probability. 
So, we must choose a smaller choice for the target acceptance probability $\bar{p}^*_{\text{acc}}$ to avoid the proposal variance from collapsing to zero.
Here we choose $\bar{p}^*_{\text{acc}}=0.1$.

% \org{[reformat this paragraph.  can add more details since in appendix and use offset equations.  also, should mention truncation for early iterations which also means the sum v2 is not correct, although it is still finite.]}
% \org{[still want to update this.  more clearly lay this out as an algorithm.  you could include the algorithm from 3.4 but add in the additional steps associated with the adaptation]}

To update the hyperparameters $\btheta_H$ associated with the variance of the regression coefficients, we use classical MH steps to update $\btheta_H$ using two random walk steps for  $\log(\tau^2_\alpha)$ and  $\log(\tau^2_\beta)$. 
% \org{[DISCUSS USING THE MARGINALIZATION DISTRIBUTIONS FOR NG]}
Here we make use of the fact that the prior on $\balpha$ and $\bbeta$ can be marginalized over $\sigma^2_{\alpha_k}$ and $\sigma^2_{\beta_k}$, respectively, in closed forms:
\[
\begin{array}{rcl}
	\pi(\alpha_k | \lambda_{\balpha}, \tau_{\balpha}^2) & = & \frac{1}{\sqrt{\pi} 2^{\lambda_{\alpha}/2-3/4} \Gamma(\lambda_{\balpha})} \left(\frac{\lambda_{\balpha}}{\tau_{\balpha}^2}\right)^{\lambda_{\balpha}/2 + 1/4} |\alpha_k|^{\lambda_{\balpha}-1/2} K_{\lambda_{\balpha}-1/2}\left( |\alpha_k| \sqrt{\frac{2\lambda_{\balpha}}{\tau_{\balpha}^2}} \right), \\
	\pi(\beta_k | \lambda_{\bbeta}, \tau_{\bbeta}^2) & = & \frac{1}{\sqrt{\pi} 2^{\lambda_{\bbeta}/2-3/4} \Gamma(\lambda_{\bbeta})} \left(\frac{\lambda_{\bbeta}}{\tau_{\bbeta}^2}\right)^{\lambda_{\bbeta}/2 + 1/4} |\beta_k|^{\lambda_{\bbeta}-1/2} K_{\lambda_{\bbeta}-1/2}\left( |\beta_k| \sqrt{\frac{2\lambda_{\bbeta}}{\tau_{\bbeta}^2}} \right).
\end{array}
\]
%where $K_\nu(\cdot)$ is the modified Bessel function of the second kind. 
The marginalized priors only depend on the hyperparameters $\tau^2_{\balpha}$ and $\tau^2_{\bbeta}$, when $\lambda_{\balpha} = \lambda_{\bbeta} = 1$.
Hence, the full conditional distribution for $\tau_{\balpha}^2$ and $\tau_{\bbeta}^2$ are proportional to $\pi(\tau_{\balpha}^2) \prod_{k=1}^d p(\alpha_k|\tau_{\balpha}^2)$ and $\pi(\tau_{\bbeta}^2) \prod_{k=1}^d p(\beta_k|\tau_{\bbeta}^2)$, respectively. 
We choose the proposal distributions $({\tau_{\balpha}^{2}}^* | \tau_{\balpha}^{2}) \sim \text{LN}(\tau_{\balpha}^2, \sigma_{\text{MH}}^2)$ for $\tau_{\balpha}^2$ and $({\tau_{\bbeta}^{2}}^* | \tau_{\bbeta}^{2}) \sim \text{LN}(\tau_{\bbeta}^2, \sigma_{\text{MH}}^2)$ for proposing $\tau_{\bbeta}^2$.

\subsection{Regression Adjustment Details} \label{appn:regression_adjustment}

We now elaborate on the regression adjustment procedure. 
The regression adjustment is typically performed on each univariate component of the parameter vector $\btheta$, by regressing it with respect to the summary statistics.
However, for adjusting a parameter that has a confined support of the form $(l,u)$, where at least one of the bounds is not $\pm \infty$, a suitable transformation is usually advisable.
In such cases, the regression correction is performed on the transformed parameter which is then inverse-transformed to obtain the adjusted samples for the parameter.
Let $\vartheta$ be a univariate estimand (after transformation when necessary), for instance a component of $\btheta$ 
%a transformed component, 
or a more complicated univariate function of $\btheta$.
Our aim is to correct the posterior samples obtained from ABC-MCMC with regression adjustment.
We first obtain an estimate of $\textrm{E}(\vartheta | \bss)$ at $\bss = \bss_{\obs}$ under conditional heteroscedasticity \citep{BlumFrancois2010}, by fitting a local-linear model: $\vartheta^{(g)} = \textrm{E}(\vartheta|\bss^{(g)}) + \sigma(\bss^{(g)})  \epsilon^{(g)}$ to the ABC-MCMC samples, where $\sigma^2(\bss^{(g)})$ denotes the conditional variance of $\vartheta$ given $\bss = \bss^{(g)}$ and $\epsilon^{(g)}$ denotes the standardized residual for the $g$-th sample.
The adjusted $\vartheta$ samples, denoted by $\ddot{\vartheta}^{(g)}$, are then computed as 
$\ddot{\vartheta}^{(g)} = \widehat{\mathrm{E}}(\vartheta | \bss_{\obs}) + \hat{\sigma}(\bss_{\obs})\hat{\epsilon}^{(g)}$, 
where the estimated standardized residual $\hat{\epsilon}^{(g)}$ is given by
\[
\hat{\epsilon}^{(g)} = \frac{1}{\hat{\sigma}(\bss^{(g)})} \left( \vartheta^{(g)} - \widehat{\E}(\vartheta | \bss^{(g)}) \right).
\]

To perform regression adjustment on a particular correlation coefficients from $\bR(\btheta_D)$, we note that each correlation is a univariate function of $\btheta_D$ with bounded support $(-1,1)$. 
For each correlation coefficient $\theta$ of interest, we apply a modified $z$-transform,  $\vartheta = z(\theta) = \log \left(\frac{ 1+\theta }{ 1-\theta } \right)$, and treat $\vartheta$ as the estimand to be regression-adjusted.
% Let the inverse z-transform be denoted by $z^{-1}(\cdot)$.
The regression-adjusted samples of $\vartheta$, denoted as $\ddot{\vartheta}^{(g)}$, are then inverse-transformed to yield the corresponding samples for $\theta$ given as $\{z^{-1}(\ddot{\vartheta}^{(g)})\}_{g=1}^G$.
% \org{[if we are only doing this with -1 and +1, we should report it in the simpler way.]}
Note that the correlation matrix $\bR$ whose elements have been regression-adjusted
% adjusting the individual correlation coefficients 
individually generally will  not give rise to samples of coherent correlation matrices;
while each component will be within $(-1, 1)$, the joint structure may not be associated with a generator $\btheta_D$ or even be positive definite.

It is worth pointing out here that an alternative implicit adjustment strategy can yield a sample of coherent correlation matrices.
Recall that $\bR = (\bI - \bB)^{-1} \bGamma (\bI - \bB)^{-1}$, where both $\bB$ and $\bGamma$ are functions of $\btheta_D$ (see Section \ref{sec:corr_structure} for details).
Therefore, we can first obtain regression-adjusted ABC-MCMC samples of $\btheta_D$, which can be used to generate samples for $\bR$.
We consider the strategy of directly adjusting $\bR$ samples as our main approach and compare it with the alternative strategy in Section \ref{appn:sim_results}.

However, adjusting multivariate $\btheta_D$ samples is challenging.
Due to the bounded support with multivariate constraints, $\btheta_D$ samples must be transformed before performing adjustment, which in turn is difficult as the support $\bTheta_D$ is not rectangular and does not have a tractable boundary.
To that end, we apply the modified $z$-transformation to each component of $\btheta_D^{(g)}$ as if the  support for each $\rho_k$ were $(-1,1)$. 
% Note that $(-1,1)^K \supset \bTheta_D$, implying that the assumed support of $\btheta_D$ vector covers its true support making the transformation of all parameter values and the subsequent regression adjustment feasible.
% \org{[this is confusing.  maybe state that you transform each componenet as if its support were [-1,1] so that you are transforming on a superset of THETA.D.?]}
The component-wise transformed $\btheta_D^{(g)}$, denoted by $\bvtheta_D^{(g)} = z(\btheta_D^{(g)})$, are regression-adjusted (component-wise) to $\ddot{\bvtheta}_D^{(g)}$ and finally inverse-transformed to obtain the corrected set of samples denoted as $\ddot{\btheta}_D^{(g)} = z^{-1}(\ddot{\bvtheta}_D)$.
However, this process does not always yield $\ddot{\btheta}_D^{(g)} \in \bTheta_D$, because for the  support for transformation space $(-1,1)^K$ is a strict superset of parameter space $\bTheta_D$.
% In these cases, the adjusted samples are further processed with a post-regression step for rectification as follows.
% Let $\dot{\btheta}_D^{(g)}$ be the transformed ABC sample, obtained from element-wise $z$-transformation of $\btheta_D^{(g)}$, and $\ddot{\btheta}_D^{(g)}$ be the corresponding regression-adjusted sample, which after transforming back, may lie outside $\bTheta_D$.
% If $\dddot{\btheta}_D\in\bTheta_D$, then we take $\dot{\btheta}_D^{*(g)}$ as the regression-adjusted sample without further modification. 
When the resulting adjusted parameters are outside the correct support $\ddot{\btheta}_D^{(g)}\notin \bTheta_D$, we then believe that $\ddot{\bvtheta}_D^{(g)}$ has been over-corrected,  and we need to backtrack toward the original sample vector $\bvtheta_D^{(g)}$ which is generated to lie in $\bTheta_D$.
Let $\bdelta = \ddot{\bvtheta}_D^{(g)} - \bvtheta_D^{(g)}$ denote the correction term.
Instead of $\bdelta$, we check if correcting $\bvtheta_D^{(g)}$ by $\bdelta/2$ 
produces a posterior sample in the support, that is, if $\ddot{\btheta}_D^{(g)} = z^{-1}(\ddot{\bvtheta}_D^{(g)})=z^{-1}(\bvtheta_D^{(g)}+\bdelta/2) \in \bTheta_D$;
if it does not, we 
further decrease the correction by selecting  the first $m=1,2,3,4,5$ such that $z^{-1}(\bvtheta_D^{(g)}+\bdelta/2^m) \in \bTheta_D$.
% whether $\bdelta/2^2$ works, and so on for maximum five times. 
If we still do not get a sample within support with $m=5$, we treat the uncorrected sample $\btheta_D^{(g)}$ as the final corrected version.
% \org{[This is getting close.  What is confusing here is that we use the same notation for $\btheta$ in the $\bTheta$-space as well as in the transformed space.  I think that we need to use different notation (for this section only) to clarify when we are working in the parameter vs the transformed space. maybe use $\vartheta$ in the transformed space so that $\vartheta=z(\theta)$?]}

\subsection{ABC  Rejection and Importance Sampling}
\label{appn:rej_imp_algo}

The standard ABC rejection algorithm is typically very inefficient in high-dimensional settings, and choosing the bandwidth $h$ in a way that balances computational resources and model fit is particularly challenging.
In our implementation of the ABC rejection algorithm, we generate and store 
a large number of 
% $G=$1,200,000 
parameter vectors $\btheta^{(g)}$
($g=1,\ldots,G$), drawn directly from the prior, along with their corresponding data $\bsy^{(g)}$ and the summary statistics $\bss^{(g)}$.
Finally, we keep 0.1\% of the parameter-data pairs that have the smallest $\Delta(\bsy^{(g)}, \bsy_{\obs})$; 
this strategy is roughly equivalent to using the uniform kernel where the bandwidth is given by this percentile of $\Delta(\bsy)$.
The samples that are retained are  regression-adjusted before using for inference.

We also implement a standard ABC-based importance sampling algorithm.
In iteration $g$ of the algorithm, we draw a $\btheta^{(g)}$ from the proposal density and use it to generate a data $\bsy^{(g)}$.
The corresponding importance weight is computed as $\tilde{w}^{(g)} \propto K_h(\Delta(\bsy^{(g)}, \bsy_\obs)) \pi(\btheta^{(g)}) / \tilde{q}(\btheta^{(g)})$, where $\pi(\cdot)$ and $\tilde{q}(\cdot)$ are the prior and the proposal.
% \org{[NEED TO THEN RENORMALIZE]}
We run $G$
% = 1,200,000$ 
such proposals, and store the weights along with the parameter-data pairs to be used for inference.
The proposal distribution for the importance sampling algorithm is chosen to be a multivariate Normal distribution $\MVN(\widetilde{\btheta}, c^2\widetilde{\bSigma})$, where $\widetilde{\btheta}$ and $\widetilde{\bSigma}$ are obtained
% given by the mean and covariance of posterior samples of $\btheta$ from misspecified model $\model{0}$, 
as described in Section \ref{appn:initial_estimation}.
The scaling factor $c^2 = 4$ is chosen to inflate the variance of the proposal distribution in order to propose parameters from a more dispersed distribution than the normal approximation to $\model{0}$.
The kernel scaling matrix $\bA$ is the same as what was used for the ABC-MCMC algorithm (see Section \ref{appn:algo_details}).
% \org{[You haven't said how you are generating the $\btheta_D$ components.]}
The importance samples are regression-adjusted to obtain the final set of weighted samples.
% \org{[Again, reorder this.  Need to introduce the algorithm before you get into details about the proposal.]}

It is worth adding some comments about the challenges that come with these algorithms.
A very large number of parameter draws is required to obtain a set of parameter samples that provides a good approximation to the posterior distribution.
This is particularly true in the rejection sampler when drawing directly from a disperse prior.
As in ABC-MCMC, the primary computational step is generating the dataset $\bsy^{(g)}$ from $p(\bsy|\btheta)$ and the summarization of the data $\bss(\bsy^{(g)})$. Consequently, the computational cost of a single iteration is roughly the same for all of these samplers.  
As ABC-MCMC is best able to propose $\btheta$ from the region of highest posterior mass, it tends to generate more useful parameter choices and datasets that better match $\bsy_\obs$.
Although regression adjustment was employed on the posterior samples, the correction will only improve estimation if the drawn $(\btheta^{(g)}, \bss^{(g)})$ captures the correct relationship locally between the parameters and the summary statistics.
But, this does not necessarily happen with rejection sampling since it is using $\bss^{(g)}$s that are very far from $\bss_\obs$.  
In our importance sampling with regression-adjustment, the local regression  model is estimated using $(\btheta^{(g)}, \bss^{(g)})$ that have approximately zero posterior weight, which may limit the utility of regression adjustment.
% With very limited sample size this is often infeasible.
Further with the importance sampler, the importance sampling  weights typically end up highly imbalanced. 
Different strategies for adjusting the tail of the weight distribution \citep[e.g.,][]{Ionides2008, Vehtari2022} could be employed to stabilize these weights, but in our experiments, these methods failed to smooth the importance weighted posterior.

\section{Additional Details on IFS Data Analysis}

In this section, we present additional details on the IFS data analysis
presented in Section \ref{sec:IFS_study} of the main manuscript. 
The adjacency matrices defining the dependence model are described in the first subsection, where we explicitly provide all the connections between tooth-time pairs.
In Section \ref{appn:IFS_computation}, we provide a discussion on ABC-MCMC convergence diagnostics.
Additional model comparison results are presented next.
This section concludes with a discussion on alternative marginal model specifications and assessment of their predictive performance compared to the proposed model.
% \org{[NOT THE RIGHT ORDER]}

\subsection{Additional model setup details} \label{appn:IFS_setup}

\begin{figure}[tb]
\centering
\includegraphics[width=0.65\textwidth]{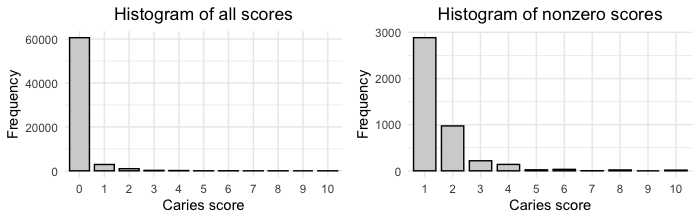}
\caption{The figure in the left panel shows the overall distribution of caries experince scores across all teeth, demonstrating clear zero inflation. The figure on the right shows distribution of the non-zero data indicating potential over-dispersion.
}
\label{fig:zero-inflation}
\end{figure}

Figure \ref{fig:zero-inflation} provides an overview of the zero-inflation and potential over-dispersion in the IFS data.
Additionally, Table \ref{tab:appn_data_summary} summarizes the IFS data at each age.
%The proportions of zeros show clear evidence of zero-inflation at every age.
%The skewness in the non-zero caries scores is also apparent from the five point summaries; the minimum, first quartile ($Q_1$) and median of the scores are same and very close to the third quartiles ($Q_3$) at every age compared to the much larger difference between $Q_3$ and the maximum scores (except at age 13).
A brief description of the predictor variables used for the IFS data analysis is contained in Table \ref{tbl:predictors}.

\begin{table}[!tb]
\footnotesize
\centering
\begin{tabular}{cccccccc}
	\hline
	Age & Number of & Number of & Proportion & Proportion & Proportion    & Proportion    & Proportion         \\
	& patients  & teeth & of $Y=0$   & of $Y>0$   & of $Y \geq 3$ & of $Y \geq 5$ & of $Y \geq 8$ \\ \hline
	all & 728 & 64,926 & 0.934 & 0.066 & 0.007 & 0.001 & 0.001 \\	
	5   & 696  & 13,751     & 0.953       & 0.047        & 0.006            & 0.002            & 0.001            \\
	9   & 629   & 14,623    & 0.948       & 0.052        & 0.006            & 0.001            & <0.001             \\
	13  & 549   & 14,523    & 0.958       & 0.042        & 0.001            & 0.000             & 0.000             \\
	17  & 463   & 12,667    & 0.881       & 0.119       & 0.011           & 0.001            & <0.001             \\
	23  & 342   & 9362    & 0.916       & 0.084        & 0.013           & 0.005            & 0.002            \\ \hline
\end{tabular}
\caption{Summary of the caries experience scores in the IFS data categorized by age.}
\label{tab:appn_data_summary}
\end{table}

\begin{table}[!tb]
\centering
\footnotesize
\begin{tabular}{ll}
	\hline
	Predictor & Interpretation \\ \hline
	Behavioral Variables &  \\
	\hspace{1em} Dental Visit (past 6 months)         &  Proportion of times a dental visit during past 6 months \\
	\hspace{1em} Daily total fluoride ingested (mgF)             &  Amount of total fluoride (mg/day) ingested from all sources\\
	\hspace{1em} Frequency of brushing (past 6 months)             &  Average daily brushing frequency during past 6 months \\
	% & (>3 times/day, 3 times/day, twice/day, once/day, < once/day) \\
	\hspace{1em} Daily amount of sugar beverages (oz)           &  Amount of total sugar-added beverages consumed (oz/day) \\ 
	\hline
	Tooth Type           &  (see Figure \ref{fig:appn_dentition})\\
	\hspace{1em} Molar                &  Reference category \\
	& (primary: A,B,I,J,K,L,S,T; permanent: 1-3,14-16,17-19,30-32)\\
	\hspace{1em} Pre-molar            &  Indicator variable for pre-molar teeth \\
	& (permanent: 4,5,12,13,20,21,28,29) \\
	\hspace{1em} Canine               &  Indicator variable for canine teeth \\
	& (primary: C,H,M,R; permanent: 6,11,23,27) \\
	\hspace{1em} Incisor              &  Indicator variable for incisor teeth \\
	&  (primary: D,E,F,G,N,O,P,Q; permanent: 7,8,9,10,23,24,25,26) \\
	Primary              &  Indicator variable to distinguish primary teeth from \\
	& permanent teeth (primary: 1, permanent: 0) \\ 
	\hline
	Observation Time     &  \\
	\hspace{1em} Age-5                &  Indicator variable for age 5 \\
	\hspace{1em} Age-9                &  Reference category \\
	\hspace{1em} Age-13               &  Indicator variable for age 13 \\
	\hspace{1em} Age-17               &  Indicator variable for age 17 \\
	\hspace{1em} Age-23               &  Indicator variable for age 23 \\ 
	\hline
\end{tabular}
\caption{\label{tbl:predictors} Description of the predictors for the IFS data.}
\end{table}

\begin{figure}[!tb]
\centering
\includegraphics[width=0.9\textwidth]{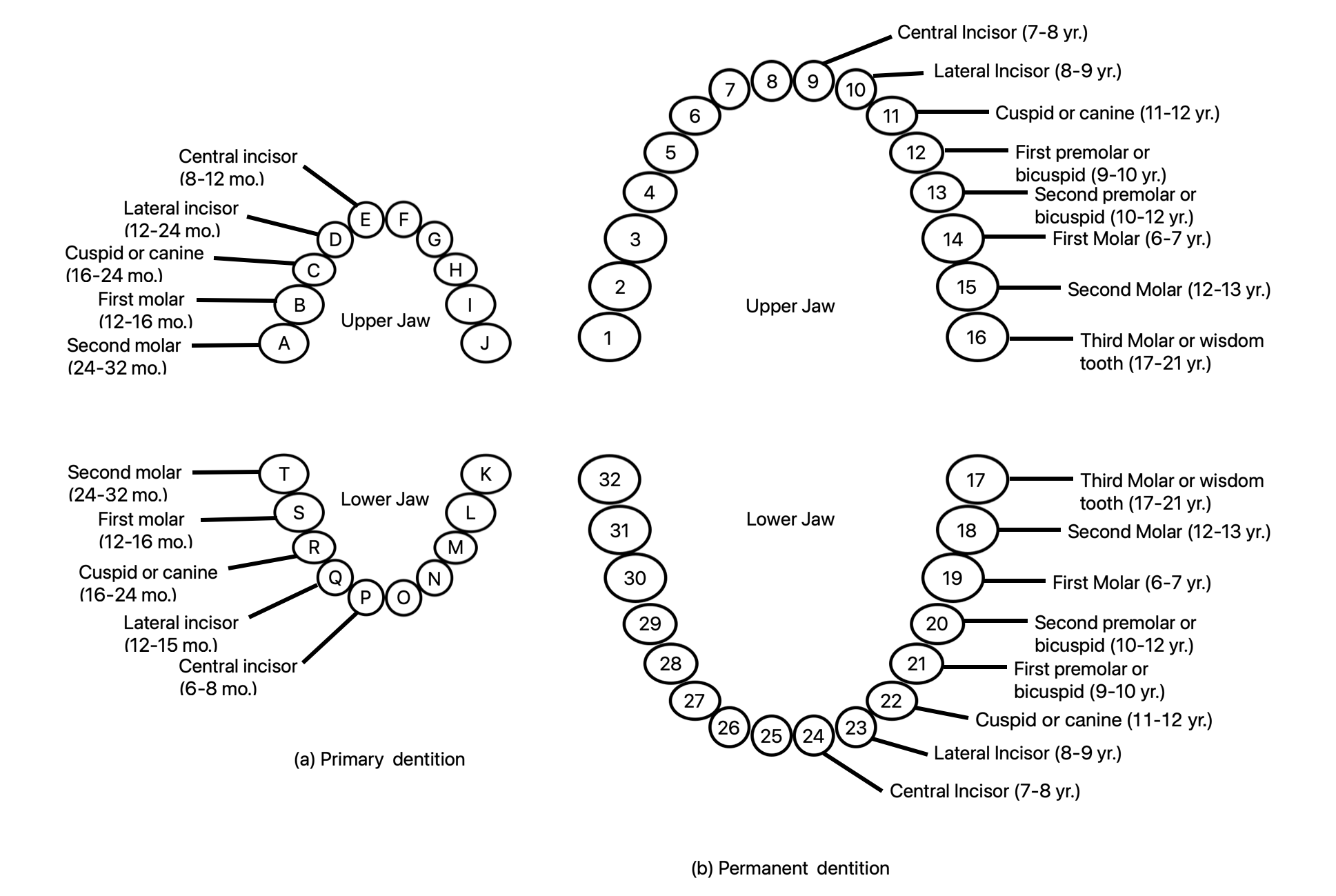}
\caption{The mixed dentition chart shows all the primary and permanent teeth with the approximate age of eruption. 
	%    The teeth on left and right side of the upper/lower jaw have same names and therefore have been labeled on one side only.
}
% \anish{(Collected from the internet, probably we will change the figure, keeping it as a quick reference for now).}
% \org{[If you can, add a citation and reference here in the caption.]}
%    }
\label{fig:appn_dentition}
\end{figure}

We now enumerate the adjacency relationships we considered for modeling the dependence structure of the IFS data.
\begin{table}[p]
\centering
\footnotesize
\begin{tabular}{ccccc}
	\hline
	% &  &  &  & \\
	\multicolumn{5}{l}{$\bW^{(t)}$: Temporal adjacency} \\
	\multicolumn{5}{l}{For each location $l \in \{1, \ldots, 32\} \cup \{A, \ldots,   T\}$} \\
	&  &  &  & \\
	$((l,5),(l,9))$ & $((l,9),(l,13))$ & $((l,13),(l,17))$ & $((l,17),(l,23))$ & \\
	\hline
	&  &  &  & \\
	\hline
	\multicolumn{5}{l}{$\bW^{(h)}$: Horizontal tooth adjacency}\\
	\multicolumn{5}{l}{For each time $t \in \{5,9,13,17,23\}$, connected pairs are:} \\
	&  &  &  & \\
	((1,$t$),(2,$t$)) & ((2,$t$),(3,$t$)) & ((3,$t$),(4,$t$)) & ((4,$t$),(5,$t$)) & ((5,$t$),(6,$t$))\\
	
	((6,$t$),(7,$t$)) & ((7,$t$),(8,$t$)) & ((8,$t$),(9,$t$)) & ((9,$t$),(10,$t$)) & ((10,$t$),(11,$t$))\\
	
	((11,$t$),(12,$t$)) & ((12,$t$),(13,$t$)) & ((13,$t$),(14,$t$)) & ((14,$t$),(15,$t$)) & ((15,$t$),(16,$t$))\\
	
	((16,$t$),(17,$t$)) & ((17,$t$),(18,$t$)) & ((18,$t$),(19,$t$)) & ((19,$t$),(20,$t$)) & ((20,$t$),(21,$t$))\\
	
	((21,$t$),(22,$t$)) & ((22,$t$),(23,$t$)) & ((23,$t$),(24,$t$)) & ((24,$t$),(25,$t$)) & ((25,$t$),(26,$t$))\\
	
	((26,$t$),(27,$t$)) & ((27,$t$),(28,$t$)) & ((28,$t$),(29,$t$)) & ((29,$t$),(30,$t$)) & ((30,$t$),(31,$t$))\\
	
	((31,$t$),(32,$t$)) &  &  &  & \\
	
	((A,$t$),(B,$t$)) & ((B,$t$),(C,$t$)) & ((C,$t$),(D,$t$)) & ((D,$t$),(E,$t$)) & ((E,$t$),(F,$t$))\\
	
	((F,$t$),(G,$t$)) & ((G,$t$),(H,$t$)) & ((H,$t$),(I,$t$)) & ((I,$t$),(J,$t$)) & ((K,$t$),(L,$t$))\\
	
	((L,$t$),(M,$t$)) & ((M,$t$),(N,$t$)) & ((N,$t$),(O,$t$)) & ((O,$t$),(P,$t$)) & ((P,$t$),(Q,$t$))\\
	
	((Q,$t$),(R,$t$)) & ((R,$t$),(S,$t$)) & ((S,$t$),(T,$t$)) &  & \\
	\hline
	&  &  &  & \\
	\hline
	\multicolumn{5}{l}{$\bW^{(v)}$: Vertical tooth adjacency}\\
	\multicolumn{5}{l}{For each time $t \in \{5,9,13,17,23\}$, connected pairs are:} \\
	&  &  &  & \\
	((1,$t$),(32,$t$)) & ((2,$t$),(31,$t$)) & ((3,$t$),(30,$t$)) & ((4,$t$),(29,$t$)) & ((5,$t$),(28,$t$))\\
	
	((6,$t$),(27,$t$)) & ((7,$t$),(26,$t$)) & ((8,$t$),(25,$t$)) & ((9,$t$),(24,$t$)) & ((10,$t$),(23,$t$))\\
	
	((11,$t$),(22,$t$)) & ((12,$t$),(21,$t$)) & ((13,$t$),(20,$t$)) & ((14,$t$),(19,$t$)) & ((15,$t$),(18,$t$))\\
	
	((16,$t$),(17,$t$)) &  &  &  & \\
	
	((A,$t$),(T,$t$)) & ((B,$t$),(S,$t$)) & ((C,$t$),(R,$t$)) & ((D,$t$),(Q,$t$)) & ((E,$t$),(P,$t$))\\
	
	((F,$t$),(O,$t$)) & ((G,$t$),(N,$t$)) & ((H,$t$),(M,$t$)) & ((I,$t$),(L,$t$)) & ((J,$t$),(K,$t$))\\
	\hline
	&  &  &  & \\
	\hline
	\multicolumn{5}{l}{$\bW^{(pp)}$: Primary-permanent adjacency}\\
	\multicolumn{5}{l}{For each time $t_k \in \{5,9,13,17,23\}$, with $t_k$ and $t_{k+1}$ being adjacent time points, connected pairs are:}\\
	&  &  &  & \\
	((2,$t_k$),(A,$t_{k+1}$)) & ((8,$t_k$),(E,$t_{k+1}$)) & ((14,$t_k$),(I,$t_{k+1}$)) & ((22,$t_k$),(M,$t_{k+1}$)) & ((26,$t_k$),(Q,$t_{k+1}$))\\
	
	((3,$t_k$),(B,$t_{k+1}$)) & ((9,$t_k$),(F,$t_{k+1}$)) & ((15,$t_k$),(J,$t_{k+1}$)) & ((23,$t_k$),(N,$t_{k+1}$)) & ((27,$t_k$),(R,$t_{k+1}$))\\
	
	((6,$t_k$),(C,$t_{k+1}$)) & ((10,$t_k$),(G,$t_{k+1}$)) & ((18,$t_k$),(K,$t_{k+1}$)) & ((24,$t_k$),(O,$t_{k+1}$)) & ((30,$t_k$),(S,$t_{k+1}$))\\
	
	((7,$t_k$),(D,$t_{k+1}$)) & ((11,$t_k$),(H,$t_{k+1}$)) & ((19,$t_k$),(L,$t_{k+1}$)) & ((25,$t_k$),(P,$t_{k+1}$)) & ((31,$t_k$),(T,$t_{k+1}$))\\
	\hline
	& & & & \\
	\hline
	\multicolumn{5}{l}{$\bW^{(ct)}$: Equal connection across location within same time} \\
	\multicolumn{5}{l}{For each location $l_1,l_2 \in \{1, \ldots, 32\} \cup \{A, \ldots,   T\} $ and $t \in \{ 5,9,13,17,23 \}$, connected pairs are:} \\
	& & & & \\
	\multicolumn{5}{c}{$((l_1, t),(l_2, t))$ } \\
	\hline
	& & & & \\
	\hline
	\multicolumn{5}{l}{$\bW^{(ce)}$: Equal connection everywhere (across location and time)} \\
	\multicolumn{5}{l}{For each location $l_1,l_2 \in \{1, \ldots, 32\} \cup \{A, \ldots,   T\} $ and $t_1,t_2 \in \{ 5,9,13,17,23 \}$, connected pairs are: } \\
	& & & & \\
	\multicolumn{5}{c}{ $((l_1, t_1),(l_2, t_2))$ } \\
	\hline
\end{tabular}
\caption{\label{tbl:adjacency}Adjacency relationships used in the IFS analysis.}
\end{table}
% \FloatBarrier
Table \ref{tbl:adjacency} presents all the tooth-time pairs that are adjacent  according to the proximity relations in  the matrices $\bW^{(t)}$, $\bW^{(h)}$, $\bW^{(v)}$, $\bW^{(pp)}$, $\bW^{(ct)}$ and $\bW^{(ce)}$.
To recall, we have five time-points in the data, based on the measurement ages of 5, 9, 13, 17 and 23.
There are 20 primary tooth locations denoted by letters A, B, $\ldots$, T and the 32 permanent tooth locations are numbered 1,2, $\ldots$, 32.
For reference, we include a pictorial representation in Figure \ref{fig:appn_dentition} of the standard naming convention for tooth locations inside mouth.
% (same as Figure \ref{fig:dentition} in Section \ref{sec:IFS_model_setup}).
% \org{[maybe include the figure again here for convenience?]}

\pagebreak

\FloatBarrier

\subsection{ABC-MCMC Mixing and Convergence} \label{appn:IFS_computation}
% \org{[including some trace plots/evidence of adequate mixing under the best fitting model would be helpful here.  maybe ESS?]}
% \org{[If this section is discussing computational details for all models it should come before the model comparisons.  ]}
For each model, we have run 3 chains of ABC-MCMC, each of length 185,000, from which the first 5000 from each chain were removed as the burn-in phase.
% \org{[Do we thin???]}
We have considered four bandwidth choices of 1, 10, 30 and 100 for the IFS data analysis.
While these samples are regression-adjusted 
% and thinned 
to yield the final set of approximate posterior samples for inference,
here we first discuss the performance and mixing of the ABC-MCMC algorithm with regard to the pre-adjustment samples.

\begin{table}[!tb]
\centering
\footnotesize
\begin{tabular}{clcccc}
\hline
Models & Dependence Structure & $h=1$ & $h=10$ & $h=30$ & $h=100$ \\ 
\hline
$\model{1}$ & $\bB = \rho_t \bW^{(t)} + \rho_h \bW^{(h)}$ & 0.22 & 0.92 & 1.06 & 1.11 \\ 
$\model{2}$ & $\bB = \rho_t \bW^{(t)} + \rho_h \bW^{(h)} + \rho_{pp} \bW^{(pp)}$ & 0.15 & 0.90 & 1.04 & 1.29 \\ 
$\model{3}$ & $\bB = \rho_t \bW^{(t)} + \rho_h \bW^{(h)} + \rho_{pp} \bW^{(pp)} + \rho_v \bW^{(v)}$ & 0.13 & 0.83 & 1.04 & 1.36 \\ 
$\model{4}$ & $\bB = \rho_{ct} \bW^{(ct)} + \rho_t \bW^{(t)}$ & 0.03 & 0.76 & 0.91 & 1.15 \\ 
$\model{5}$ & $\bB = \rho_{ct} \bW^{(ct)} + \rho_t \bW^{(t)} + \rho_h \bW^{(h)}$ & 0.02 & 0.76 & 0.86 & 1.20 \\ 
$\model{6}$ & $\bB = \rho_{ce} \bW^{(ce)}$ & 0.04 & 0.71 & 1.01 & 1.25 \\ 
$\model{7}$ & $\bB = \rho_{ce} \bW^{(ce)} + \rho_{t} \bW^{(t)}$ & 0.04 & 0.74 & 0.99 & 1.17 \\ 
$\model{8}$ & $\bB = \rho_{ce} \bW^{(ce)} + \rho_{t} \bW^{(t)} + \rho_{h} \bW^{(h)}$ & 0.09 & 0.77 & 1.20 & 1.17 \\ 
\hline
\end{tabular}
\caption{Acceptance rate per 100 ABC-MCMC samples across different models and bandwidth choices for the IFS analysis.}
\label{tbl:acceptance_rates_IFS}
\end{table}

Table \ref{tbl:acceptance_rates_IFS} reports the MH acceptance rate across different bandwidths and models.
Since the ABC-MCMC algorithm accepts or rejects the parameter and the generated dataset jointly, the acceptance rate is low compared to standard Metropolis algorithms.
As expected when the bandwidth increases, so does the percentages of accepted samples for each model.
This is because higher bandwidth allows more disagreement between the proposed $\bss$ and the observed $\bss_\obs$.
While we may wish to use $h=1$, since smaller $h$ means that $\pi_{\ABC}(\btheta\,|\,\bss_\obs)$ in equation (\ref{eq:posterior_ABC}) is closer to the target posterior $\pi(\btheta\,|\,\bsy_\obs)$ in (\ref{eq:posterior}), we get much poorer computational performance with this small bandwidth, yielding very few unique samples during ABC-MCMC.
This indicates that the ABC algorithm has failed to move around the parameter space effectively and is potentially stuck at a local mode. 
% We have attempted to use $h=0.1$ but the number of unique samples is too small to draw any reasonable inference when $h = 0.1$, making the bandwidth choice infeasible.
% It is also important to observe that 
Under $h=10$ we get a manageable acceptance rate, which
% the acceptance probability 
does not increase considerably as the bandwidth increases from $h = 10$ to $h = 100$.
% The acceptance rate is still very small when $h=1$ considering the total number of samples generated using 3 chains.
% This indicates that the ABC algorithm has fail to cover the parameter space effectively, being potentially stuck at a local optimum. 
Therefore, we  consider at least $h=10$ in this case.

\begin{figure}[tb]
\centering
\includegraphics[width=0.5\textwidth]{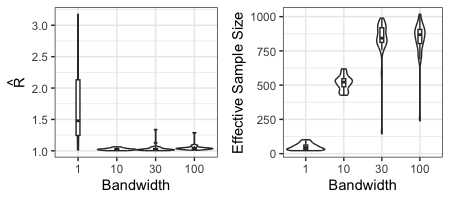}
\caption{Boxplots with $\hat{R}$ and effective sample sizes for all the parameters across different bandwidth choices under the selected model $\model{8}$.}
\label{fig:rhat_ess_IFS}
\end{figure}

\begin{figure}[tb]
\centering
\includegraphics[width=\textwidth]{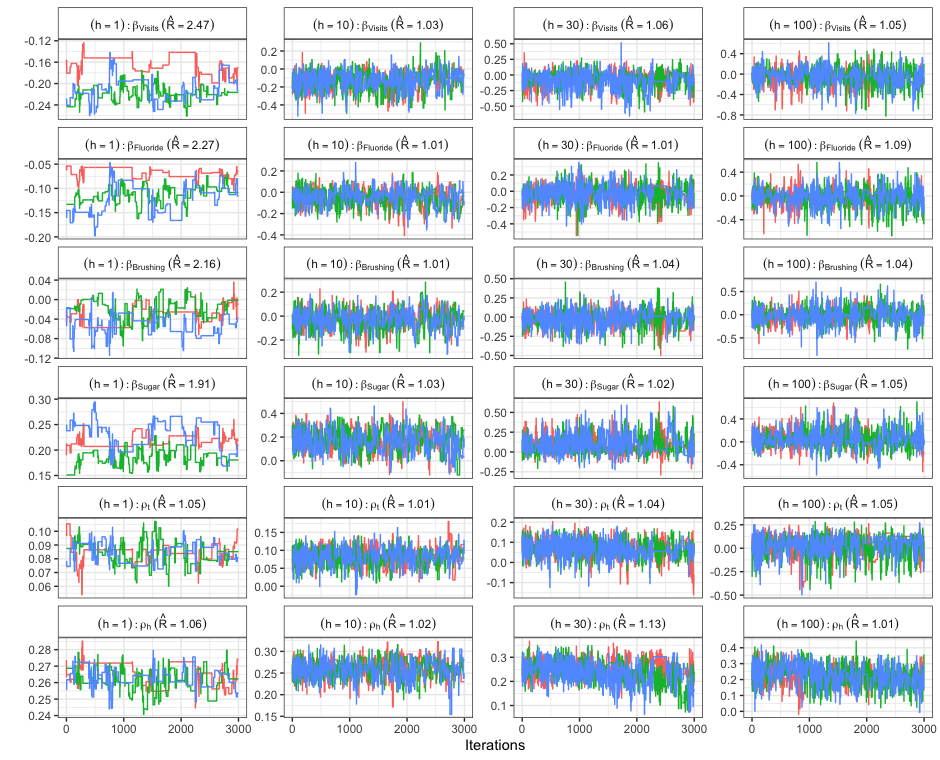}
\caption{Comparisons of the mixing of the ABC-MCMC chains from model $\model{8}$ 
for few selected parameters under different bandwidth choices. The retained values (pre-adjustment) after thinning are depicted in different colors (red, green, blue) for each chain.
% $\hat{R}$, the Gelman-Rubin statistic, are reported upto 2 decimal120s.
}
\label{fig:abc_mixing_IFS_compare}
\end{figure}

\begin{figure}[tb]
\centering
\includegraphics[width=\textwidth]{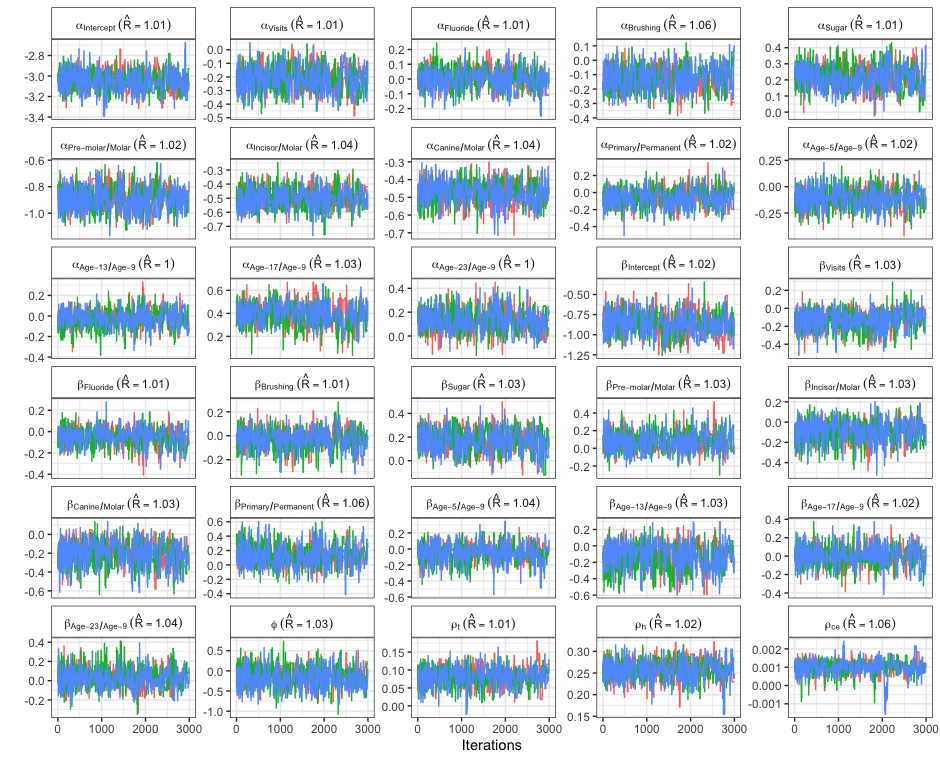}
\caption{Mixing of the ABC-MCMC chains ($h=10$) for different parameters under model $\model{8}$. 
The retained values (pre-adjustment) after thinning for each chain are depicted in different colors (red, green, blue).
% $\hat{R}$, the Gelman-Rubin statistic, are reported upto 2 decimals.
}
\label{fig:abc_mixing_IFS}
\end{figure}

We compare the MCMC convergence and mixing across different bandwidth choices in Figure \ref{fig:rhat_ess_IFS}.
To generate the plots, we combine the samples from all the chains (without thinning) and obtain a total sample size of 540,000.
% Figure \ref{fig:rhat_ess_IFS} shows how the bandwidth selection impacts convergence of the MCMC chains and the effective sample size.
% \org{[ARE THESE RHAT AND ESS COMPUTED AFTER THINNING OR BEFORE?  IT ISN'T CLEAR.  IF IT IS AFTER, CAN YOU RECOMPUTE TO BEFORE? ALTHOUGH IF REMOVING THINNING MESSES UP THE RHATS, THEN IT IS FINE TO LEAVE IT AS THINNED.]}
For each choice of $h$, we compute the Gelman-Rubin $\hat{R}$ statistic for all the parameters and create a boxplot in the left panel of Figure \ref{fig:rhat_ess_IFS}.
These comparisions are based on the selected model $\model{8}$, although results are qualitatively similar when comparing within other model choices.
We observe poor convergence for most of the parameters when $h=1$ with $\hat{R} > 1.1$ for almost all parameters.
% as expected from a very low acceptance rate.
Convergence is much better in the case of $h=10$, as also shown in the trace plots in Figures \ref{fig:abc_mixing_IFS_compare}--\ref{fig:abc_mixing_IFS}. 
Mixing for the larger bandwidths is also generally good. 
%except for a few parameters which did not converge well.

The effective sample sizes (ESS) also portray a similar picture in the right panel of Figure \ref{fig:rhat_ess_IFS}.
We calculate the ESS for each MCMC chain using the standard definition $N/(1 + \sum_{t=1}^\infty \varrho_t)$, where $N$ is the chain length and $\varrho_t$ represents the estimated autocorrelation at lag $t$ for the chain.
The combined ESS is found by summarizing across the three MCMC chains.
%, that henceforth we will refer to simply as ESS, is determined by the sum of the effective sample sizes for each chain.
% \org{[NEED TO DEFINE ESS]}
% Each boxplot is made out of these ESS for all the parameters.
% , where the ESS for a parameter is determined by the sum of the ESS 
% from each of the three chains.
% from the concatenation of the samples of the parameter from the three chains.
% the samples are obtained combining three chains of ABC-MCMC.
Average ESS across all parameters grows as $h$ increases from 1 to 30, but does not increase further when $h=100$.
Additionally, the range of ESS across the different parameters is quite wide at $h=30$ and  100. 
%while much lower variability is observed at $h=10$.
This suggests that, for larger $h$, ABC-MCMC is not moving as efficiently in some directions of the joint parameter space.
This indicates that increasing the bandwidth, thereby allowing the sampler to accept $\bss$ that differ more from $\bss_{\obs}$, beyond a certain threshold has not helped increase the effective sample size.
Moreover, with larger $h$ we allow for more approximation error in the ABC posterior.
These observations indicates that $h=10$ is an effective choice.

% \org{[Note that I swapped the order here to match the order in which we should consider them.  Remove the final row from Figure B3]}

We investigate the MCMC mixing/convergence for a subset of parameters
% these severity model parameters along with the SAR model parameters $\btheta_D = (\rho_{ce}, \rho_t, \rho_h)$ for different bandwidth choices 
in Figure \ref{fig:abc_mixing_IFS_compare}, where each column represents the MCMC chains corresponding to a bandwidth choice.
As noted previously, we run three chains for each method, and these traceplots depict the samples from each chain in different color, after thinning to 3000 samples per chain.
It is very clear that the $\bbeta$ parameters do not mix well when $h=1$, with $\hat{R}$ values around 2.
% , whereas $\btheta_D$ appears to have better mixing.
For the other bandwidths, convergence appears to be achieved and mixing is acceptable.  As $h=10$ is the smallest bandwidth considered that has acceptable computational performance, this is the choice we use for inference of the IFS data.
% For larger bandwidths $h=30,100$, $\bbeta$ chains mix fairly well while $\rho_{ce}$ chains diverge.
% These observations go fairly well with the acceptance rates and effective sample sizes observed in Table \ref{tbl:acceptance_rates_IFS} and Figure \ref{fig:rhat_ess_IFS} respectively.
% \org{[How are we seeing that here?  What is the point you want to make with this?]}
%  The acceptance rate for the algorithm turns out to be \anish{??}.

Having chosen $h=10$ as the bandwidth for analysis, we now
% To visually inspect the mixing, we 
create trace plots from the ABC-MCMC chains for all model  parameters $\btheta$ in Figure \ref{fig:abc_mixing_IFS} to ensure no other diagnostic concerns. 
The chains appear to have good mixing and to have reached convergence as reflected by the Gelman-Rubin statistic ($\hat{R} < 1.1$ for each parameter).
% \org{[based on $R<1.2$? if this is the criteria, you are using report that.  Really, comparing to 1.1 is probably more common.]}
Note that we achieve fairly good convergence for the severity model parameters $\bbeta$ and $\phi$, despite having much smaller amounts of data available to fit the severity model due to zero-inflation.

We would also like to further comment on the applicability of the importance sampler for analysis of the IFS data.
The importance sampling strategy (Appendix \ref{appn:rej_imp_algo}) was employed to generate $G$=1,200,000 posterior samples of $\btheta$ and their corresponding weights for the same bandwidth choices ($h=1,10,30,100$).
Generating 1.2 million samples required approximately double the computing time as was required for running the corresponding ABC-MCMC.
The importance weights were highly imbalanced regardless of the bandwidth used; with the largest $h = 100$, the top five weights turned out to be 0.23, 0.19, 0.17, 0.06 and 0.05, accounting for more 70\% of the posterior distribution.
The posterior was even more tightly concentrated on a handful of samples for the other bandwidths.
Pareto smoothing the importance weights (\cite{Vehtari2022}) was not useful, as the Pareto $\hat{k}$ diagnostics were more 1, indicating that a reliable finite variance point estimate can not be obtained.
Hence, we can not pursue any further interpretation of the IFS analysis under an  ABC importance sampling estimation strategy.
While not included  here, additional experiments using a 
sequential Monte Carlo (SMC)  algorithm and its extensions for ABC \citep{Sisson2007, bonassi2015} also failed to achieve  balanced weights in our experiments.

% \org{[If useful, could include this a portion of this plot (maybe first row) for each choice of bandwidth.]}

% \begin{figure}[tb]
%     \centering
%     \includegraphics[width=\textwidth]{pics/abc_mixing_IFS_uniform.png}
%     \caption{Mixing of the ABC-MCMC chains for different parameters under model $\model{8}$ for $h=150$. 
%     The retained values (pre-adjustment) after thinning for each chain are depicted in different colors (red, green, blue).
%     % $\hat{R}$, the Gelman-Rubin statistic, are reported upto 2 decimals.
%     }
%     \label{fig:abc_mixing_IFS_uniform}
% \end{figure}

We also investigated the performance of the proposed ABC-MCMC algorithm under a uniform kernel. 
When using a uniform kernel rather than the Gaussian kernel, we find that the initial parameter value is a more important choice to ensure MCMC mixing.  For one, we must initialize using a dataset $\bsy$ such that $\Delta(\bss)<h$, otherwise the joint likelihood is not defined. 
But, finding a $\btheta^{(0)}$ that yields such a $\bsy$ may require hundreds or thousands of draws from $p(\bsy|\btheta)$ when $h$ is small. 
Even after finding an initial value, it will continue to be rare to generate a proposed $\bsy$ with $\Delta(\bss)<h$ as the chain runs. 
Alternatively, using a large $h$ can facilitate accepting new data, this tended to lead to high autocorrelation in the posterior samples $\btheta$. 
While this can be corrected through careful tuning of the adaptive MH parameters, we find that running ABC-MCMC with the uniform kernel is more sensitive to algorithmic choices such as the initial data and parameter values and the adaptation parameters $\eta_0$ and $(\nu_g)$. 
Consequently, we recommend the Gaussian kernel and limit our analyses to this choice.

%%While the estimation performance turned out to be similar with those from the Gaussian kernel, finding good initial values of $\btheta$ and choosing hyper-parameters for the covariate adaptation were more tricky. 
%With a uniform kernel, it required more effort to find the first sample $\btheta^{(0)}$ that would be accepted with a small $h$, compared to a Gaussian kernel.
%Starting off the algorithm with a large $h$ uniform kernel, on the other hand, resulted in very slow mixing, as some chosen sequence $(\nu_g)$ did not result in fast enough adaptation.
%Finding correct choices for the covariance-adaptation hyperparameters $\eta_0$ and $(\nu_g)$, therefore, becomes more tricky in case of uniform kernel.
%With Gaussian kernel having small $h$, finding an acceptable $\btheta^{(0)}$ was easier, which made tuning the adaptation scheme more manageable.
%%However, it is worth mentioning that $\Delta(\bsy, \bsy_\obs)$ for the accepted sammples under Gaussian kernel ($h=10$) were close to 120, being effectively similar to those from an uniform kernel. 
%Running a Gaussian kernel based ABC-MCMC sampler first and investigating the accepted samples helped us adapt the sampler for a uniform kernel.

\subsection{Additional Model Comparison Results for the IFS Data} \label{appn:IFS_modeling}

%\begin{table}[!tb]
%	\centering
%	\footnotesize
%	\begin{tabular}{ll}
%		\hline
%		Statistic ($\bss_M$) & Interpretation \\
%		\hline
%		$\bar{Y}$ &  Overall Mean of $Y$             \\
%		$\textrm{Var}(Y)$ & Overall Variance of $Y$  \\
%		$\overline{Y \geq k}$ & Proportion of $Y \geq k$               \\
%		$\overline{Y(Y \geq k)}$ &  Mean of $Y$ such that $Y \geq k$ \\
%		$\textrm{Var}(Y(Y \geq k))$ &  Variance of $Y$ such that $Y \geq k$   \\
%		$\overline{Y(X \leq X_{1/3})} - \overline{Y(X > X_{2/3})}$  &  Difference between mean $Y$ where corresponding \\
%		&  covariate $X$ is $\leq$ its first tertile $X_{1/3}$ and mean $Y$ \\
%		& where $X$ is $> X_{2/3}$, its second tertile. \\
%		\hline          
%	\end{tabular}
%	\label{tbl:pp_stat_M}
%	\caption{Statistics for posterior predictive checks of the marginal model fit.}
%\end{table}

\begin{figure}[!bt]
\centering
\includegraphics[width=\textwidth]{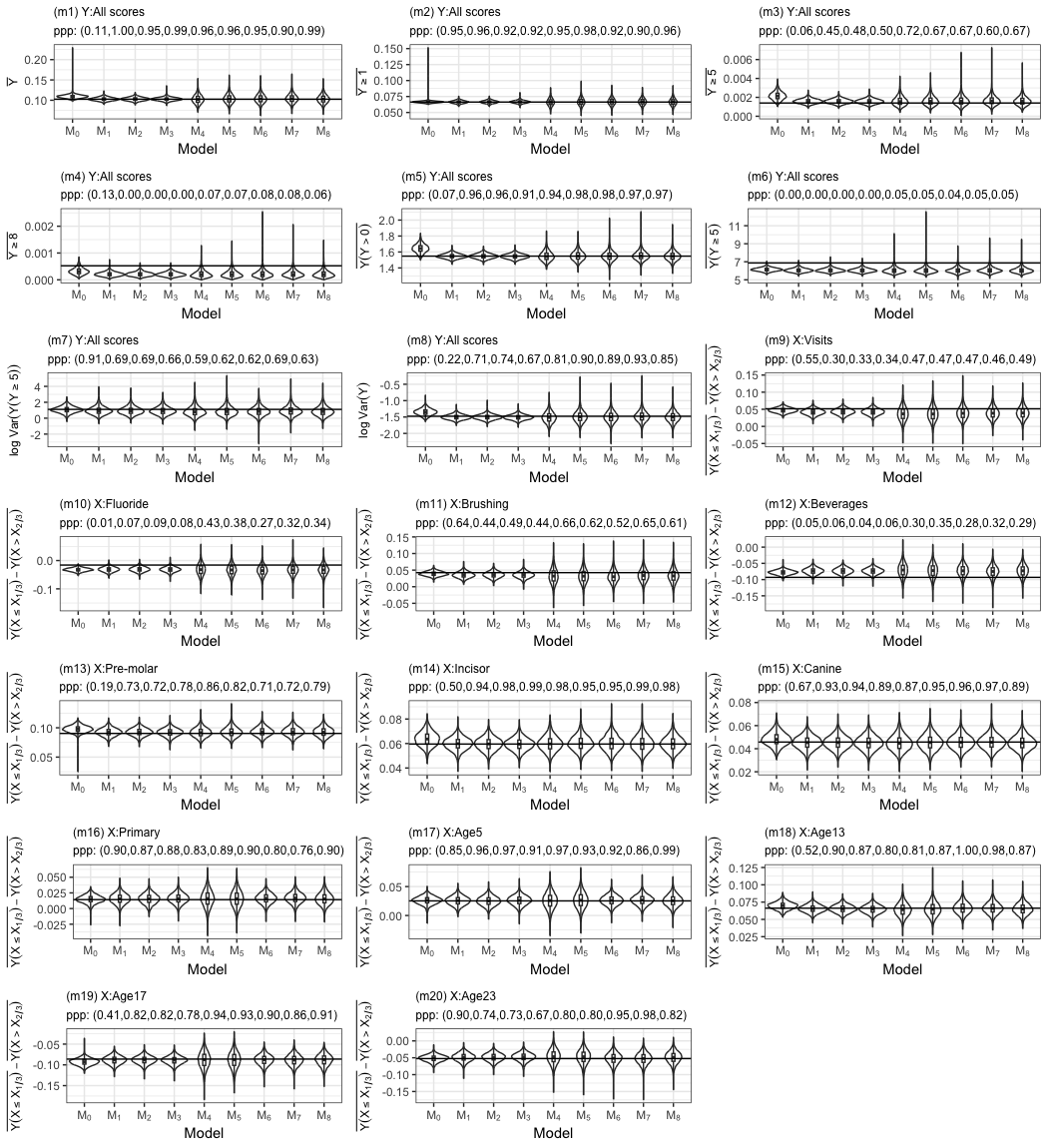}
\caption{Comparisons of the posterior predictive plots for $\bst_M$ from different models based on IFS data. The title of each plot shows the $ppp$ values ($ppp = 0.00$ indicates $ppp < 0.01$) for the corresponding summary statistic obtained from $\model{0}$--$\model{8}$
%  , $\model{2}$, $\model{3}$, $\model{4}$, $\model{5}$ and $\model{6}$ 
consecutively and the horizontal line indicates the observed summary statistics.}
\label{fig:pp_M_IFS}
\end{figure}

We now present additional model comparison results for the IFS data.
First, we present the full set of posterior predictive plots for assessing the marginal model fit in Figure \ref{fig:pp_M_IFS}.
In addition to the overall mean and variance of $Y$, we consider $\overline{Y \geq k}$ denoting proportion of $Y \geq k$ for different choices of $k$, 
we let $\overline{Y(Y \geq k)}$ be the mean of $Y$ over those $Y \geq k$, and 
$\textrm{Var}(Y(Y \geq k))$ indicates the variance of $Y$ restricted to $Y \geq k$. 
In particular, we choose $k=8$ to evaluate the predictive ability of the model in the tail of the $Y$ distribution.
Moreover, for different predictors $X$, we considered statistics of the form $\overline{Y(X \leq X_{1/3})} - \overline{Y(X > X_{2/3})}$, that captures the difference between the mean of $Y$ corresponding to observations where the predictor value is less than or equal to its first tertile $X_{1/3}$ and those with $X$ is above its second tertile $X_{2/3}$.
This will inform if the model is capturing the association between $X$ with $Y$.
%when $X$ lies toward the tail regions of its distribution.
When $X$ is binary, such as, if the predictor is the indicator of premolar tooth, the statistic will be the mean difference between the caries scores among all premolar teeth to the mean score of other tooth types.

As summarized in the main manuscript, we observe from panels (m1)--(m20) that the SAR models $\model{1}$--$\model{8}$ have comparable performance. 
Model $\model{0}$ with its independence misspecification does not account for the correlation present in the data and results in overestimation of the overall mean (m1) and the mean of the non-zero counts (m5).
None of the models could explain the larger counts very well (see Figure \ref{fig:pp_M_IFS}(m4,m6)).
This is potentially due to the fact that the largest count score available in IFS data is 10, with much smaller frequencies for the counts $\geq 8$ compared to that for zeros.
Models $\model{4}$--$\model{8}$, however, performed better in this regard compared to $\model{1}$--$\model{3}$.
They also performed better when characterizing the covariate effects on the count scores.
For example, the effect of total fluoride consumption is underestimated (m10) by $\model{0}$--$\model{3}$, while that of the sugary beverage is overestimated (m12).
It is important to note here that $\model{4}$--$\model{8}$ have wider predictive distributions compared to the rest, as the dependence structure for the former models include the overall connectivity relationships $\bW^{(ct)}$ and $\bW^{(ce)}$, these assume higher amounts of correlation (less information from the data) than $\model{1}$--$\model{3}$ with their sparser connectivity.

\begin{figure}[!tb]
\centering
\includegraphics[width=\textwidth] {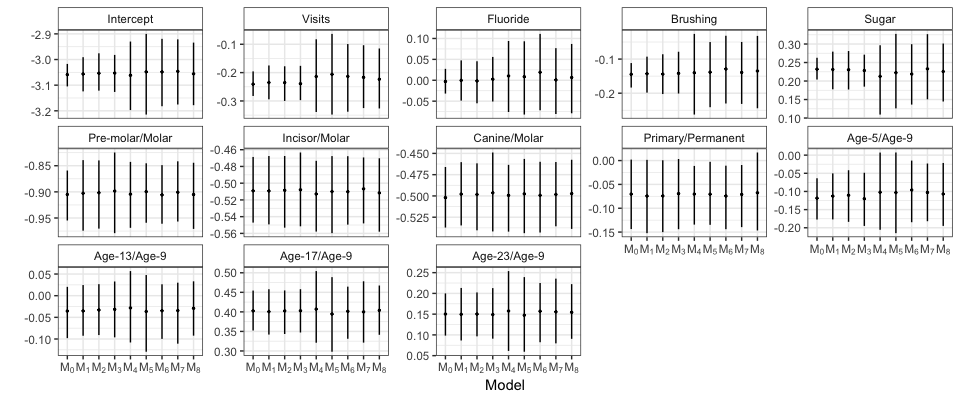}
\caption{Comparisons of the regression-adjusted $\balpha$ estimates from different models. 
The point estimate for each model is denoted by the dot, while the line shows the 95\% confidence interval.
}
\label{fig:alpha_IFS_compare} \quad
\includegraphics[width=\textwidth] {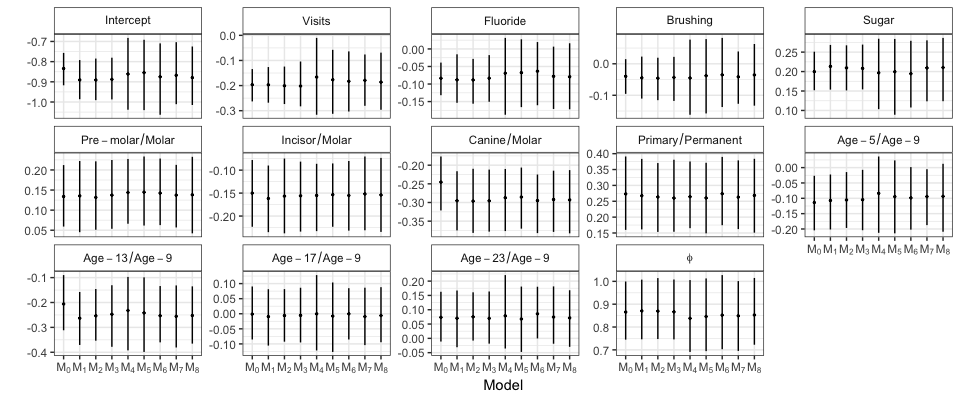}
\caption{Comparisons of the regression-adjusted $\bbeta$ and $\phi$ estimates from different models.
The point estimate for each model is denoted by the dot, while the line shows the 95\% confidence interval.
}
\label{fig:beta_phi_IFS_compare}
\end{figure}

To compare parameter estimation of $\btheta_M$ across the different models, we turn to
% based on credible interval widths.
% \org{[You should first comment on differences in mean estimates.  Notably, there aren't any meaningful differences except for the NB intercept and age 13 effect for M0 vs the others.]}
Figure \ref{fig:alpha_IFS_compare} that displays the posterior mean and 95\% credible interval under each model choice for each of the parameters in the logistic prevalence model.
% Looking into Figure \ref{fig:alpha_IFS_compare} 
We observe that model $\model{0}$ has the narrowest credible intervals throughout.
Recall that model fit for $\model{0}$ was performed using a Gibbs sampler targeting the exact posterior, under the independence misspecfication.
The point estimate for $\balpha$ does not seem to be sensitive to the misspecified dependence structure, while the uncertainty does vary across models.
Models $\model{1}$--$\model{3}$ specify more structured dependence (see Table \ref{tbl:dependence_structures}),
while models $\model{4}$--$\model{8}$ 
%include equal connection based adjacency relations, defined in terms of $\bW^{(ct)}$ and $\bW^{(ce)}$, that 
directly provide an overall connectivity.
% \org{[should this be 1--3 and 4--8???]}
Hence, consistent with the variability in the prediction distributions, 
many parameters in $\balpha$ have wider credible intervals for these less structured SAR models.
% first set of models compared to the second set.
% \FloatBarrier

We observe similar behavior in the NB parameters as well, as shown in Figure 
\ref{fig:beta_phi_IFS_compare}.
%Consistent with the conclusions from the predictive distributions,
The point estimates for $\bbeta$ are more sensitive to the independence misspecification than the $\balpha$ parameter; in particular the estimates for the tooth indicator ``Canine/Molar'' and time indicator ``Age-13/Age-9'' differ for model $\model{0}$ compared to the rest of the models.

Overall, hese results show relatively minimal differences in the models for the marginal parameters consistent with the posterior predictive plots in Figure \ref{fig:pp_M_IFS}. 
However, we saw more meaningful differences in the posterior predictive plots of the selected 24 Spearman correlations (Figure \ref{fig:pp_M_R_IFS} of main manuscript).  
To complement these comparisons and further investigate the much larger collection of correlations that our model specifies, we consider the univariate $ppp$ values for a larger number of correlations (instead of the full predictive distribution for a small number of correlations).
% To provide an overall assessment for the the adequacy of the dependence structures for different models, we perform posterior predictive tests across all the available correlation coefficients in $\bR$. 
For each model, we compute the corresponding $ppp$ values of the Spearman correlation of all pairs $(l_j,t_j)$ and $(l_{j'},t_{j'})$ and plot the histogram across these $ppp$s.
Note that we include only those $ppp$ values which correspond to correlation coefficients obtained from at least 100 pairs of caries scores. 
While looking into the posterior predictive distributions for a few representative elements from $\bR$ as in Figure \ref{fig:pp_M_R_IFS} helps identify  those specific features of the dependence model that are being mis-modeled, considering the histogram consisting of all  $ppp$ values  provides a more holistic view into  the  model fit.

\begin{figure}[!tb]
\centering
\includegraphics[width=0.9\textwidth]{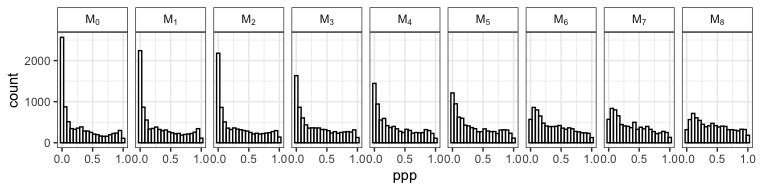}
\caption{Histogram of all available $ppp$-values across all model choices.
}
\label{fig:ppp_histogram}
\end{figure}
As shown in Figure \ref{fig:ppp_histogram},
the histogram for $\model{0}$ has a very high peak near zero, implying that the posterior predictive distributions of a large number of Spearman correlations  poorly represent the corresponding values observed in the IFS data.
This is also observed in most of the alternative model choices 
based on the anatomic SAR models $\model{1}$--$\model{3}$ and to a lesser extent the equal connection within time based $\model{4}$ and $\model{5}$. 
The models with $\bW^{(ce)}$ included have the best performance and mostly avoid the peak near zero.
% except for the equicorrelation based models $\model{6}$ - $\model{8}$, with $\model{8}$ as the best model.
% showing the least number of extreme mismatches (leading to $ppp$ value very close to zero).
% The distribution for $\model{3}$ have the lowest peak near zero.
% We acknowledge here that, each $ppp$ value may not have a uniform distribution \citep{Hjort2006}, and these $ppp$ values are computed based on the Spearman correlations corresponding to each element in $\bR$ which are not independent.
% We would intuitively argue here that looking into the overall histogram of the $ppp$ values, in addition to studying the posterior predictive distributions for a few representative elements from $\bR$, will portray a more complete picture of the model fit.
% The framework for the dependence model being same across all the models considered here, the higher peak at zero in one model compared to other will reflect an inferior fit for the model.
% Judging by this intuition along with other observations mentioned before, we choose $\model{3}$ to be most preferable among all the alternatives.
As noted in Section \ref{sec:IFS_model_comparison}, we take this as evidence that $\model{8}$ is the best performing model among our choices.

We note that, as the distribution of $ppp$s for $\model{8}$ displays some skewness toward zero, this potentially suggests some deficiencies in our dependence modeling.  
This is not fully surprising since we only use between one and four parameters in $\btheta_D$ 
(three parameters for $\model{8}$)
to describe this set of 8128 Spearman correlations. 
Further, due to the high level of zero-inflation and the discreteness of the non-zero caries scores, there is relatively limited information about the correlation between most pairs of teeth, despite the overall large number of observations.
%leading to remaining uncertainty between similar correlation models.  
Consequently, there remains meaningful uncertainty in the correlation estimates of a given model and in the comparisons between similar correlation models.
We will further consider the utility of this strategy of comparing $ppp$ histograms in the  simulation analysis in Section \ref{appn:sim_model_misspecification}.

\begin{figure}[!tb]
\centering
\includegraphics[width=\textwidth] {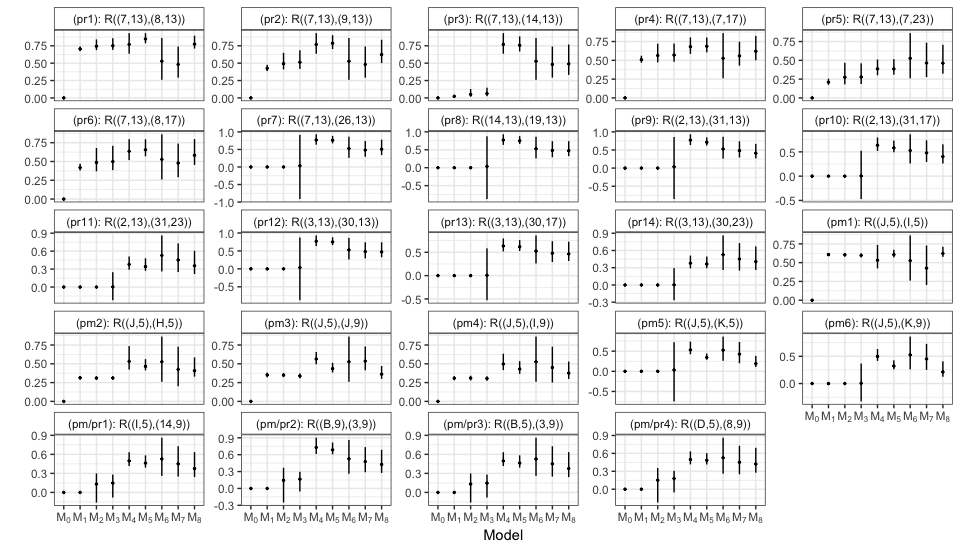}
\caption{Comparisons of the estimates of $\bR$ from different models based on a specific set of elements. The panel titles include the panel name along with the tooth-time pair. The point estimate for each model is denoted by the dot, while the line shows the 95\% confidence interval.}
\label{fig:compare_R_IFS}
\end{figure}

A comparison of the estimated correlation coefficients from $\bR$ 
for the set of correlations from Table \ref{tbl:theta_R}
is presented in Figure \ref{fig:compare_R_IFS}.
Due to the independence misspecification,  all the correlation elements in $\model{0}$ are at zero.
Models $\model{1}$ and $\model{2}$
%, having incomplete description for the marginal correlation structure, cannot estimate the correlation coefficients 
have no connections
between vertically-adjacent tooth-time pairs (panels: pr7-14), and the estimates and intervals are estimated to be zero. 
Model $\model{1}$ additionally fails to estimate correlations between primary-permanent teeth in panels pm/pr1-4.
We observe very wide credible intervals for $\model{3}$ for some of correlation coefficients corresponding to vertically-adjacent tooth pairs.
Overall, model $\model{6}$, that assumes equicorrelation structure, seems to have the widest confidence intervals (except for the above $\model{3}$ vertical adjacency issues).
This seems reasonable as $\btheta_D$ consists of only one parameter specifying the complete correlation matrix $\bR$, and, therefore, will have lower precision in estimation.

\begin{table}[!tb]
\centering
\footnotesize
\begin{tabular}{rccl}
\hline
Tooth-location pair & Mean  & 95\% CI        & Interpretations                                  \\ \hline
((7,13),(8,13))     & 0.774                    & (0.710, 0.896)              & Same time, horizontally adjacent teeth                        \\
((7,13),(9,13))     & 0.623                    & (0.500, 0.837)              & Same time, horizontally non-adjacent teeth                         \\
((7,13),(14,13))    & 0.489                    & (0.331, 0.767)              & Same time, horizontally even farther teeth                    \\
((7,13),(7,17))     & 0.620                    & (0.500, 0.826)              & Adjacent time, same tooth                        \\
((7,13),(7,23))     & 0.461                    & (0.323, 0.706)              & Non-adjacent time, same tooth                         \\
((7,13),(8,17))     & 0.582                    & (0.454, 0.798)              & Adjacent time, horizontally adjacent teeth                    \\
((7,13),(26,13))    & 0.505                    & (0.340, 0.779)              & Adjacent time, vertically adjacent incisor teeth            \\
((14,13),(19,13))   & 0.473                    & (0.323, 0.743)              & Same time, vertically adjacent molar teeth                  \\
((2,13),(31,13))    & 0.409                    & (0.266, 0.669)              & Same time, vertically adjacent molar teeth                  \\
((2,13),(31,17))    & 0.403                    & (0.260, 0.655)              & Adjacent time, vertically adjacent molar teeth                    \\
((2,13),(31,23))    & 0.356                    & (0.218, 0.609)              & Non-adjacent time, vertically adjacent molar teeth               \\
((3,13),(30,13))    & 0.473                    & (0.323, 0.743)              & Same time, vertically adjacent molar teeth                  \\
((3,13),(30,17))    & 0.465                    & (0.313, 0.729)              & Adjacent time, vertically adjacent molar teeth                    \\
((3,13),(30,23))    & 0.407                    & (0.259, 0.670)              & Non-adjacent time, vertically adjacent molar teeth                     \\
((J,5),(I,5))       & 0.622                    & (0.577, 0.711)              & Same time, horizontally adjacent primary teeth                \\
((J,5),(`H,5))       & 0.408                    & (0.327, 0.585)              & Same time, horizontally non-adjacent primary teeth                 \\
((J,5),(J,9))       & 0.362                    & (0.300, 0.471)              & Adjacent time, same primary tooth                \\
((J,5),(I,9))       & 0.375                    & (0.297, 0.530)              & Adjacent time, adjacent primary teeth            \\
((J,5),(K,5))       & 0.191                    & (0.108, 0.383)              & Same time, vertically adjacent primary teeth     \\
((J,5),(K,9))       & 0.212                    & (0.125, 0.410)              & Adjacent time, vertically adjacent primary teeth \\
((I,5),(14,9))      & 0.376                    & (0.241, 0.636)              & Adjacent time, primary/permanent molar teeth     \\
((B,9),(3,9))       & 0.426                    & (0.276, 0.687)              & Same time, primary/permanent molar teeth         \\
((B,5),(3,9))       & 0.376                    & (0.241, 0.636)              & Adjacent time, primary/permanent molar teeth     \\
((D,5),(8,9))       & 0.421                    & (0.276, 0.694)              & Adjacent time, primary/permanent incisor teeth   \\ \hline
\end{tabular}
\caption{Approximate posterior estimates of specific representative elements from $\bR(\btheta_D)$ for model $\model{8}$. Tooth labels are found in Figure \ref{fig:appn_dentition}.}
\label{tbl:theta_R}
\end{table}

\subsection{Assessing Marginal Model Specification} \label{appn:IFS_compare_marginal_models}
We now describe two alternative marginal model specifications and consider similar validation strategies.
% that we considered to validate our model choice.
The Poisson-hurdle model is often a standard choice for modeling count data when one does not expect the non-zero counts to be  over-dispersed.
The CDF for the marginal distribution of $Y_{ij}$ following Poisson hurdle model is given as,
\begin{equation} \label{eq:cdf_pois_hurdle}
\begin{array}{rcl}
F_{ij}(y)=P(Y_{ij} \leq y) & = & 
\frac{1}{1 + \exp(\bsx_{ij}' \balpha)} + \\& & \left\{ \frac{\exp(\boldsymbol{x}_{ij}'\balpha)}{1 + \exp(\boldsymbol{x}_{ij}'\balpha)} \sum_{u=1}^{y} \mathrm{Pois} \left(u-1 \mid \mu = \exp(\boldsymbol{x}_{ij}' \bbeta) \right) \right\} \mathbb{I}(y > 0),
\end{array}
\end{equation}
The marginal parameters will consist of $\balpha$ and $\bbeta$.
We consider same Gaussian copula based dependence structure with the latent Gaussian vectors following the SAR model specifications described in Table \ref{tbl:dependence_structures}. 
We first obtain the initial estimate for $\btheta$ under independence misspecification following an approach similar to the one described in Section \ref{appn:initial_estimation}, but derived for Poisson hurdle model.
An ABC-MCMC algorithm, modified accordingly, is then employed starting with the initial estimate of $\btheta$ and a set of posterior samples are obtained, which are then regression-adjusted to yield the final set of samples.

Another alternative choice for the marginal model is to simply consider the Negative Binomial model without any modeling of the zero inflation.
% Due to the zero-inflation, the IFS data may appear to be over-dispersed.
That is, unlike our hurdle models, the excess zeros are not directly accounted for by the model and must be explained as part of the NB distribution.
% Therefore having a Negative Binomial distribution, that allows for over-dispersion in the count data, will make sense.
% The CDF for the marginal distribution of $Y_{ij}$ will simply be the CDF for the Negative Binomial distribution, with the marginal parameters consisting of $\bbeta$ amd $\phi$ only.
The dependence model specification remains the same with the corresponding NB CDF. 
All other computational details are as before.
% as discussed previously for the Poisson hurdle model.
% An initial estimation of $\btheta$ is first obtained followed by the employment of a suitably designed ABC-MCMC algorithm along with regression adjustment that yields the final set of posterior samples.

For both of these models, we choose similar approach to parameter estimation.
We obtain the initial estimates for the marginal parameters and their covariance by employing a Gibbs sampler under independence misspecification, and the initial estimates for the dependence parameters are obtained using an empirical approach as described in Section \ref{appn:initial_estimation}.
The summary statistics for the models come from the MLEs of their respective independence models as proposed in Section \ref{sec:ABC-summary-statistics}, and the kernel scaling matrix is also obtained similarly.
For each model, we run three ABC-MCMC chains of length 185,000 each, and burnin the first 5000 samples.
The remaining samples are regression-adjusted based on unique samples and are thinned to yield 9000 adjusted samples to be used for inference.
All models are fit using the SAR structure $\model{8}$.

% \org{[For these methods, you should clarify if you update the sM statistics.  That is, does sM come from the MLE of the corresponding independence model?]}

%\anish{Will add more text once I get the results from the other marginal model choices and add Figure \ref{fig:comparison_marginal_models}}
%\org{you will also want some kind of predictive plot and/or ppp plot based on these}

\begin{figure}[!tb]
\centering
\includegraphics[width=\textwidth]{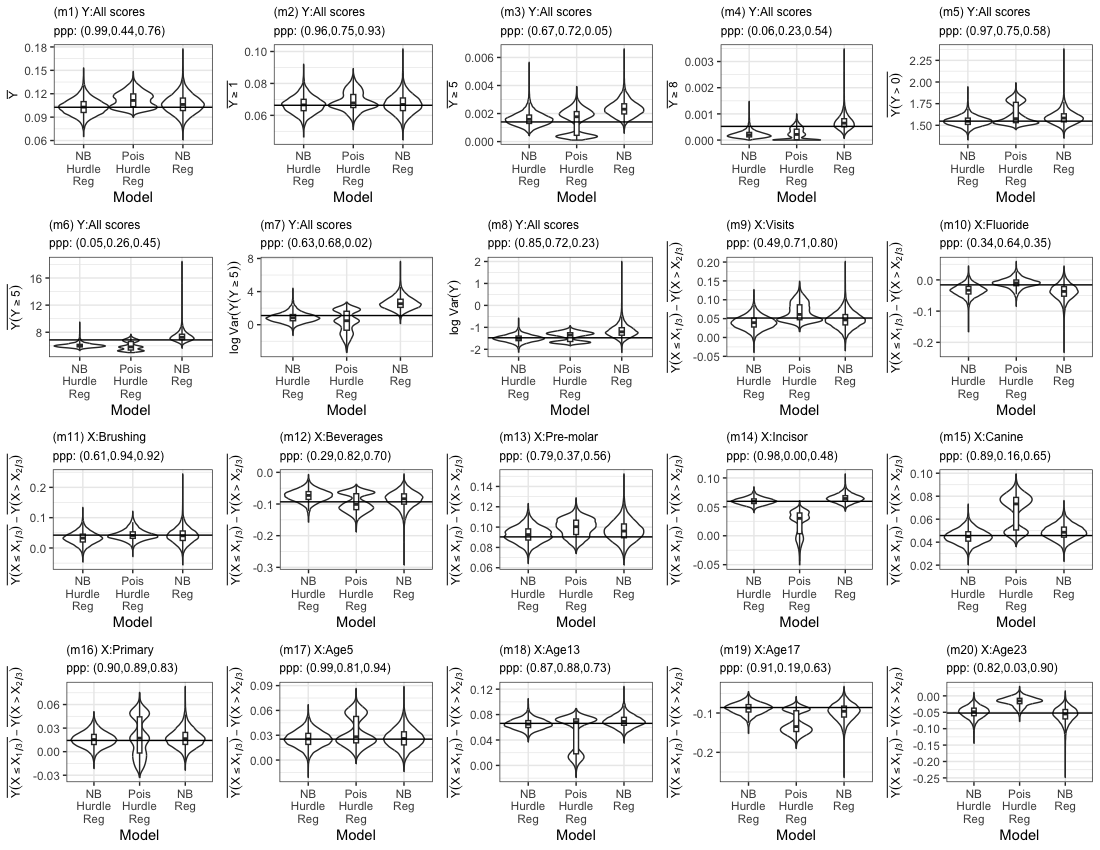}
% \vspace{1in}
\caption{Comparisons of the posterior predictive plots for marginal parameters from models with different marginal model specifications. The title of each plot shows the corresponding $ppp$ values ($ppp = 0.00$ indicates $ppp < 0.01$). $X_{1/3}$ and $X_{2/3}$ represent the first and second tertiles of predictor $X$, respectively.}.
\label{fig:comparison_marginal_models}
\end{figure}

Figure \ref{fig:comparison_marginal_models} compares the posterior predictive distributions for the NB-Hurdle model against the two alternative choices to assess their marginal model fit using the same set of summary statistics as in Figure \ref{fig:pp_M_IFS}.
We observe that the NB model, which has to explain zero-inflation and over-dispersion with the same parameters overestimates the proportion of larger counts (panel: m3). 
The variance is also overestimated (panels: m7-m8).
The Poisson-hurdle model shows quite competitve performance compared to NB-hurdle model.
It is worth noting, however, that the effects of incisor teeth (panel: m14) and age 17 (panel: m19) are underestimated by the Poisson-hurdle model, while the age 23 effect (panel: 20) is overestimated.
Overall NB-hurdle model turns out to be the best among all the marginal model choices.

\FloatBarrier

\newpage

\section{Further Details on Simulation Analysis} 
\label{appn:simulation}

Here we present further details and analyses regarding our simulation study.
First, we describe the true parameter values we chose to generate the simulated data.
A brief account on the computation of the overall empirical coverage score is also included, followed by a discussion of the rank aggregation approach.
%used to rank different sampling strategies with a range of bandwidth choices.
In Section \ref{appn:sim_results}, we investigate estimation of the parameters individually for different ABC-MCMC sampling strategies, including the classical rejection and importance sampling. 
The following subsection provides an alternative regression adjustment method and compares it against our proposed approach. 
Sensitivity of the model fit to the dependence structure specification is studied in Section \ref{appn:sim_model_misspecification}, where we evaluate estimation of the correct model against other models with misspecified dependence structures.

\subsection{Additional Notes on Simulation Setup} 
\label{appn:simulation_setup}

We present the true data-generating 
parameter values used for the simulation setup in Table \ref{tbl:sim_setup}. 
To replicate the relevant features of the IFS data in simulation setup, the true values of $\btheta_M$ and $\btheta_D$ are chosen to be close to the their respective point estimates obtained from fitting the IFS data with model $\model{4}$.

\begin{table}[tb]
\centering
\footnotesize
\begin{tabular}{lccc}
\hline
Marginal Model Predictor ($\btheta_M$)                              &                      & Presence($\balpha$)       & Severity($\bbeta$)        \\ \hline
Intercept                              &                      & -3.059               & -0.831               \\
Dental Appointments                    &                      & -0.241               & -0.196               \\
Total fluoride ingested (mgF)          &                      & -0.003               & -0.084               \\
Frequency of brushing                  &                      & -0.145               & -0.040               \\
Amount of sugar beverage (oz)          &                      & 0.232                & 0.200                \\ \hline
Tooth Type                             &                      &                      &                      \\
\hspace{1em}Molar     &                      & Ref                  & --                   \\
\hspace{1em}Pre-Molar &                      & -0.905               & 0.134                \\
\hspace{1em}Incisor   &                      & -0.509               & -0.150               \\
\hspace{1em}Canine    &                      & -0.502               & -0.245               \\
Primary                                &                      & -0.070               & 0.273                \\ \hline
Observation Time                       &                      &                      &                      \\
\hspace{1em} Age 5    &                      & -0.119               & -0.114               \\
\hspace{1em} Age 9    &                      & Ref                  & --                   \\
\hspace{1em} Age 13   &                      & -0.035               & -0.206               \\
\hspace{1em} Age 17   &                      & 0.403                & -0.001               \\
\hspace{1em} Age 23   &                      & 0.150                & 0.073                \\ \hline
NB size $\phi$                         &                      &                      & 0.865 \\ 
\hline \\
\hline \hline
SAR Model Parameters ($\btheta_D)$ & & Values &  \\ \hline
$\rho_{t}$ & & 0.12 &  \\
$\rho_{ct}$ & & 0.012 & \\
\hline
\end{tabular}
\caption{True values for $\btheta_M$ and $\btheta_D$ for the simulation data generation.}
\label{tbl:sim_setup}
\end{table}

\begin{figure}[!tb]
\centering
\includegraphics[width=0.8\textwidth]{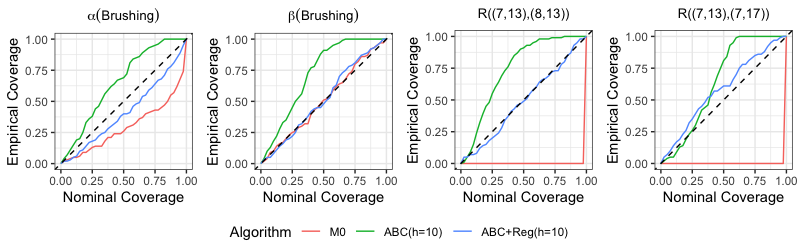}
\caption{Comparisons of the empirical coverage rates across different sampling strategies.}
\label{fig:OECS}
\end{figure}

Before looking into the parameter estimation performance across different sampling strategies in the next section, we provide further clarity on the computation of the overall empirical coverage score (OECS).
To that end, we consider two components from $\btheta_M$ and two from $\btheta_\bR$ and show their empirical coverage curves for different model-fitting algorithms in Figure \ref{fig:OECS}.
For simplicity, we only show the results for the best bandwidth, $h=10$.
% We consider bandwidth $h = 10$ and the regression adjustment is based on unique samples.
Observe that the coefficient for
% the effects of
brushing frequency in the presence model is under-covered by $\model{0}$, as its empirical coverage curve falls below the diagonal reference line.  
In contrast, ABC-MCMC over-covers the parameter, as its empirical coverage is consistently above the nominal rate.
% while ABC-MCMC algorithm over-covers it.
The regression adjustment, however, improves the coverage, with the true value of the parameter being covered at nearly the nominal coverage level across all levels $\zeta$ from 0 to 1.
The OEC score, defined as the area under the empirical coverage curve, summarizes these observations and gives the best score (closest to 0.5) to the regression-adjusted ABC samples.
In the case of the effect of brushing frequency in the severity model, the empirical coverage rates for both $\model{0}$ 
and the regression-adjusted sample
match the target coverage well.
% , and regression-adjusted ABC-MCMC samples appear to be very similar as well.
However, the unadjusted ABC samples 
over-cover the true coefficient.
% provide over-coverage.
Regarding the correlation coefficients, $\model{0}$ assumes all the correlations to be zero, and coverage rates are not estimated.
For the purpose of comparison, we have shown the corresponding OEC scores indicating zero coverage.
% performs worst as its independence misspecification. 
% \org{[want to be careful with these.  we should avoid treating these are estimates when using M0.]}
While ABC-MCMC samples perform better by learning the correlation structure from the data, they are plagued by some over-coverage which gets corrected by regression adjustment.

% \anish{Provide more details on the rank aggregation technique}
% \org{[Move further rank aggregiation details here.]}
% \subsubsection{Rank aggregation}

\begin{figure}[!tb]
\centering
\includegraphics[width=0.7\linewidth]{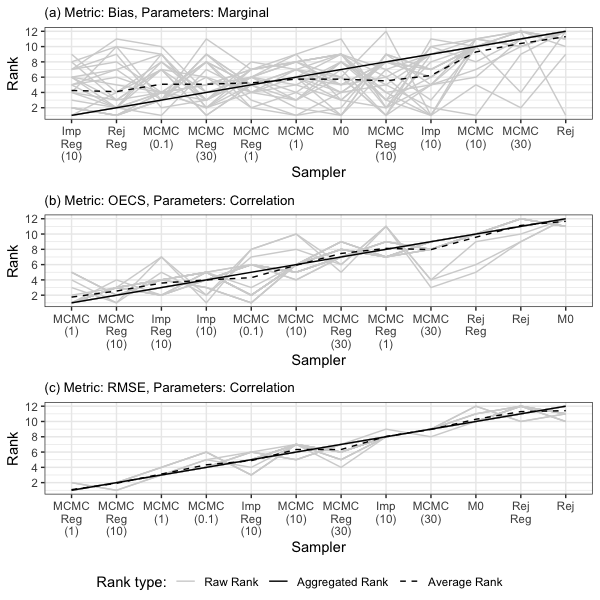}
\caption{Raw, Aggregated and Average ranks of the samplers based on three different sets of ranking criteria described in the panel titles.}
\label{fig:rank_aggregation}
\end{figure}

We now briefly provide some background on the rank aggregation strategy used to provide an overall ordering of the samplers.
%based on different performance metrics.  
Let $\mfR_{j}$ represent the ranked list of the samplers (the first sampler in $\mfR_{j}$ being the ``best’’) for 
some collection $j=1,\ldots,J$ of ranked lists.  For instance, in Table \ref{tbl:rank_aggregation}, the first collection considered is the (absolute) bias for the $j=27$ marginal parameters in $\btheta_M.$
% $\mfp$-th parameter in $\btheta$ (27 marginal, 24 correlation) and $\mfm$-th performance metric (3 metrics: absolute bias, RMSE, OECS).
% Let $\mfR_\mfc$ denote the ranked list of the samplers (the first sampler in $\mfR_\mfc$ being the ``best’’) with respect to $\mfc$-th criteria, where $\mfc=1, \ldots, \mfC$.
% with $\mfC$ being the total number of criteria.
% \org{[What is being indexed by $\mfc$? What is $\mfC$? ]}
While these individual lists/rankings are informative about a single parameter, ``aggregating'' them into a single consensus ranked list $\mfR$   provides insight into the samplers’ performances overall; in this case, we use an aggregrated list to rank the methods in terms of the bias of the full set of $\btheta_M$ point estimators.
% For example, we may be interested to know aggregated sampler ranks in terms of marginal model fit when bias is the performance metric.
To achieve this, we aggregate all $\mfR_{j}$, 
% where $\mfp$ varies over 27 marginal parameters and metric $\mfm$ is bias, 
using a Cross-Entropy Monte Carlo stochastic search proposed by \cite{Pihur2007}. 
The idea is to find an optimal list $\mfR$ that is closest to all the $\mfR_{j}$, by minimizing the objective function $O(\mfR) = \sum_{j=1}^J d(\mfR,\mfR_{j})$, where the sum is over the collection of lists considered.  
Here,  $d(\cdot,\cdot)$ is chosen to be the Spearman footrule distance given by $d(\mfR,\mfR_{j}) = \sum_{t} |\mfr(t) - \mfr_{j}(t)|$, where $t$ represents the sampling method and $\mfr(t)$ and $\mfr_{j}(t)$ denote the ranks of method $t$ in $\mfR$ and $\mfR_{j}$, respectively.
That is, the aggregated list $\mfR$ minimizes the absolute differences of the rank of method $t$ in the consensus list $\mfR$ and in the list $\mfR_{j}$ across all methods $t$ and all lists $j$.
Figure \ref{fig:rank_aggregation}(a) illustrates the rank aggregation visually for the bias of the marginal parameters.
The raw ranks of the samplers are clearly highly variable across parameters, without any apparent overall ordering, since most methods had bias of approximately zero for all parameters. 
This lack of distinction among the performance of the samplers was also noted previously in Figure \ref{fig:model_fit_M_compare_sim}(m1) and in the corresponding discussion.
The aggregated rank, marked with the solid black line, deviates from the dotted average rank line, suggesting that the average ranks do not differ much among the first 9 methods and that the average performance of these methods is similar among this set of lists.  
% for most of the samplers, we do get some ordering with the rank aggregation approach. 

% In our context, let $\mfR_{\mfp \mfm}$ represent the ranks of the samplers for $\mfp$-th parameter in $\btheta$ (27 marginal, 24 correlation) and $\mfm$-th performance metric (3 metrics).
% We obtain the sampler ranks for the marginal (or dependence) model fit with regard to metric $\mfm$ by aggregating $\mfR_{\mfp \mfm}$, where $\mfp$ varies over 27 marginal (or 24 dependence) parameters, and a metric $\mfm$ specific combined rank is obtained by varying $\mfp$ over all the marginal as well as correlation parameters.
% For illustration, panel (a) in Figure \ref{fig:rank_aggregation} considers rank aggregation across marginal parameters in $\btheta_M$ based on their bias. The raw ranks of the samplers appear to be highly variable across parameters without any apparent ordering. This lack of distinction across samplers was also noted in panel (m1) of Figure \ref{fig:model_fit_M_compare_sim} and in the corresponding discussion.
% The aggregated ranks marked with the solid black line deviates from the dotted line showing the average ranks of the samplers implying that while the average ranks do not differ much for most of the samplers, we do get some ordering with the rank aggregation approach.  
In contrast, there is more consistency in the individual rankings  in Figure \ref{fig:rank_aggregation}(b), where we show sampler ranks for how well they cover the dependence model parameters.
In this setting, the average rank is approximately the same as the aggregrate rank.
Here, we have aggregated  the collection $\mfR_{1},\ldots,\mfR_J$ representing the ranked list of OECS (absolute difference between OECS and $1/2$) for each of  the $J=24$ correlation parameters in $\btheta_\bR$.
% and the metric $\mfm$ is OECS.
The corresponding boxplots in Figure \ref{fig:model_fit_R_compare_sim}(r3) have less overlap between methods, leading to a  more clear ordering.  Hence,  the average ranks match more closely with the aggregated ranks.
Similarly, Figure \ref{fig:rank_aggregation}(c) shows the raw, average and aggregated ranks based on RMSE for the correlation parameters.
As with Figure \ref{fig:model_fit_R_compare_sim}(b), we have similar ordering across all constituent lists, the average rank, and the aggregated rank.
% ) the raw ranks explicitly show the ordering with the average ranks matches almost exactly with the aggregated ranks.

% are first ranked by absolute bias, yielding 27 ranked lists of sampler, each  corresponding to one parameter in $\btheta_{M}$. Rank aggregation is carried out on these multiple lists of ranks to obtain an overall ``super''-list. 
% \org{[CONNECT TO NOTATION FROM ABOVE]}
% This would best represent the consensus of the individual ranked lists (each corresponding to a parameter in $\btheta_M$) for the bias metric. Figure \ref{fig:RankAggreg_1a_bias} provides a visual representation of the convergence path of the algorithm; as can be seen in the figure, the optimal ranking is obtained in 27 iterations.

In Table \ref{tbl:rank_aggregation} of the main manuscript, we have reported various aggregrated ranks by choosing different collections of $\mfR_1,\ldots,\mfR_J$.  
The  rankings are determined by ranking the performance of each parameter under each estimation criterion: the absolute bias for each parameter, RMSE for each parameter and the absolute difference between OECS and 0.5 for each parameter.
%, and the average width of the 80\% credible interval for each parameters.  
The sets of rankings for each criterion are considered using only those parameters in the marginal model $\btheta_M$, using only the subset of correlation parameters $\btheta_\bR$, and combining these two collections of rankings.  
In addition to obtaining a ranking under each criterion separately, we obtain an overall ranking by combining the per-parameter lists across the bias, RMSE and OECS criteria for $\btheta_M$, for $\btheta_\bR$, and using the combined set of parameters.
% Additionally, overall sampler ranks for the marginal (or dependence) model fit across all the performance metrics is obtained by aggregating $\mfR_{\mfp \mfm}$ over all $\mfm$ and over all marginal (or correlation) parameters $\mfp$. 
% An overall combined rank across parameters and performance metrics is also be obtained similarly.
% A rank is also obtained aggregating all the ranks across all metrics where each pair of parameter and performance metric is considered as a criterion.
All these ranks are presented in Table \ref{tbl:rank_aggregation}.

\subsection{Further Simulation Analysis} 
\label{appn:sim_results}

Here we compare the point estimates of the marginal and correlation parameters for the different sampling strategies.
In Figures \ref{fig:model_fit_M_compare_sim} and \ref{fig:model_fit_R_compare_sim}, we included violin plots showing the estimation accuracy summarized across the set of parameters in $\btheta_M$ and $\btheta_\bR$, respectively.  Here, we report the accuracy for each parameter individually in 
% We now investigate estimation performance of the individual parameters in terms of bias, RMSE, overall empirical coverage score (OECS) obtained using different ABC based sampling strategies (
Tables \ref{tbl:sim_compare_M} and \ref{tbl:sim_compare_R}.
% The results for ABC-MCMC and ABC importance sampling are based on kernel bandwidth $h=10$. 
For brevity, we exclude results for all bandwidths except the optimal $h=10$, and we exclude the non-regression-adjusted rejection and importance sampling methods.
% The choice of the bandwidth has been discussed in Section \ref{sec:simulation}.

As discussed already in Section \ref{sec:simulation}, we observe that estimation under the (unadjusted) ABC-MCMC sampler is not as good as $\model{0}$, with the coverage for the severity model parameters $\bbeta$ being particularly worse compared to $\model{0}$. 
However, regression adjustment provides much better results across all the metrics.
Overall, across all the sampling strategies, estimation of $\bbeta$ is less accurate compared to $\balpha$ of the presence model, which is reasonable considering there is less information available for model fit due to zero inflation.
The Negative Binomial size parameter has the largest RMSE across the parameters.
The ABC rejection sampler with regression adjustment also performs fairly well overall for the marginal parameters.
% We observe in table \ref{tbl:sim_compare_M} that the bias for $\model{0}$ across all the $\btheta_M$ parameters are less compared to those for the ABC-MCMC sampler.
% However, the bias improves when we perform the regression adjustment.
% The RMSE also is worse in the case of ABC-MCMC sampler compared to $\model{0}$, however regression-adjustment makes the RMSE more comparable with $\model{0}$.
% Regarding coverage, the OEC scores are less than 0.4 for some of the $\balpha$ and $\bbeta$ parameters while the target score is 0.5, indicating under-coverage.
% The ABC-MCMC, having wider credible intervals, over-covers all of the $\balpha$ parameters, with OEC scores more than 0.6, while $\bbeta$ parameters have mixed coverage.
% The regression-adjustment brings all the OEC scores much closer to 0.5, improving overall coverage.

While investigating Table \ref{tbl:sim_compare_R} with the estimation performance for the selected elements of $\bR$, we 
also consider estimation accuracy for the two individual SAR parameters.
We observe that ABC-MCMC samples typically overcover  both the SAR model parameters $\btheta_D$, which is  corrected in regression adjustment, achieving the target coverage and increased accuracy in terms of bias and RMSE. 
The estimation of the individual $\bR$ correlations is also improved from ABC-MCMC by the regression adjustment, while unadjusted samples have worse bias, worse RMSE and wider credible intervals providing overcoverage.
The ABC rejection sampler does not work very well for $\btheta_D$ even after regression adjustment, with the estimation of $\rho_{ct}$ being particularly worse.
This may be expected as the rejection sampler draws parameters from a non-informative prior with the true value of $\rho_{ct}$ lying near the boundary, resulting in a small number of accepted samples to work with.

% \org{[Probably just don't include M0 at all here.]}

% \org{[We should include some discussion of the accuracy of the $\btheta_D$ parameters somewhere.  You could add them here to the end of the table.  Even if we focus on $\btheta_\bR$ mostly, we need to show that we get accurate estimation of these parameters when we have the correct model.]}

\begin{landscape}
\begin{table}[!tb]
\centering
\scriptsize
\begin{tabular}{llllllllllllllll}
	\hline
	& \multicolumn{3}{c}{$\model{0}$}                                                   & \multicolumn{3}{c}{ABC-MCMC}                                                      & \multicolumn{3}{c}{ABC-MCMC + Regression}                                         & \multicolumn{3}{c}{ABC-Rej + Regression}                                          & \multicolumn{3}{c}{ABC-Imp + Regression}                                          \\
	Parameters          & \multicolumn{1}{c}{Bias} & \multicolumn{1}{c}{RMSE} & \multicolumn{1}{c}{$\OECS$} & \multicolumn{1}{c}{Bias} & \multicolumn{1}{c}{RMSE} & \multicolumn{1}{c}{$\OECS$} & \multicolumn{1}{c}{Bias} & \multicolumn{1}{c}{RMSE} & \multicolumn{1}{c}{$\OECS$} & \multicolumn{1}{c}{Bias} & \multicolumn{1}{c}{RMSE} & \multicolumn{1}{c}{$\OECS$} & \multicolumn{1}{c}{Bias} & \multicolumn{1}{c}{RMSE} & \multicolumn{1}{c}{$\OECS$} \\ \hline
	$\btheta_M$         & \multicolumn{1}{c}{}     & \multicolumn{1}{c}{}     & \multicolumn{1}{c}{}        & \multicolumn{1}{c}{}     & \multicolumn{1}{c}{}     & \multicolumn{1}{c}{}        & \multicolumn{1}{c}{}     & \multicolumn{1}{c}{}     & \multicolumn{1}{c}{}        & \multicolumn{1}{c}{}     & \multicolumn{1}{c}{}     & \multicolumn{1}{c}{}        & \multicolumn{1}{c}{}     & \multicolumn{1}{c}{}     & \multicolumn{1}{c}{}        \\
	$\alpha$(Intercept) & -0.006                   & 0.037                    & 0.329                       & 0.006                    & 0.079                    & 0.677                       & -0.002                   & 0.046                    & 0.467                       & -0.007                   & 0.043                    & 0.392                       & -0.006                   & 0.002                    & 0.332                       \\
	$\alpha$(Visits)    & 0.000                    & 0.038                    & 0.284                       & 0.011                    & 0.081                    & 0.628                       & 0.006                    & 0.047                    & 0.416                       & -0.002                   & 0.037                    & 0.240                       & 0.000                    & 0.002                    & 0.339                       \\
	$\alpha$(Fluoride)  & 0.002                    & 0.031                    & 0.282                       & 0.009                    & 0.059                    & 0.661                       & 0.005                    & 0.038                    & 0.434                       & 0.002                    & 0.031                    & 0.250                       & 0.002                    & 0.001                    & 0.343                       \\
	$\alpha$(Brushing)  & 0.003                    & 0.036                    & 0.273                       & 0.009                    & 0.071                    & 0.627                       & 0.006                    & 0.043                    & 0.390                       & 0.003                    & 0.035                    & 0.225                       & 0.004                    & 0.002                    & 0.293                       \\
	$\alpha$(Beverages) & 0.001                    & 0.028                    & 0.321                       & 0.003                    & 0.064                    & 0.675                       & -0.001                   & 0.036                    & 0.468                       & 0.001                    & 0.029                    & 0.312                       & 0.003                    & 0.001                    & 0.346                       \\
	$\alpha$(Pre-molar) & -0.004                   & 0.033                    & 0.462                       & 0.013                    & 0.078                    & 0.729                       & -0.003                   & 0.033                    & 0.465                       & -0.004                   & 0.032                    & 0.435                       & -0.003                   & 0.001                    & 0.346                       \\
	$\alpha$(Incisor)   & 0.002                    & 0.027                    & 0.527                       & 0.016                    & 0.067                    & 0.733                       & 0.004                    & 0.026                    & 0.529                       & 0.001                    & 0.027                    & 0.522                       & 0.000                    & 0.001                    & 0.351                       \\
	$\alpha$(Canine)    & -0.002                   & 0.026                    & 0.438                       & 0.012                    & 0.062                    & 0.717                       & 0.000                    & 0.026                    & 0.442                       & -0.003                   & 0.028                    & 0.437                       & -0.003                   & 0.001                    & 0.300                       \\
	$\alpha$(Primary)   & 0.000                    & 0.048                    & 0.465                       & 0.006                    & 0.095                    & 0.763                       & -0.002                   & 0.046                    & 0.442                       & -0.001                   & 0.049                    & 0.437                       & 0.002                    & 0.002                    & 0.381                       \\
	$\alpha$(Age5)      & 0.000                    & 0.046                    & 0.372                       & 0.016                    & 0.092                    & 0.650                       & 0.004                    & 0.050                    & 0.415                       & -0.002                   & 0.047                    & 0.377                       & 0.000                    & 0.003                    & 0.325                       \\
	$\alpha$(Age13)     & 0.001                    & 0.039                    & 0.518                       & 0.003                    & 0.082                    & 0.787                       & 0.000                    & 0.041                    & 0.542                       & 0.000                    & 0.038                    & 0.464                       & 0.002                    & 0.002                    & 0.411                       \\
	$\alpha$(Age17)     & 0.005                    & 0.039                    & 0.400                       & -0.012                   & 0.085                    & 0.732                       & 0.002                    & 0.041                    & 0.463                       & 0.005                    & 0.038                    & 0.361                       & 0.005                    & 0.002                    & 0.400                       \\
	$\alpha$(Age23)     & 0.001                    & 0.040                    & 0.412                       & -0.015                   & 0.084                    & 0.674                       & -0.001                   & 0.045                    & 0.463                       & 0.001                    & 0.039                    & 0.379                       & 0.000                    & 0.002                    & 0.364                       \\ \hline
	$\beta$(Intercept)  & 0.049                    & 0.069                    & 0.268                       & 0.027                    & 0.124                    & 0.719                       & 0.012                    & 0.056                    & 0.470                       & 0.020                    & 0.107                    & 0.634                       & 0.008                    & 0.003                    & 0.341                       \\
	$\beta$(Visits)     & 0.019                    & 0.047                    & 0.411                       & 0.079                    & 0.121                    & 0.367                       & 0.016                    & 0.049                    & 0.451                       & 0.004                    & 0.051                    & 0.506                       & 0.007                    & 0.002                    & 0.368                       \\
	$\beta$(Fluoride)   & 0.010                    & 0.038                    & 0.396                       & 0.041                    & 0.079                    & 0.474                       & 0.011                    & 0.041                    & 0.411                       & 0.002                    & 0.043                    & 0.443                       & 0.002                    & 0.002                    & 0.322                       \\
	$\beta$(Brushing)   & 0.007                    & 0.039                    & 0.483                       & 0.017                    & 0.070                    & 0.695                       & 0.003                    & 0.041                    & 0.491                       & 0.002                    & 0.045                    & 0.463                       & 0.004                    & 0.002                    & 0.371                       \\
	$\beta$(Beverages)  & -0.015                   & 0.033                    & 0.457                       & -0.054                   & 0.092                    & 0.447                       & -0.008                   & 0.035                    & 0.516                       & -0.004                   & 0.041                    & 0.557                       & -0.001                   & 0.001                    & 0.379                       \\
	$\beta$(Pre-molar)  & -0.011                   & 0.052                    & 0.486                       & -0.075                   & 0.115                    & 0.350                       & -0.012                   & 0.051                    & 0.461                       & 0.001                    & 0.067                    & 0.534                       & -0.004                   & 0.003                    & 0.350                       \\
	$\beta$(Incisor)    & 0.012                    & 0.048                    & 0.496                       & 0.074                    & 0.117                    & 0.375                       & 0.006                    & 0.048                    & 0.472                       & 0.002                    & 0.068                    & 0.541                       & -0.003                   & 0.003                    & 0.377                       \\
	$\beta$(Canine)     & 0.044                    & 0.062                    & 0.295                       & 0.106                    & 0.149                    & 0.301                       & 0.013                    & 0.054                    & 0.424                       & 0.008                    & 0.071                    & 0.503                       & 0.008                    & 0.003                    & 0.337                       \\
	$\beta$(Primary)    & -0.025                   & 0.080                    & 0.443                       & -0.173                   & 0.211                    & 0.159                       & -0.033                   & 0.078                    & 0.405                       & -0.016                   & 0.121                    & 0.576                       & -0.010                   & 0.006                    & 0.361                       \\
	$\beta$(Age5)       & 0.023                    & 0.064                    & 0.439                       & 0.082                    & 0.121                    & 0.306                       & 0.032                    & 0.064                    & 0.412                       & 0.009                    & 0.098                    & 0.610                       & 0.010                    & 0.004                    & 0.378                       \\
	$\beta$(Age13)      & 0.039                    & 0.079                    & 0.414                       & 0.122                    & 0.168                    & 0.268                       & 0.003                    & 0.071                    & 0.461                       & -0.005                   & 0.107                    & 0.519                       & 0.005                    & 0.007                    & 0.364                       \\
	$\beta$(Age17)      & -0.007                   & 0.058                    & 0.487                       & -0.006                   & 0.082                    & 0.826                       & -0.013                   & 0.056                    & 0.457                       & -0.013                   & 0.084                    & 0.529                       & -0.006                   & 0.004                    & 0.338                       \\
	$\beta$(Age23)      & -0.010                   & 0.059                    & 0.449                       & -0.046                   & 0.096                    & 0.490                       & -0.013                   & 0.057                    & 0.449                       & -0.009                   & 0.085                    & 0.499                       & -0.007                   & 0.005                    & 0.357                       \\
	$\log \phi$         & -0.013                   & 0.092                    & 0.536                       & -0.004                   & 0.225                    & 0.784                       & -0.011                   & 0.091                    & 0.517                       & 0.032                    & 0.206                    & 0.640                       & -0.015                   & 0.008                    & 0.420                       \\ \hline
\end{tabular}
\caption{Average bias, RMSE and $\OECS$ for $\btheta_M$ from different sampling strategies. The bandwidth for ABC-MCMC is chosen to be 10.}
\label{tbl:sim_compare_M}
\end{table}
\end{landscape}

\begin{landscape}
\begin{table}[!tb]
\centering
\scriptsize
\begin{tabular}{lcccccccccccc}
	\hline
	& \multicolumn{3}{c}{ABC-MCMC} & \multicolumn{3}{c}{ABC-MCMC + Regression} & \multicolumn{3}{c}{ABC-Rej + Regression} & \multicolumn{3}{c}{ABC-Imp + Regression} \\
	Parameters           & Bias     & RMSE   & $\OECS$  & Bias         & RMSE        & $\OECS$      & Bias         & RMSE       & $\OECS$      & Bias         & RMSE       & $\OECS$      \\ \hline
	$\btheta_D$ elements &          &        &          &              &             &              &              &            &              &              &            &              \\
	$\rho_t$             & -0.005   & 0.028  & 0.716    & -0.002       & 0.012       & 0.463        & -0.081       & 0.125      & 0.229        & -0.001       & 0.000      & 0.349        \\
	$\rho_{ct}$          & -0.001   & 0.002  & 0.684    & 0.000        & 0.001       & 0.475        & -0.346       & 0.412      & 0.000        & 0.000        & 0.000      & 0.552        \\ \hline
	$\btheta_R$ elements       &          &        &          &              &             &              &              &            &              &              &            &              \\
	((7,13),(8,13))      & -0.022   & 0.082  & 0.620    & 0.004        & 0.031       & 0.557        & -0.197       & 0.197      & 0.000        & 0.008        & 0.001      & 0.421        \\
	((7,13),(9,13))      & -0.022   & 0.082  & 0.620    & 0.004        & 0.031       & 0.557        & -0.197       & 0.197      & 0.000        & 0.008        & 0.001      & 0.421        \\
	((7,13),(14,13))     & -0.022   & 0.082  & 0.620    & 0.004        & 0.031       & 0.557        & -0.197       & 0.197      & 0.000        & 0.008        & 0.001      & 0.421        \\
	((7,13),(7,17))      & -0.022   & 0.079  & 0.670    & 0.000        & 0.033       & 0.488        & -0.156       & 0.286      & 0.474        & 0.003        & 0.001      & 0.386        \\
	((7,13),(7,23))      & -0.008   & 0.042  & 0.649    & 0.004        & 0.019       & 0.509        & 0.185        & 0.291      & 0.401        & 0.006        & 0.001      & 0.406        \\
	((7,13),(8,17))      & -0.018   & 0.066  & 0.582    & 0.003        & 0.028       & 0.561        & -0.119       & 0.120      & 0.000        & 0.005        & 0.001      & 0.419        \\
	((7,13),(26,13))     & -0.022   & 0.082  & 0.620    & 0.004        & 0.031       & 0.557        & -0.197       & 0.197      & 0.000        & 0.008        & 0.001      & 0.421        \\
	((14,13),(19,13))    & -0.022   & 0.082  & 0.620    & 0.004        & 0.031       & 0.557        & -0.197       & 0.197      & 0.000        & 0.008        & 0.001      & 0.421        \\
	((2,13),(31,13))     & -0.022   & 0.082  & 0.620    & 0.004        & 0.031       & 0.557        & -0.197       & 0.197      & 0.000        & 0.008        & 0.001      & 0.421        \\
	((2,13),(31,17))     & -0.018   & 0.066  & 0.582    & 0.003        & 0.028       & 0.561        & -0.119       & 0.120      & 0.000        & 0.005        & 0.001      & 0.419        \\
	((2,13),(31,23))     & -0.008   & 0.036  & 0.574    & 0.003        & 0.017       & 0.568        & -0.055       & 0.056      & 0.000        & 0.004        & 0.000      & 0.419        \\
	((3,13),(30,13))     & -0.022   & 0.082  & 0.620    & 0.004        & 0.031       & 0.557        & -0.197       & 0.197      & 0.000        & 0.008        & 0.001      & 0.421        \\
	((3,13),(30,17))     & -0.018   & 0.066  & 0.582    & 0.003        & 0.028       & 0.561        & -0.119       & 0.120      & 0.000        & 0.005        & 0.001      & 0.419        \\
	((3,13),(30,23))     & -0.008   & 0.036  & 0.574    & 0.003        & 0.017       & 0.568        & -0.055       & 0.056      & 0.000        & 0.004        & 0.000      & 0.419        \\
	((J,5),(I,5))        & -0.015   & 0.053  & 0.654    & 0.000        & 0.017       & 0.543        & -0.151       & 0.152      & 0.000        & 0.002        & 0.000      & 0.404        \\
	((J,5),(H,5))        & -0.015   & 0.053  & 0.654    & 0.000        & 0.017       & 0.543        & -0.151       & 0.152      & 0.000        & 0.002        & 0.000      & 0.404        \\
	((J,5),(J,9))        & -0.020   & 0.069  & 0.684    & -0.003       & 0.027       & 0.475        & -0.152       & 0.270      & 0.433        & 0.000        & 0.001      & 0.363        \\
	((J,5),(I,9))        & -0.015   & 0.050  & 0.581    & 0.000        & 0.019       & 0.550        & -0.100       & 0.101      & 0.000        & 0.001        & 0.000      & 0.398        \\
	((J,5),(K,5))        & -0.015   & 0.053  & 0.654    & 0.000        & 0.017       & 0.543        & -0.151       & 0.152      & 0.000        & 0.002        & 0.000      & 0.404        \\
	((J,5),(K,9))        & -0.015   & 0.050  & 0.581    & 0.000        & 0.019       & 0.550        & -0.100       & 0.101      & 0.000        & 0.001        & 0.000      & 0.398        \\
	((I,5),(14,9))       & -0.015   & 0.050  & 0.581    & 0.000        & 0.019       & 0.550        & -0.100       & 0.101      & 0.000        & 0.001        & 0.000      & 0.398        \\
	((B,9),(3,9))        & -0.022   & 0.077  & 0.622    & 0.002        & 0.027       & 0.551        & -0.191       & 0.191      & 0.000        & 0.005        & 0.001      & 0.416        \\
	((B,5),(3,9))        & -0.015   & 0.050  & 0.581    & 0.000        & 0.019       & 0.550        & -0.100       & 0.101      & 0.000        & 0.001        & 0.000      & 0.398        \\
	((D,5),(8,9))        & -0.015   & 0.050  & 0.581    & 0.000        & 0.019       & 0.550        & -0.100       & 0.101      & 0.000        & 0.001        & 0.000      & 0.398        \\ \hline
\end{tabular}
\caption{Average bias, RMSE and $\OECS$ for $\btheta_D$ and $\btheta_R$ from different sampling strategies. The bandwidth for ABC-MCMC is chosen to be 10.}
\label{tbl:sim_compare_R}
\end{table}
\end{landscape}

\begin{figure}
\centering
\includegraphics[width=\linewidth]{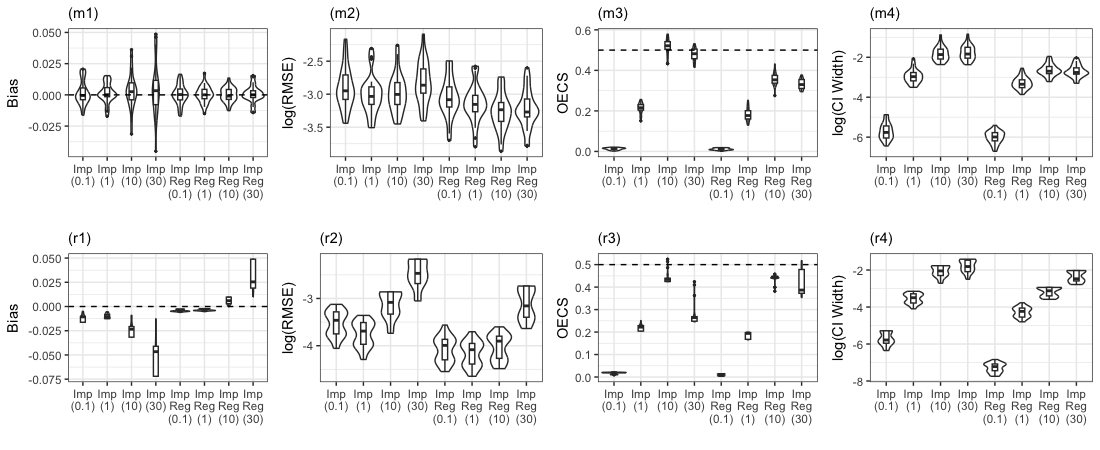}
\caption{Comparisons of performances of different ABC importance samplers across marginal parameters in panels (m1-m4) and across correlation parameters in panels (r1-r4).}
\label{fig:importance_compare_sim}
\end{figure}

To understand sensitivity of the ABC importance sampler to the bandwidth choice, we now compare its estimation performance without/with regression adjustment for different choices of $h$.
Figure \ref{fig:importance_compare_sim} illustrates that, as the bandwidth increases, so does the bias, caused by the larger error in posterior approximation.
% This is due to the fact that as bandwidth decreases, the importance weights become more unbalanced.
% \org{[I don't see how this is driving the bias.  I would assume that the increased bias is due to the differences between the true and ABC posteriors.]}
Investigating the importance samples with smaller bandwidths $h=0.1$ and 1 for a few of the simulated datasets, we observed that the point estimate of $\btheta$ is dominated by fewer than 4 out of the 250,000 posterior samples, accounting for more than 95\% of the total weight.
Consequently, the credible intervals will be unreasonably narrow to be concentrated on these few points, leading to poor coverage, even if the point estimate is close to the target.
Although larger bandwidths resulted in better overall coverage for the marginal parameters, this behavior was not consistent with the correlation parameters.
% To summarize, while the importance sampler with smaller bandwidths were completely impractical, larger bandwidths generate more usable samples which coupled with regression adjustment achieved improved bias, RMSE, and reasonable OEC scores.
% \org{[this sounds like we think this is an acceptable method.  Try something like:]} 
To summarize,  the importance sampler with small bandwidths is found to be completely impractical, as the posterior is based on a small collection of sample.  While using a large bandwidth spreads the posterior weight across a larger collection of samples, the overall coverage is consistently below the normal rate after regression and the overall performance is inferior to the ABC-MCMC estimation approach, as discussed in Section \ref{sec:simulation}.

\subsection{Comparison of Regression Adjustment Strategies} \label{appn:sim_compare_regression_strategies}

We recall from Section \ref{sec:regression_adjustment} that the posterior samples $(\btheta^{(g)}, \bss^{(g)})$ generated by the ABC-MCMC algorithm will have many repetitions. The design matrix $\bS$ consisting of the $\bss^{(g)}$ as rows is such that $\bS'\bS$ is non-singular, which makes fitting the regression model problematic.
% \org{[Need to provide some more details to remind the reader how you get to this issue.  Not clear what $\bS$ is here or why it matters.]}
We have avoided this issue by only using the unique samples from ABC-MCMC when fitting the regression model. 
We apply applying the regression adjustment to the (unweighted) sample of unique values and then reweighting them according to their frequency in the original ABC-MCMC chain.
However, this strategy may not be optimal since we fit the heteroskedastic, locally linear regression model without using the weights from MCMC sampling (the abc function in the R library does not accomodate weighted samples).

% We now describe an alternative strategy that can bypass the problem of singularity.
We describe an alternative strategy that introduces noise to the original MCMC samples (with repeated values) to produce a sample with non-singular $\bS'\bS$,
% We add a small amount of random noise to the $\bS$ matrix 
% which is small enough that
but such  that it does not affect the relationship between $\btheta$ and $\bss$.
To that end, let $\sigma_{s_k}^2$ denote the variance of the $k$-th summary statistic obtained from the full set of posterior samples $\bss^{(g)}$ including replications.
The $(g,k)$-th element $\tilde{s}^{(g)}_k$ of the modified design matrix $\widetilde{\bS}$ is obtained as $\tilde{s}^{(g)}_k = s^{(g)}_k + \varepsilon_k$, where $\varepsilon_k \sim \text{Unif}(-0.001\sigma_{s_k}, 0.001\sigma_{s_k})$.
% \org{[Is the endpt $\sigma$ or $0.01\sigma$? or something else along those lines?]}
We perform regression adjustment with the modified design matrix $\widetilde{\bS}$.
% Note that, when the number of unique accepted samples is very small, we cannot perform regression adjustment (based on heteroscedastic local linear regression) with our main approach.
% The noise based alternative strategy can still be used in this case, which nevertheless may turn out to be not very helpful due to the lack of available information.

\begin{figure}[!tb]
\centering
\includegraphics[width=\textwidth]{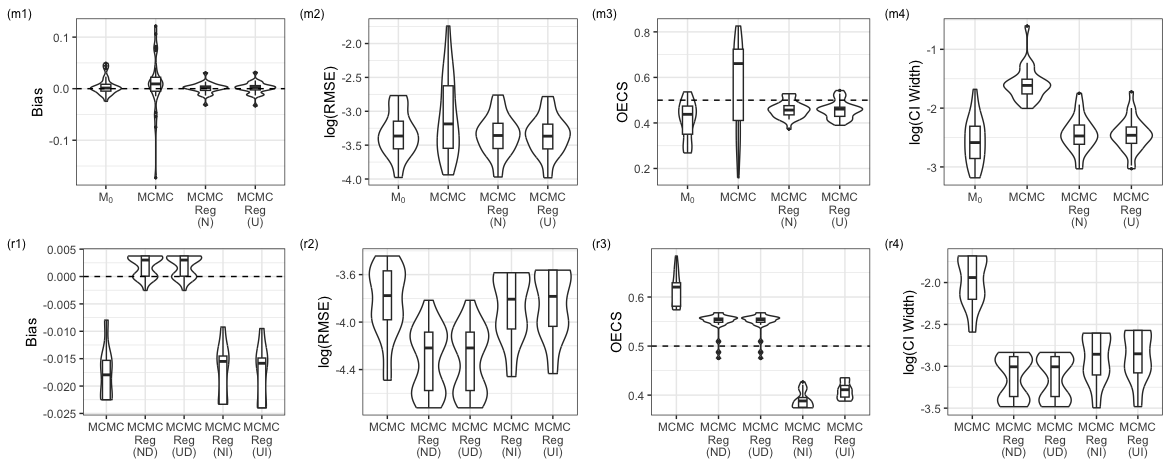}
\caption{Comparisons of estimation of $\btheta_M$ (in panels m1-m4) and $\btheta_\bR$ (in panels r1-r4) across regression strategies. `N' within braces indicates regression adjustment considering only unique samples, while `U' indicates regression with addition of noise. Additionally for $\btheta_\bR$, `D' and `I' stand for direct and indirect regression adjustment for the correlation coefficients, respectively.}
\label{fig:compare_regressions}
% \label{fig:compare_regressions_R}
\end{figure}
\FloatBarrier

The two regression adjustment strategies are now compared under our simulation setting.
We use \textit{Reg(U)} to represent 
our primary regression strategy of fitting the
regression based on the unique samples, and \textit{Reg(N)} represents regression adjustment with addition of noise.
In the case of the marginal parameters $\btheta_M$, we compare estimation performance against the misspecified model $\model{0}$, considered as the baseline, and with the data-generating model $\model{4}$ using ABC-MCMC.
The correlation coefficients $\btheta_\bR$ can be adjusted in two ways, as discussed earlier in Section \ref{appn:regression_adjustment}: (D) direct adjustment of each correlation coefficient treating each component as a univariate estimand, or (I) indirect adjustment, achieved by adjusting $\btheta_D$ samples and using those to estimate the corresponding samples for the correlation coefficients.
Note that the results reported previously in Section \ref{sec:simulation} are based on the (D) direct adjustment.
% For each of these approaches, we perform \textit{Reg(U)} and \textit{Reg(N)} regression strategies.
% These four regression strategies are compared against samples from $\model{0}$ and those obtained using ABC-MCMC.

% \org{[Combine into one figure with two rows.  Also, remove M0 form corr plots.  Also are these log-RMSE or RMSE?  The ylim suggests maybe no log.]}
Figure \ref{fig:compare_regressions}(m1-m4) shows that both the regression adjustment approaches yield very similar bias, RMSE, overall empirical coverage score (OECS), as well as credible interval width, for the marginal parameters.  Using either is clearly  superior in comparison to the unadjusted ABC samples and the independence model $\model{0}$.
% Inspite of being generated from an approximately correct target posterior, the regression-adjusted samples perform at per with those from $\model{0}$, which are Gibbs samples from the exact posterior (although with misspecified dependence structure).
Investigating the results for the correlation coefficients in Figure \ref{fig:compare_regressions} panels (r1-r4), we again observe minimal difference between the unique sample adjustment and its noisy version.
However, bias and RMSE for the direct adjustment are meaningfully better than those for the indirect approach.
The OEC scores are also noticably better under the direct adjustment, while the indirect approach shows under-coverage, with many OEC scores lying below 0.4.
The credible intervals also seem to be narrower, showing superiority of the direct adjustment strategy.
To understand this better, we recall from Appendix \ref{appn:regression_adjustment} that indirect adjustment of $\btheta_D$ is challenging, as the support $\bTheta_D$ is not rectangular, requiring an additional post-regression adjustment step.
Many of the regression-adjusted $\btheta_D$ samples fall outside the support, forcing shrinkage toward the corresponding unadjusted samples.
The resulting bias for the indirectly adjusted samples resembles that for the ABC samples.
This shrinkage also appears to cause undercoverage.
% \org{[Can we provide more intuition about why?  Presumably with (I) we struggle to adjust within $\bTheta_D$.  Maybe that means we end up with $\btheta_D$ biased towards zero, yieling correlations closer to zero (negative bias) and intervals shifted towards zero (undercovers with larger intervals).  But not sure if this is right.]}

Ultimately, this investigation suggests that, at our optimal bandwidth $h=10$, our original regression strategy based on using only the unique samples from ABC-MCMC is basically equivalent to performing adjustment on the full sample after adding random noise.  
Investigating the differences in regression adjustment strategies using other bandwidths yields similar results under regression adjustment based on the unique sample or based on the noise-added samples.
Further, we note that the indirect adjustment which was designed to respect the complex geometry of $\bTheta_D$ yields meaningfully worse estimation compared to applying direct univariate transformations for each parameter.  
Strategies for multivariate adjustments of non-rectangular parameter spaces requires further work in the future.

\FloatBarrier
\pagebreak

\subsection{SAR Model Misspecification} \label{appn:sim_model_misspecification}

\begin{figure}[!tb]
\centering
\includegraphics[width=0.9\textwidth]{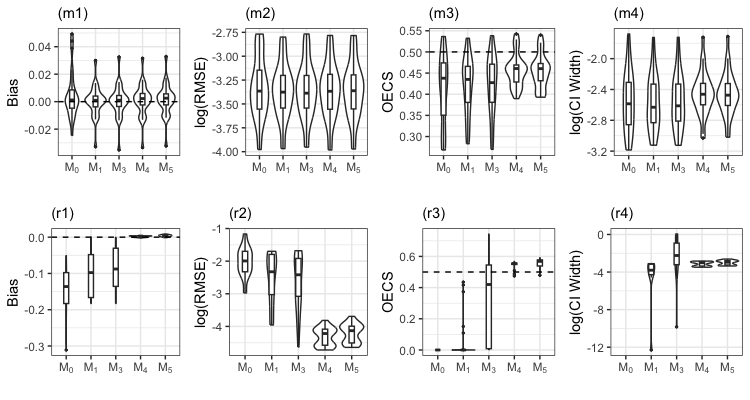}
\caption{Comparisons of $\btheta_M$ and $\btheta_\bR$ across $\model{0}$, $\model{1}$, $\model{3}$, $\model{4}$ and $\model{5}$. Recall $\model{4}$ is the true data-generating model.
% \org{[FIX FIGURE LABEL!!!!!!!!!!!]}
}
\label{fig:compare_models_sim}
% \includegraphics[width=\textwidth]{diagnostics_sim_R}
% \caption{Compare $\bR$ across $\model{1}$, $\model{2}$, $\model{3}$ and $\model{4}$.}
% \label{fig:compare_sim_models_R}
\end{figure}

To investigate the sensitivity of the model fit to the dependence model specification, we have fit a subset of alternative model choices to the simulation data sets, including $\model{1}$, $\model{3}$, $\model{4}$ and $\model{5}$ described in Table \ref{tbl:dependence_structures} (also Table \ref{tbl:acceptance_rates_IFS}).
Recall that the true data-generating model is $\model{4}$, so analysis under these choices illustrates the impact of misspecifying the SAR form.  Of note, the  data-generating model $\model{4}$ is nested within $\model{5}$, so we anticipate $\model{5}$ to give reasonable performance as well.
We compute the same estimation performance metrics and compare against  the true model $\model{4}$ shown in Figure \ref{fig:compare_models_sim}.
% \org{[combine into one figure.]}

It is important to first note that all the models share the same marginal structure and, therefore, are expected to perform similarly in $\btheta_M$.
We observe in Figure \ref{fig:compare_models_sim} panels (m1-m4) that the biases across all the parameters are not very sensitive to the dependence model.
The RMSEs are also similar across all the parameters.
% The median RMSE over all the parameters is, however, slightly higher for the true model compared to $\model{0}$, $\model{1}$ and $\model{2}$.
In terms of the overall empirical coverage score (OECS), the true model performs fairly well as expected, as does $\model{5}$, with the scores across all the parameters close to the target 0.5.  Models $\model{0}$, $\model{1}$ and $\model{3}$ are farther away from the target, with somewhat higher variability among the scores.
% The credible intervals for $\model{1}$ and $\model{3}$ turn out to be too narrow to provide the correct level of coverage, compared to the model $\model{4}$.
It is not surprising that $\model{5}$ would have quite similar performance to  $\model{4}$ since this true data-generating model is contained within the parameter space of $\model{5}$.
Overall, estimation of the marginal parameters do not appear to be very sensitive to the dependence model.

Regarding the correlation parameters in Figure \ref{fig:compare_models_sim} panels (r1-r4), 
% $\model{0}$ with independence misspecification performs worst among all the models.
the true model $\model{4}$ estimates all the correlation parameters we considered with approximately zero bias, indicating that correlation estimation is accurate when the data are fit to the true model.
% while the bias is non-zero in other models and vary considerably across parameters.
It is not surprising that bias for $\model{5}$ is close to zero, while we have meaningful bias under $\model{1}$ and $\model{3}$, as there is no parameter value within the parameter space of these models that can capture the $\btheta_\bR$ from the data-generating model. 
Regardless, we do see closer estimation from these incorrect SAR models than under $\model{0}$, which is ``estimating'' all correlations to be zero.
We see similar trends in RMSE and OECS, with the true $\model{4}$ having best performance and the other SAR models beating the independence model.
We observe slightly higher RMSE and overall coverage rate for $\model{5}$ due to the excess variability introduced by the extra parameter $\rho_{h}$.
%corresponding to the horizontal connectivity.
Note that the 95\% equal-tailed credible interval for $\rho_h$ covered the true value ($\rho_h = 0$) in 93 out of 100 simulated datas sets.
% \org{[Can you say something about estimation of $\rho_h$ in M5?  It is correctly estimated to zero?  In how many of the 100 datasets does its interval contain zero?]}
% The RMSE is also least for $\model{4}$ among all the models, with $\model{2}$ having the widest variability across parameters.
% The OEC scores are also very close to the target 0.5 in the case of the true model, whereas the rest of the models provide much poor coverage for the correlation parameters.
% Since all the correlation elements in $\bR$ are zero in the case of $\model{0}$, we don't have a credible interval width.
% In terms of credible interval widths, recall that $\model{3}$ yields an equicorrelation structure, so all elements of $\bR$ have the same estimate and the same interval. 
Model $\model{3}$ involves four parameters for the different adjacency relationships and is the most complex of the models. 
With these additional parameters, we see very large differences in the interval widths across the correlations considered. 
It is unclear whether this is associated with inconsistencies between fitting $\model{3}$ when data are generated under $\model{4}$ or whether this is associated with estimation of additional $\rho_k$s.
% Overall, $\model{4}$ and $\model{5}$ provide best performance while $\model{4}$ does so with parsimony.
Overall, the models that contain the data generation ($\model{4}$ and $\model{5}$) accurately estimate the marginal models and the dependence, while all SAR models have acceptable estimation of the $\btheta_M$ parameters even under misspecification of the dependence.
% In case of $\model{3}$, on the other hand, all the correlation parameters are defined in terms of a univariate $\btheta$ and therefore have the same estimate.
% The resulting credible intervals will also have the same width across all the parameters.
% Model $\model{2}$ again have the widest variable credible interval widths implying poor estimation.

% \org{[Need more here.  What is this trying to show?  Connect it to IFS discussion from B.3.  ]}

\begin{figure}
\centering
\includegraphics[width=0.8 \textwidth]{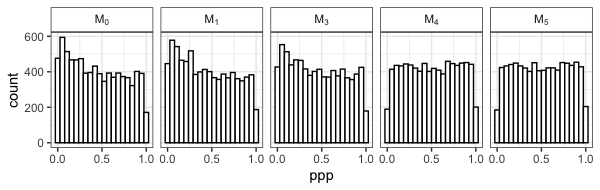}
\caption{Histogram of all available $ppp$-values across all model choices.}
\label{fig:ppp_histogram_sim}
\end{figure}
To conclude this section, we now briefly describe the process of using the $ppp$ values for model comparison across SAR choices as was utilized for model selection in the IFS case (see Appendix \ref{appn:IFS_modeling}).  
While we show these posterior predictive checks for only one of the simulated datasets, we note that the behavior is similar across the other simulated datasets used in the simulation.
% the checks for the rest of the datasets look very similar.
The histograms of $ppp$ values across the complete set of Spearman correlations are shown in Figure \ref{fig:ppp_histogram_sim}.
We see that $\model{0}$ has the highest skewness toward zero, which is expected due to its independence misspecification.
As model $\model{1}$ assumes independence among vertically-adjacent tooth pairs or primary/permanent teeth, it also involves many exact zeros in the correlation matrix and is inadequate for modeling the dependence of model $\model{4}$. This is apparent from the higher concentration of $ppp$ near zero. 
While $\model{3}$ does not have exact zeros in its inverse, it is still sparse and a misspecified model here, as indicated by another peak near zero.  In contrast, the true data-generating model $\model{4}$ and the model encompassing the truth $\model{5}$ have no evidence of inadequency, as the distribution of $ppp$ is not skewed toward zero.  If performing model selection here, one would immediately limit attention to $\model{4}$ and $\model{5}$ as the two best performing models; as $\model{4}$ is more parsimonious, it is likely to be the model that would be chosen.
Thus, following the same strategy as was used in the IFS data analysis, we have determined the true data-generating model to be the best model and  the model to use for inference in this simulated example.
% Note that the $ppp$ histograms in the IFS data analysis (see Figure \ref{fig:ppp_histogram}) also appeared to have some degree of a  peak near zero for most of the models indicating adequacy.
% While the best model $\model{8}$ for IFS data still showed some imbalance in the distribution in Figure \ref{fig:ppp_histogram}, in these simulation experiments (Figure \ref{fig:ppp_histogram_sim}) both models $\model{4}$ and $\model{5}$, appear to achieve much more balanced distribution of the $ppp$ values.
% This further emphasizes the fact that under our modeling setup, an adequate model would be expected to have balanced $ppp$ distribution which strengthens the arguments we made in Section \ref{appn:IFS_modeling} in favor of choosing $\model{8}$ as the best fit for IFS data.

% \begin{figure}[!bt]
%   \centering
%   \includegraphics[width=\textwidth]{pp_R_plots_sim.png}
%   \caption{Comparing posterior predictive plots for $\bR$ elements from different models based on a simulated data. The title of each plot shows the $ppp$ values for the corresponding summary statistic obtained from $\model{1}$, $\model{2}$, $\model{3}$ and $\model{4}$ consecutively and the horizontal line indicates the observed summary statistics. \anish{(plot will be updated)}}
%   \label{fig:pp_R_sim}

\end{document}